\documentclass[fleqn,usenatbib]{mnras}

\usepackage{newtxtext,newtxmath}
\usepackage[T1]{fontenc}

\DeclareRobustCommand{\VAN}[3]{#2}
\let\VANthebibliography\thebibliography
\def\thebibliography{\DeclareRobustCommand{\VAN}[3]{##3}\VANthebibliography}

\usepackage{graphicx}	% Including figure files
\usepackage{amsmath}	% Advanced maths commands
\usepackage{bm}

\title[Polarized Thomson-slab transport]{Polarized transport in a Thomson slab: Stokes recursion, asymptotic eigenmodes, and Comptonization applications}

\author[H. Xu]{
Haichao Xu,$^{1}$\thanks{E-mail: Haichao\_Xu@zju.edu.cn}
\\
$^{1}$Institute for Astronomy, School of Physics, Zhejiang University, 866 Yuhangtang Rd, Hangzhou 310058, People’s Republic of China
}

\date{Accepted XXX. Received YYY; in original form ZZZ}

\pubyear{\the\year{}}

\begin{document}
\label{firstpage}
\pagerange{\pageref{firstpage}--\pageref{lastpage}}
\maketitle

\begin{abstract}
We develop a scattering-order-resolved Stokes-vector recursion for axisymmetric polarized transport in a finite plane-parallel Thomson slab. The post-scattering radiation field closes exactly within two depth-dependent coefficients, yielding escape probabilities, angular distributions and linear polarization while conserving photon number. At high scattering orders, the principal eigenmode of the weighted self-adjoint transport operator controls the escape statistics, and the normalized angular distribution and polarization become independent of the seed source. The subdominant spectrum quantifies convergence, with the characteristic scattering order scaling as the square of the optical depth in thick slabs. We derive logarithmic corrections to the thin-slab complete-polarization limit away from the slab normal, together with a finite-optical-depth Galerkin approximation. In optically thick slabs, matching bulk diffusion to a polarized Milne surface layer yields finite-thickness corrections to the Chandrasekhar solution. We provide compact formulae for the principal and leading subdominant eigenvalues and the asymptotic angular and polarization distributions across both regimes. Combining the recursion with an angle-averaged thermal-Compton redistribution kernel gives approximate energy-resolved Stokes spectra. Comparisons with full polarized Compton calculations show that this separable reconstruction is most reliable for cool, optically thick media. Applied to the non-dipping observations of neutron-star low-mass X-ray binary system 4U 1624–49, the model provides acceptable joint fits to the broadband intensity and \textit{IXPE} Stokes spectra, showing that an extended corona overlying the accretion disc can account for the observed X-ray polarization.
\end{abstract}

\begin{keywords}
polarization -- radiative transfer -- scattering -- X-rays: binaries -- stars: neutron
\end{keywords}

%%%%%%%%%%%%%%%%%%%%%%%%%%%%%%%%%%%%%%%%%%%%%%%%%%

%%%%%%%%%%%%%%%%% BODY OF PAPER %%%%%%%%%%%%%%%%%%

\section{Introduction}
\label{sec:introduction}

Inverse Compton scattering of soft photons by energetic electrons is one of the fundamental radiative processes in high-energy astrophysics, and is widely invoked to explain the hard X-ray emission from accreting compact objects \citep[e.g.,][]{1979rpa..book.....R, 1983ASPRv...2..189P}. In compact accretion systems, seed photons originating from the accretion disc, neutron-star surface or boundary layer, and other radiation sources may undergo multiple scatterings before escaping from the Comptonizing medium \citep[e.g.,][]{1977ApJ...218..247L, 1980A&A....86..121S, 1988ApJ...324..363W, 1989PASJ...41...97M}. The emergent radiation is determined not only by the energy exchange in individual scattering events, but also by the propagation and angular redistribution of photons through the scattering medium \citep[e.g.,][]{1980A&A....86..121S, 1994ApJ...434..570T, 1996ApJ...470..249P}. In particular, the polarization produced by scattering is sensitive to the geometry and anisotropy of the radiation field, offering an additional diagnostic of the Comptonization process \citep[e.g.,][]{2010ApJ...712..908S, 2017ApJ...850...14B}.

The advent of the Imaging X-ray Polarimetry Explorer \citep[\textit{IXPE},][]{2022JATIS...8b6002W} has opened a new observational window on the geometry of Comptonizing regions in accreting compact objects. \textit{IXPE} enables energy-resolved measurements of the Stokes $I$, $Q$, and $U$ spectra, from which the polarization degree and position angle can be reconstructed across the X-ray band. Recent X-ray polarimetric observations of black hole X-ray binaries (XRBs) and active galactic nuclei (AGNs)  favour a spatially extended geometry in the accretion disc plane for the X-ray emitting region, rather than a compact spherical geometry \citep[e.g.,][]{2022Sci...378..650K, 2023MNRAS.523.4468G}. The sandwich disc--corona model provides a natural theoretical basis for this geometry, with a hot Comptonizing layer extending above the cold accretion disc \citep[e.g.,][]{1991ApJ...380L..51H, 1993ApJ...413..507H}.

For neutron-star XRBs, the interpretation of the X-ray emission and polarization is generally more complicated because multiple components may contribute simultaneously, including the disc, stellar surface, boundary layer, Comptonizing region, and reflection \citep[e.g.,][]{2007ApJ...667.1073L, 2007A&ARv..15....1D}. Nevertheless, \textit{IXPE} observations of several neutron-star systems have revealed polarization at the level of a few per cent \citep[e.g.,][]{2023MNRAS.519.3681F, 2023ApJ...953L..22D}. If a substantial fraction of this polarization originates in the Comptonized emission, it would favour an anisotropic Comptonizing region and make flattened geometries physically plausible. A plane-parallel Comptonizing layer therefore provides a useful candidate geometry for exploring the polarized transfer in these systems.

Theoretically, a Comptonized continuum can be decomposed into a superposition of contributions from photons of different orders \citep[e.g.,][]{1996ApJ...470..249P, 2000ApJ...536..788K}. Different scattering orders trace different stages of the radiative-transfer process and contribute differently to the emerging spectrum. The low-order components remain sensitive to the geometry and angular distribution of the seed photons, whereas photons surviving to high scattering orders progressively lose this memory and approach a transport eigenstate \citep[e.g.,][]{1985A&A...143..374S, 2026MNRAS.550g1392X}. A scattering-order-resolved description provides direct information on how the angular distribution and polarization of the escaping radiation are progressively established through repeated scatterings.

Polarized radiative transfer in plane-parallel scattering media has long been studied. The classical semi-infinite electron-scattering atmosphere was solved by \citet{1960ratr.book.....C}. For finite slabs, \citet{1985A&A...143..374S} developed a successive-scattering treatment in the Thomson limit and investigated the angular distribution and polarization at individual scattering orders, including their high-order behaviour. \citet{1996ApJ...470..249P} subsequently developed a fully energy-, angle-, and polarization-dependent iterative Compton transfer scheme in which the Stokes radiation field is evolved explicitly order by order. More recently, \citet{2025A&A...704A..87T} revisited the high-scattering-order slab-polarization solutions of \citet{1985A&A...143..374S} and combined their inclination- and optical-depth-dependent polarization predictions with \textit{IXPE} measurements and X-ray spectral indices to constrain the properties of Comptonizing regions in black-hole systems.

\citet{2026MNRAS.550g1392X} (hereafter Paper I) developed a scattering-order-resolved recursive formalism for photon transport in a finite Thomson slab. By evolving the post-scattering depth--direction distribution, the method gives the boundary- and angle-resolved escape probabilities at each scattering order. The scalar recursion also revealed a dominant high-order eigenmode that controls both the asymptotic escape ratio and the limiting angular distribution. However, Paper I employed the polarization-averaged Thomson phase function, neglecting the feedback of scattering-induced polarization on subsequent intensity redistribution. 

In this work, we generalize the scalar recursion of Paper I to the axisymmetric linear-Stokes problem in a finite Thomson slab. The resulting formulation follows the intensity and polarization self-consistently from one scattering order to the next and provides a compact operator description of the high-order transport. We derive inverse-logarithmic corrections to the thin-slab polarization limit and a finite-optical-depth Galerkin approximation, and construct finite-thickness corrections to the Chandrasekhar solution by matching bulk diffusion to a polarized surface layer. The leading subdominant modes quantify the approach to the high-order state, while compact approximations for the principal and leading subdominant eigenvalues provide practical estimates of the associated escape and convergence behaviour.

We further combine the scattering-order-resolved transport with an angle-averaged thermal-Compton redistribution kernel to construct approximate energy-resolved Stokes spectra. Comparisons with full polarized Compton calculations assess the validity and limitations of this separation and show that agreement in the intensity spectrum alone does not guarantee an accurate polarization spectrum. We illustrate the application of the method through joint spectro-polarimetric modelling of the non-dipping observations of 4U 1624–49.

The paper is organized as follows. Section~\ref{sec:stokes_recursion} presents the Stokes recursion, Section~\ref{sec:high_order_polarization} analyses its high-order asymptotic behaviour, and Section~\ref{sec:band_polarization} discusses energy-resolved polarized Comptonization and the application to 4U~1624--49. Appendices~\ref{app:mcrt} and \ref{app:mean_number} describe the Monte Carlo implementation and scattering-number statistics, respectively. Appendix~\ref{app:galerkin_thin} derives the finite-optical-depth Galerkin approximation and evaluates its accuracy, while Appendix~\ref{app:surface_asymptotics} presents the half-space construction of the finite-thickness surface corrections.

\section{Stokes-vector scattering-order-resolved recursion in a slab medium}
\label{sec:stokes_recursion}

In the slab corona geometry, we assume that Compton scattering is confined predominantly to the local coronal region in which the photons are produced. In other words, before escaping, photons do not propagate horizontally into regions with substantially different physical conditions. Under this local-Comptonization assumption, radiative coupling between different coronal regions can be neglected, and we can idealize each local coronal region as a horizontally homogeneous, infinitely extended, plane-parallel slab, as commonly adopted in disc-corona models \citep{1991ApJ...380L..51H, 1993ApJ...413..507H}. The optical-depth coordinate $t\in[0,\tau]$ is measured upwards from the lower surface, such that $t=0$ and $t=\tau$ correspond to the lower and upper boundaries, respectively. The photon propagation direction is denoted by $\Omega=(\theta,\varphi)$, with $\mu=\cos\theta$; hence, $\mu>0$ and $\mu<0$ describe photons propagating towards the upper and lower surfaces, respectively. Free-escape conditions are imposed at both boundaries. If the local seed-photon source and the boundary conditions are also axisymmetric about the slab normal, the azimuthally averaged radiation field depends only on $t$ and $\mu$.

\subsection{General formalism for axisymmetric polarized scattering}

Polarized radiation considered here is generally described by the Stokes vector,
\begin{equation}
    \bm{S}=(I,Q,U,V)^{\rm T},
\end{equation}
where $I=I_\perp+I_\parallel$ and $Q=I_\perp-I_\parallel$. The perpendicular and parallel components are defined with respect to the local meridian plane spanned by the photon propagation direction and the slab normal. The corresponding linear polarization degree is
\begin{equation}
    \Pi_{\rm L}=\frac{\sqrt{Q^2+U^2}}{I}.
\end{equation}

For a general scattering geometry, the incident and outgoing Stokes vectors must first be transformed between their local meridian bases and the scattering-plane basis. Hereafter, unprimed and primed quantities denote the incident and outgoing photon directions, respectively. In the Thomson limit, the scattered Stokes vector is given by
\begin{equation}
    \bm{S}_{\rm out}(\Omega')
    =\int
    \mathbfss{R}(-\chi_{\rm out})
    \mathbfss{M}_{\rm T}(\Theta)
    \mathbfss{R}(\chi_{\rm in})
    \bm{S}_{\rm in}(\Omega)\,{\rm d}\Omega,
    \label{eq:full_pol_rec}
\end{equation}
where $\Theta$ is the scattering angle between the incident and outgoing photon directions, which satisfies
\begin{equation}
    \cos\Theta=\mu\mu'+
    \sqrt{1-\mu^2}\sqrt{1-\mu'^2}
    \cos(\varphi'-\varphi),
\end{equation}
and $\mathbfss{M}_{\rm T}(\Theta)$ is the Thomson Mueller matrix \citep{1960ratr.book.....C}. In a Stokes basis referenced to the scattering plane, the normalized Mueller phase matrix is
\begin{equation}
    \mathbfss{M}_{\rm T} (\Theta) = \frac{3}{16\pi}
\begin{pmatrix}
1+\cos^2\Theta & 1-\cos^2\Theta & 0 & 0\\
1-\cos^2\Theta & 1+\cos^2\Theta & 0 & 0\\
0 & 0 & 2\cos\Theta & 0\\
0 & 0 & 0 & 2\cos\Theta
\end{pmatrix}.
\end{equation}
In (\ref{eq:full_pol_rec}), $\chi_{\rm in}$ and $\chi_{\rm out}$ specify the rotations between the corresponding meridian planes and the scattering plane. Specifically, $\mathbfss{R}(\chi_{\rm in})$ transforms the incident Stokes vector from the incident local meridian basis to the scattering-plane basis, whereas $\mathbfss{R}(-\chi_{\rm out})$ transforms the scattered Stokes vector from the scattering-plane basis to the outgoing local meridian basis. The Stokes rotation matrix is defined by
\begin{equation}
    \mathbfss{R}(\chi)
    =
    \begin{pmatrix}
        1 & 0 & 0 & 0\\
        0 & \cos 2\chi & \sin 2\chi & 0\\
        0 & -\sin 2\chi & \cos 2\chi & 0\\
        0 & 0 & 0 & 1
    \end{pmatrix}.
\end{equation}

For completeness, the rotation angles can be defined geometrically. Let $\hat{\bm z}$ and $\hat{\bm n}$ denote the slab normal and the photon propagation direction. The normals to the incident and outgoing meridian planes are then
\begin{equation}
    \hat{\bm b} = \frac{\hat{\bm z}\times\hat{\bm n}}{\sqrt{1-\mu^2}},\qquad \hat{\bm b}' = \frac{\hat{\bm z}\times\hat{\bm n}'}{\sqrt{1-\mu'^2}}.
\end{equation}
The unit vector perpendicular to the scattering plane is
\begin{equation}
    \hat{\bm b}_{\rm s}
    =
    \frac{\hat{\bm n}\times\hat{\bm n}'}
    {\sin\Theta}.
\end{equation}
The signed rotation angles are then fixed by
\begin{equation}
    \cos\chi_{\rm in}
    =
    \hat{\bm b}\cdot\hat{\bm b}_{\rm s},
    \qquad
    \sin\chi_{\rm in}
    =
    \hat{\bm n}\cdot(\hat{\bm b}\times\hat{\bm b}_{\rm s}),
\end{equation}
and
\begin{equation}
    \cos\chi_{\rm out}
    =
    \hat{\bm b}'\cdot\hat{\bm b}_{\rm s},
    \qquad
    \sin\chi_{\rm out}
    =
    \hat{\bm n}'\cdot(\hat{\bm b}'\times\hat{\bm b}_{\rm s}).
\end{equation}

Reflection symmetry about any meridian plane requires $U$ to vanish, while Thomson scattering does not generate circular polarization from $I$, $Q$, or $U$, so that $V$ remains zero in the absence of incident circular polarization. The axisymmetric polarized radiation field can therefore be reduced to the two components $(I,Q)$. Related polarized successive-scattering treatments have been considered by \citet{1985A&A...143..374S} and \citet{1996ApJ...470..249P}.

In keeping with the notation of Paper I, we define the azimuth-integrated reduced post-scattering Stokes state as
\begin{equation}
    \bm{\Psi}_n(t,\mu)
    =
    \begin{pmatrix}
        \Psi_{I,n}(t,\mu)\\
        \Psi_{Q,n}(t,\mu)
    \end{pmatrix},
    \label{eq:reduced_stokes_state}
\end{equation}
where $\bm{\Psi}_n(t,\mu)$ describes the photon population immediately after the $n$-th scattering. The first component $\Psi_{I,n}$ is the scattering-order-resolved photon-number density in the present polarized transport problem and is the natural Stokes-vector counterpart of the scalar state distribution introduced in Paper I, whereas $\Psi_{Q,n}$ is its signed linearly polarized component. 

For the axisymmetric seed-photon distributions and boundary conditions, the incident Stokes field is independent of the absolute azimuth. The scattering kernel, however, still depends on the relative azimuth $\phi=\varphi'-\varphi$. The azimuth-integrated phase matrix acting on the reduced Stokes state $\bm{\Psi}$ is defined by
\begin{equation}
    \mathbfss{P}(\mu',\mu)
    =
    \int_{0}^{2\pi}
    \left[
        \mathbfss{R}(-\chi_{\rm out})
        \mathbfss{M}_{\rm T}(\Theta)
        \mathbfss{R}(\chi_{\rm in})
    \right]_{IQ}
    {\rm d}\phi,
    \label{eq:azimuthally_averaged_phase_matrix}
\end{equation}
where the subscript $IQ$ denotes the block acting on the $(I,Q)$ components. Carrying out the azimuthal integration gives
\begin{equation}
    \mathbfss{P}(\mu',\mu)
    =
    \frac{3}{16}
    \begin{pmatrix}
        3-\mu'^2-\mu^2+3\mu'^2\mu^2
        &
        (3\mu'^2-1)(1-\mu^2)
        \\
        (1-\mu'^2)(3\mu^2-1)
        &
        3(1-\mu'^2)(1-\mu^2)
    \end{pmatrix}.
    \label{eq:axisymmetric_polarized_phase_matrix}
\end{equation}
Here $\mu$ and $\mu'$ denote the incident and outgoing direction cosines, respectively. As a normalization check, the elements in the intensity row of the reduced phase matrix satisfy
\begin{equation}
    \int_{-1}^1 P_{II}(\mu', \mu) \,{\rm d}\mu' = 1,\qquad \int_{-1}^1 P_{IQ}(\mu', \mu) \,{\rm d}\mu' = 0
    \label{eq:phase_matrix_normalization}
\end{equation}
for any incident direction $\mu$. Hence the total number of scattered photons equals the number of incident photons, independently of the incident linear polarization. This kernel is equivalent to the two-polarization scattering matrix employed by \citet{1985A&A...143..374S}, after transforming from their $(I_l,I_r)$ representation to the present $(I,Q)$ convention. 

Given the reduced post-scattering state $\bm{\Psi}_n(t,\mu)$, its contribution to the radiation escaping through either boundary can be obtained by applying the corresponding free-escape projection. For $0<\mu<1$, the Stokes parameters of the radiation through the upper or the lower boundary are given by 
\begin{equation}
    \begin{aligned}
    &I_n^+(\mu)
    =
    \int_{0}^{\tau}
    \Psi_{I, n}(t,\mu)
    \exp\left(
        -\frac{\tau-t}{\mu}
    \right)
    {\rm d}t\\
    &Q_n^+(\mu)
    =
    \int_{0}^{\tau}
    \Psi_{Q, n}(t,\mu)
    \exp\left(
        -\frac{\tau-t}{\mu}
    \right)
    {\rm d}t
    \end{aligned},
    \label{eq:upper_stokes_escape}
\end{equation}
and
\begin{equation}
    \begin{aligned}
    &I_n^-(\mu)
    =
    \int_{0}^{\tau}
    \Psi_{I, n}(t,-\mu)
    \exp\left(
        -\frac{t}{\mu}
    \right)
    {\rm d}t\\
    &Q_n^-(\mu)
    =
    \int_{0}^{\tau}
    \Psi_{Q, n}(t,-\mu)
    \exp\left(
        -\frac{t}{\mu}
    \right)
    {\rm d}t
    \end{aligned}.
    \label{eq:lower_stokes_escape}
\end{equation}
These expressions generalize the boundary-resolved angular escape distributions obtained in Paper I to the Stokes-vector case. The signed linear-polarization fraction at the $n$-th scattering order is therefore
\begin{equation}
    \Pi_n^\pm(\mu)
    =
    \frac{Q_n^\pm(\mu)}
         {I_n^\pm(\mu)}.
    \label{eq:order_resolved_escape_polarization}
\end{equation}

With the adopted convention, $\Pi_n^\pm>0$ corresponds to an electric vector preferentially parallel to the slab surface, whereas $\Pi_n^\pm<0$ corresponds to an electric vector preferentially parallel to the projected slab normal.

Paper I evolved the scattering-order-resolved scalar photon-number distribution using the polarization-averaged Thomson phase function. In the present formulation, the transported state is instead the reduced Stokes vector, and the corresponding recursion operator differs from its scalar counterpart. Although the Stokes phase matrix remains exactly photon-number conserving, linear polarization generated at one scattering feeds back into the subsequent redistribution of the intensity through the $I-Q$ coupling. For an initially unpolarized seed source, the angular and scattering-order distributions obtained in the current framework generally deviate from the scalar result.

Since the intensity component of the reduced Stokes state is the azimuth-integrated photon-number density, the intensity escape projections in (\ref{eq:upper_stokes_escape}) and (\ref{eq:lower_stokes_escape}) directly give the differential escape probabilities, their boundary-integrated values, and the normalized angular distributions:
\begin{equation}
    \frac{{\rm d}P_n^\pm}{{\rm d}\mu}
    =I_n^\pm(\mu),\qquad
    P_n^\pm=\int_0^1 I_n^\pm(\mu){\rm d}\mu,\qquad
    \Phi_n^\pm(\mu)=\frac{I_n^\pm(\mu)}{P_n^\pm}.
    \label{eq:normalized_angular}
\end{equation}
The total probability of escaping after exactly $n$ scatterings is $P_n=P_n^++P_n^-$.

\subsection{Scattering-order recursion and the two-coefficient closure}

To construct the scattering-order recursion, the local scattering kernel in (\ref{eq:axisymmetric_polarized_phase_matrix}) must be combined with the free propagation between two consecutive scattering events. Following Paper I, we define the propagation kernel
\begin{equation}
    G(t',t;\mu)
    =
    \frac{1}{|\mu|}
    \exp\left(
        -\frac{|t'-t|}{|\mu|}
    \right)
    H\left[(t'-t)\mu\right],
    \label{eq:polarized_propagation_kernel}
\end{equation}
where $H$ is the Heaviside step function. The quantity $G(t',t;\mu)\,{\rm d}t'$ gives the probability that a photon leaving a scattering event at depth $t$ in direction $\mu$ undergoes its next scattering within ${\rm d}t'$ around $t'$. The Heaviside function enforces the directional constraint: $t'>t$ for $\mu>0$ and $t'<t$ for $\mu<0$. Because the slab has free boundaries, the propagation kernel satisfies the normalization condition:
\begin{equation}
    \int_0^\tau G(t',t;\mu)\,{\rm d}t'
    =
    \begin{cases}
        1-\exp[-(\tau-t)/\mu], & \mu>0,\\
        1-\exp[-t/|\mu|], & \mu<0.
    \end{cases}
\end{equation}
The missing probability is the probability of escaping through the corresponding free boundary before another scattering occurs.

Combining the azimuth-integrated phase matrix with the free-propagation kernel gives the Stokes-state recursion
\begin{equation}
    \bm{\Psi}_{n+1}(t',\mu')
    =
    \int_{0}^{\tau}{\rm d}t
    \int_{-1}^{1}{\rm d}\mu\,
    \mathbfss{P}(\mu',\mu)
    G(t',t;\mu)
    \bm{\Psi}_n(t,\mu),
    \label{eq:polarized_recursive_relation}
\end{equation}
which is the direct two-component generalization of the scalar recursion for Thomson scattering introduced by Paper I. It evolves the post-scattering state at $(t,\mu)$ through free propagation to the next scattering depth $t'$, followed by angular and polarization redistribution into the outgoing direction $\mu'$.

An important consequence of the azimuthally averaged Thomson kernel is that the post-scattering Stokes state closes exactly within an angular function space specified by only two depth-dependent coefficients. Substituting (\ref{eq:axisymmetric_polarized_phase_matrix}) into (\ref{eq:polarized_recursive_relation}) and collecting the terms independent of $\mu'$ and proportional to $\mu'^2$, we obtain the general expression
\begin{equation}
    \Psi_{I,n}(t,\mu)=a_n(t)+b_n(t)\mu^2,\qquad
    \Psi_{Q,n}(t,\mu)=b_n(t)(1-\mu^2).
    \label{eq:polarized_IQ_ab}
\end{equation}
The two coefficients $a_{n+1}$ and $b_{n+1}$ are given directly in terms of the previous post-scattering state by
\begin{equation}
\begin{aligned}
    a_{n+1}(t')
    =
    \frac{3}{16}
    &\int_0^\tau {\rm d}t
    \int_{-1}^{1}{\rm d}\mu\,
    G(t',t;\mu)\\
    &\times\Big[(3-\mu^2)\Psi_{I,n}(t,\mu)-
        (1-\mu^2)\Psi_{Q,n}(t,\mu)\Big],
\end{aligned}
\label{eq:an_general}
\end{equation}
and
\begin{equation}
\begin{aligned}
    b_{n+1}(t')
    =
    \frac{3}{16}
    &\int_0^\tau {\rm d}t
    \int_{-1}^{1}{\rm d}\mu\,
    G(t',t;\mu)\\
    &\times\Big[(3\mu^2-1)\Psi_{I,n}(t,\mu)+
        3(1-\mu^2)\Psi_{Q,n}(t,\mu)\Big].
\end{aligned}
\label{eq:bn_general}
\end{equation}
Substituting (\ref{eq:polarized_IQ_ab}) into (\ref{eq:an_general}) and (\ref{eq:bn_general}) explicitly gives the two-component depth recursion relation:
\begin{equation}
\begin{aligned}
    a_{n+1}(t')
    =
    \frac{3}{16}
    &\int_0^\tau {\rm d}t
    \int_{-1}^{1}{\rm d}\mu\,
    G(t',t;\mu)\\
    &\times
    \Big[(3-\mu^2)a_n(t)+
        (-1+5\mu^2-2\mu^4)b_n(t)\Big],
\end{aligned}
\label{eq:an_direct_recursion}
\end{equation}
and
\begin{equation}
\begin{aligned}
    b_{n+1}(t')
    =
    \frac{3}{16}
    &\int_0^\tau {\rm d}t
    \int_{-1}^{1}{\rm d}\mu\,
    G(t',t;\mu)\\
    &\times\Big[(-1+3\mu^2)a_n(t)+
        (3-7\mu^2+6\mu^4)b_n(t)\Big].
\end{aligned}
\label{eq:bn_direct_recursion}
\end{equation}
The axisymmetric polarization problem can thus be formulated as a recursion for the two depth-dependent coefficients $a_n(t)$ and $b_n(t)$.

To express these recursions entirely as depth integrals, we introduce the generalized exponential integrals
\begin{equation}
    E_m(x)
    =
    \int_0^1
    u^{m-2}
    \exp\left(-\frac{x}{u}\right)
    {\rm d}u
\end{equation}
and the associated depth operators
\begin{equation}
    \left(\mathcal K_m f\right)(t')
    =
    \int_0^\tau
    E_m(|t'-t|)
    f(t)\,
    {\rm d}t.
    \label{eq:Km_definition}
\end{equation}
Consequently, (\ref{eq:an_direct_recursion}) and (\ref{eq:bn_direct_recursion}) reduce to
\begin{equation}
    \bm{y}_{n+1}
    =
    \mathbfss K_{\rm pol}\bm{y}_n,
    \qquad n\geq1.
    \label{eq:compact_polarized_recursion}
\end{equation}
where
\begin{equation}
    \bm{y}_n(t)
    =
    \begin{pmatrix}
        a_n(t)\\
        b_n(t)
    \end{pmatrix}
    \label{eq:polarized_recursion_state}
\end{equation}
and
\begin{equation}
    \mathbfss K_{\rm pol}
    =
    \frac{3}{16}
    \begin{pmatrix}
        3\mathcal K_1-\mathcal K_3
        &
        -\mathcal K_1+5\mathcal K_3-2\mathcal K_5
        \\
        -\mathcal K_1+3\mathcal K_3
        &
        3\mathcal K_1-7\mathcal K_3+6\mathcal K_5
    \end{pmatrix}.
    \label{eq:polarized_depth_kernel}
\end{equation}
The numerical algorithm for evaluating the recursion efficiently was described in detail in Paper I.

Once the two depth-dependent coefficients $a_n(t)$ and $b_n(t)$ have been obtained, the Stokes parameters of the radiation through the free boundaries defined by (\ref{eq:upper_stokes_escape}) and (\ref{eq:lower_stokes_escape}) can therefore be written directly in terms of the two recursion coefficients as
\begin{equation}
    \begin{aligned}
        &I_n^+ = \int_0^\tau\, [a_n(t) + \mu^2 b_n(t)] \exp[-(\tau-t) / \mu]\,{\rm d} t\\
        &Q_n^+ = (1-\mu^2)\int_0^\tau\, b_n(t) \exp[-(\tau-t) / \mu]\,{\rm d} t
    \end{aligned},
    \label{eq:IQ_upper}
\end{equation}
and 
\begin{equation}
    \begin{aligned}
        &I_n^- = \int_0^\tau\, [a_n(t) + \mu^2 b_n(t)] \exp(-t / \mu)\,{\rm d} t\\
        &Q_n^- = (1-\mu^2)\int_0^\tau\, b_n(t) \exp(-t / \mu)\,{\rm d} t
    \end{aligned}.
    \label{eq:IQ_lower}
\end{equation}
The corresponding signed linear-polarization fraction then follows directly from $Q_n^\pm / I_n^\pm$. In addition, integrating the intensity over the escaping direction gives the boundary-resolved escape probabilities, i.e.,
\begin{equation}
    \begin{aligned}
        &P_n^+ = \int_0^\tau {\rm d}t \, \left[a_n(t)E_2(\tau - t) + b_n(t) E_4(\tau - t)\right],\\
        &P_n^- = \int_0^\tau {\rm d}t \, \left[a_n(t)E_2(t) + b_n(t) E_4(t)\right].
    \end{aligned}
\end{equation}

\subsection{Seed-photon sources and first-scattering initialization}

For any seed-photon source, the zeroth-order state can be represented by
\begin{equation}
    \bm{\Psi}_0(t,\mu)
    =
    \begin{pmatrix}
        I_0(t,\mu)\\
        Q_0(t,\mu)
    \end{pmatrix}.
    \label{eq:polarized_seed_state}
\end{equation}
The coefficients describing the radiation immediately after the first
scattering are then
\begin{equation}
\begin{aligned}
    a_1(t')
    =
    \frac{3}{16}
    &\int_0^\tau {\rm d}t
    \int_{-1}^{1}{\rm d}\mu\,
    G(t',t;\mu)\\
    &\times\Big[
        (3-\mu^2)I_0(t,\mu)-
        (1-\mu^2)Q_0(t,\mu)
    \Big],
\end{aligned}
\label{eq:a1_general_source}
\end{equation}
and
\begin{equation}
\begin{aligned}
    b_1(t')
    =
    \frac{3}{16}
    &\int_0^\tau {\rm d}t
    \int_{-1}^{1}{\rm d}\mu\,
    G(t',t;\mu)\\
    &\times\Big[
        (3\mu^2-1)I_0(t,\mu)+
        3(1-\mu^2)Q_0(t,\mu)
    \Big].
\end{aligned}
\label{eq:b1_general_source}
\end{equation}
For an unpolarized seed-photon source, the corresponding expressions are obtained by setting $Q_0=0$. Starting from any prescribed source, (\ref{eq:a1_general_source}) and (\ref{eq:b1_general_source}) determine $(a_1,b_1)$, after which (\ref{eq:compact_polarized_recursion}) generates all higher scattering orders.

\subsubsection{Seed photons injected through the lower boundary}
\label{sec:bottom_seed_injection}

A general axisymmetric seed-photon distribution injected through the lower boundary can be written as
\begin{equation}
    \bm{\Psi}_0^{\rm b}(t,\mu)
    =
    \delta(t)H(\mu)
    \begin{pmatrix}
        p_I(\mu)\\
        p_Q(\mu)
    \end{pmatrix},
    \label{eq:general_bottom_seed}
\end{equation}
where $\delta(t)$ is the delta function. The functions $p_I(\mu)$ and $p_Q(\mu)$ are the normalized intensity and signed linear-polarization distributions per unit direction cosine for photons crossing the lower boundary.

Substituting (\ref{eq:general_bottom_seed}) into (\ref{eq:a1_general_source}) and (\ref{eq:b1_general_source}) gives
\begin{equation}
    a_1^{\rm b}(t')
    =
    \frac{3}{16}
    \int_0^1
    \frac{{\rm d}\mu}{\mu}
    \exp\left(-\frac{t'}{\mu}\right)
    \Big[
        (3-\mu^2)p_I(\mu)
        -(1-\mu^2)p_Q(\mu)
    \Big],
\label{eq:a1_general_bottom_source}
\end{equation}
and
\begin{equation}
    b_1^{\rm b}(t')
    =
    \frac{3}{16}
    \int_0^1
    \frac{{\rm d}\mu}{\mu}
    \exp\left(-\frac{t'}{\mu}\right)
    \Big[
        (3\mu^2-1)p_I(\mu)
        +3(1-\mu^2)p_Q(\mu)
    \Big].
\label{eq:b1_general_bottom_source}
\end{equation}
These expressions allow the angular distribution and intrinsic polarization of the seed photons to be specified independently.

An important special case is the normally incident, axisymmetric beam injection described by
\begin{equation}
    p_I^{\rm B}(\mu)
    =
    \delta(\mu-1).
    \label{eq:vertical_beam_source}
\end{equation}
If the source is unpolarized, its first-scattering coefficients are
\begin{equation}
    a_1^{\rm B}(t')
    =
    b_1^{\rm B}(t')
    =
    \frac{3}{8}
    \exp(-t').
    \label{eq:vertical_beam_ab}
\end{equation}

Another important special case is an unpolarized Lambert-law source, which provides a more realistic description of blackbody photons entering the corona from the disc surface. Its angular distribution is
\begin{equation}
    p_I^{\rm L}(\mu)
    =
    2\mu.
\end{equation}
Using the generalized exponential integrals defined above, the first-scattering coefficients become
\begin{equation}
    a_1^{\rm L}(t')
    =
    \frac{3}{8}
    \left[
        3E_2(t')-E_4(t')
    \right],\qquad
    b_1^{\rm L}(t')
    =
    \frac{3}{8}
    \left[
        3E_4(t')-E_2(t')
    \right].
    \label{eq:b1_lambert_source}
\end{equation}

\subsubsection{Seed photons generated within the slab medium}
\label{sec:volume_seed_source}

An arbitrary axisymmetric seed-photon source generated within the slab
can be written as
\begin{equation}
    \bm{\Psi}_0^{\rm vol}(t,\mu)
    =
    q(t)
    \begin{pmatrix}
        p_I(t,\mu)\\
        p_Q(t,\mu)
    \end{pmatrix},
    \label{eq:general_volume_seed}
\end{equation}
where $q(t)$ specifies the depth distribution of the seed photons, while $p_I(t,\mu)$ and $p_Q(t,\mu)$ specify the angular and polarization distributions at depth $t$. These functions are normalized as
\begin{equation}
    \int_0^\tau q(t)\,{\rm d}t=1,
    \qquad
    \int_{-1}^{1}p_I(t,\mu)\,{\rm d}\mu=1.
    \label{eq:volume_seed_normalization}
\end{equation}

The first-scattering coefficients generated by this source are
\begin{equation}
\begin{aligned}
    a_1^{\rm vol}(t')
    =
    \frac{3}{16}
    &\int_0^\tau{\rm d}t\,q(t)\,
    \int_{-1}^{1}{\rm d}\mu\,
    G(t',t;\mu)\\
    &\times\Big[
        (3-\mu^2)p_I(t,\mu)
        -(1-\mu^2)p_Q(t,\mu)
    \Big],
\end{aligned}
\label{eq:a1_general_volume_source}
\end{equation}
and
\begin{equation}
\begin{aligned}
    b_1^{\rm vol}(t')
    =
    \frac{3}{16}
    &\int_0^\tau{\rm d}t\,q(t)
    \int_{-1}^{1}{\rm d}\mu\,
    G(t',t;\mu)\\
    &\times\Big[
        (3\mu^2-1)p_I(t,\mu)
        +3(1-\mu^2)p_Q(t,\mu)
    \Big].
\end{aligned}
\label{eq:b1_general_volume_source}
\end{equation}
This class includes, for example, internally generated synchrotron, bremsstrahlung, or thermal seed photons with an arbitrary prescribed depth distribution.

A useful special case is a vertically uniform, isotropic internal source, for which
\begin{equation}
    q(t)=\frac{1}{\tau},
    \qquad
    p_I(t,\mu)=\frac{1}{2}.
    \label{eq:homogeneous_volume_source}
\end{equation}
For the unpolarized seed photons, we define
\begin{equation}
    J_m(t')
    =
    \int_0^\tau
    E_m(|t'-t|)\,{\rm d}t
    =
    \frac{2}{m}
    -
    E_{m+1}(t')
    -
    E_{m+1}(\tau-t').
    \label{eq:Jm_uniform_source}
\end{equation}
The first-scattering coefficients then reduce to
\begin{equation}
\begin{aligned}
        a_1^{\rm iso}(t')
    =
    \frac{3}{32\tau}
    \left[
        3J_1(t')-J_3(t')
    \right],
    b_1^{\rm iso}(t')
    =
    \frac{3}{32\tau}
    \left[
        3J_3(t')-J_1(t')
    \right].
\end{aligned}
    \label{eq:ab1_uniform_isotropic_source}
\end{equation}

\subsection{Finite-order escape statistics and polarization}

In the previous subsections, we have established the Stokes-vector recursion and the corresponding first-scattering states for different seed-photon sources. In this subsection, we illustrate the finite-order escape statistics. We use three representative source configurations: a normally incident beam, a Lambert-law source injected through the lower boundary, and a vertically uniform isotropic source generated within the slab medium. The seed photons are assumed to be initially unpolarized. As in Paper I, we calculate the scattering-order-resolved escape probabilities and normalized angular distributions, and we additionally evaluate the corresponding surface polarization. The scattering-order-resolved survival probabilities and the resulting mean scattering number, whose scalar formulation was developed in Paper I, are deferred to Appendix~\ref{app:mean_number}.

\begin{figure*}
	\includegraphics[width = 0.75\textwidth]{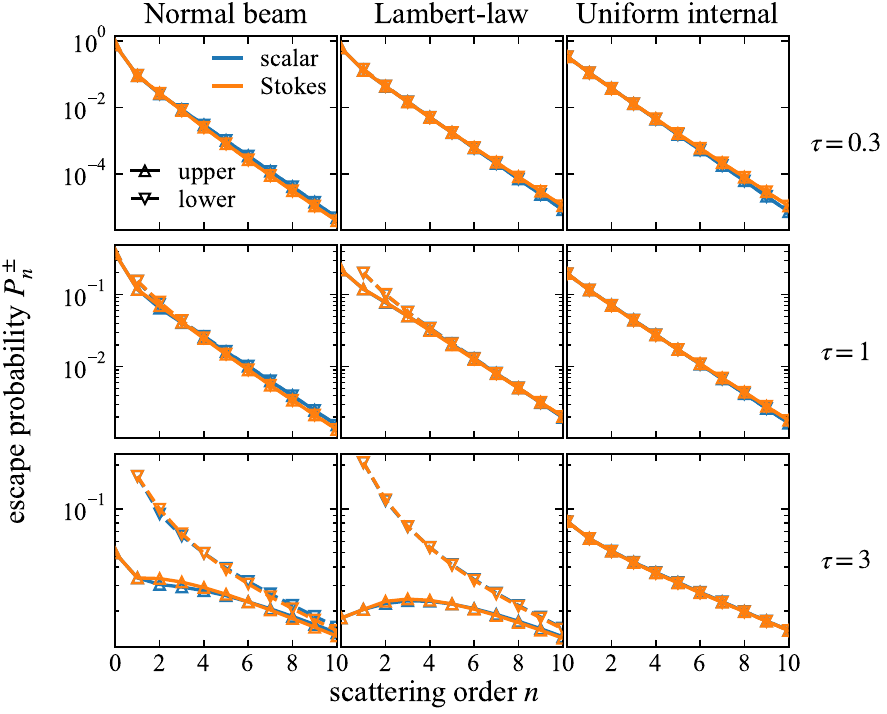}
    \caption{ Boundary-resolved escape probabilities $P_n^\pm$ as functions of scattering order for the normally incident beam, Lambert-law boundary source, and vertically uniform isotropic internal source. Rows correspond to slab optical depths $\tau=0.3$, $1$, and $3$. Curves with different colours correspond to different recursive schemes. Curves with upper and lower triangular markers denote escape through the upper and lower boundaries, respectively.} 
    \label{fig:finite_order_escape_probability}
\end{figure*}

Figure~\ref{fig:finite_order_escape_probability} compares the boundary- and scattering-order-resolved escape probabilities $P_n^\pm$ obtained from the present Stokes recursion with those calculated using the scalar recursion of Paper I. The two calculations remain close over most of the parameter range and exhibit the same qualitative behaviour, showing that polarization feedback produces only a modest correction to these integrated escape statistics. This should not be interpreted as a change in the underlying microscopic Thomson scattering process; rather, the difference arises because the polarization feeds back into the redistribution of the intensity through the $I-Q$ coupling in the Stokes recursion. 

To validate the recursive formalism derived here, we also compared the recursive results with polarized Monte Carlo radiative-transfer (MCRT) simulations and found agreement within the Monte Carlo statistical uncertainties. The MCRT calculation follows the full polarized scattering process independently of the recursion developed above; details of the implementation are given in Appendix~\ref{app:mcrt}. For clarity, the MCRT results are not overplotted in Figure~\ref{fig:finite_order_escape_probability}; a reproducible notebook is provided with the data and code accompanying this work.

The numerical results show that both recursive schemes satisfy photon-number conservation. For the Stokes formulation, this follows from the normalization of the reduced phase matrix together with the same survival-probability argument used in Paper I; the corresponding global proof is not repeated here. 

For the two lower-boundary injection configurations, the upper and lower escape probabilities differ at low scattering orders because the initial source breaks reflection symmetry about the slab mid-plane. Repeated scattering progressively reduces this boundary asymmetry. As the optical depth increases, the boundary asymmetry persists to higher scattering orders, indicating that more scatterings are required to erase the memory of the initial injection geometry. For the vertically uniform isotropic source, the source is symmetric initially, and the upper and lower escape probabilities are therefore identical at every scattering order.

\begin{figure*}
    \includegraphics[width = 0.75\textwidth]{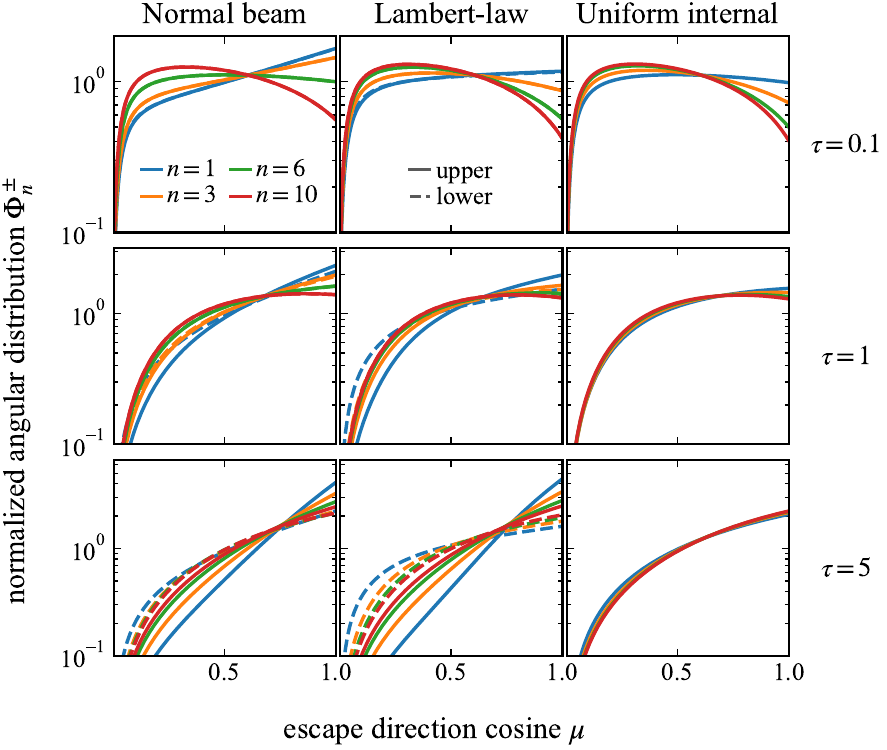}
    \caption{ Normalized angular distributions $\Phi_n^\pm(\mu)$ corresponding to different scattering orders. Columns show the normally incident beam, Lambert-law boundary source, and vertically uniform isotropic internal source, while rows correspond to $\tau=0.1$, $1$, and $5$. Colours indicate $n=1$, $3$, $6$, and $10$, and solid and dashed curves denote escape through the upper and lower boundaries, respectively.} 
    \label{fig:finite_order_angular}
\end{figure*}

The normalized angular distributions are shown in Figure~\ref{fig:finite_order_angular}. At low scattering orders, $\Phi_n^\pm(\mu)$ retains a strong memory of both the seed-photon geometry and the escape boundary. As the scattering order increases, these source- and boundary-dependent differences become progressively weaker. For example, at $\tau=0.1$ and $n=10$, the angular distributions for the Lambert-law boundary source and the vertically uniform isotropic source are nearly identical, while that for the normally incident beam has not yet fully converged but is clearly approaching the same limiting form. The angular distributions are also consistent with the polarized MCRT simulations. Compared with the escape probabilities, the normalized angular distributions obtained from the Stokes and scalar recursions show somewhat larger differences, particularly at high scattering orders. This reflects the fact that the two recursion operators have different high-order asymptotic behaviour, even though their integrated escape statistics remain relatively similar.

The angular distribution also varies systematically with the total optical depth. In a thin slab, multiply scattered photons preferentially emerge over a broad range of oblique directions, with the high-order distribution suppressed toward the slab normal. In a thick slab, by contrast, repeated spatial diffusion produces an emergent distribution increasingly concentrated toward the slab normal. These trends indicate that repeated application of $\mathbfss K_{\rm pol}$ drives the radiation toward a common asymptotic state whose detailed form depends on $\tau$ but is independent of the initial seed-photon source.

\begin{figure*}
	\includegraphics[width = 0.75\textwidth]{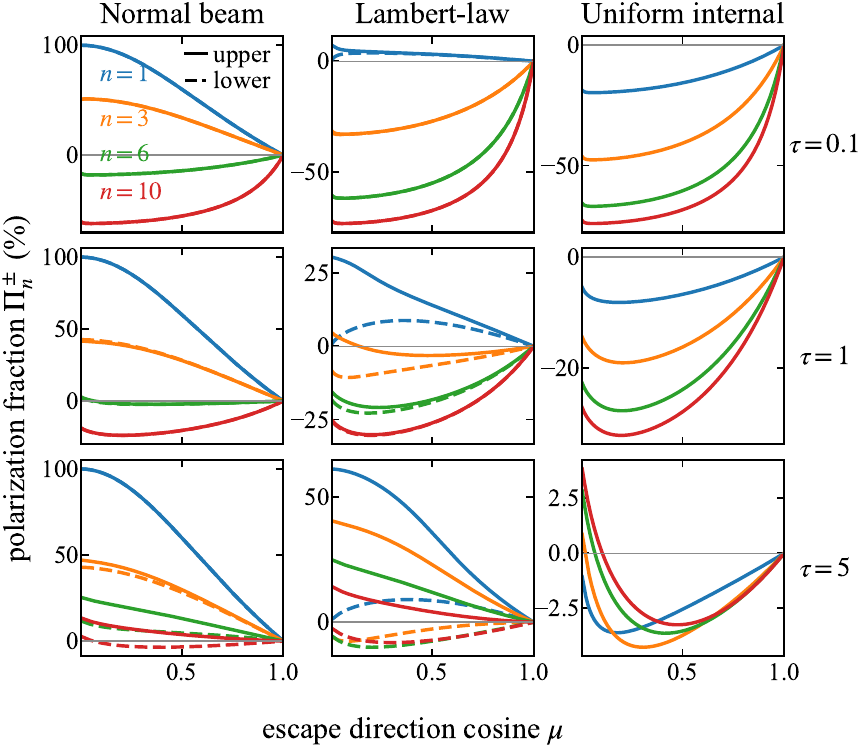}
    \caption{ Signed linear-polarization fractions $\Pi_n^\pm(\mu)$ corresponding to different scattering orders. Columns show the normally incident beam, Lambert-law boundary source, and vertically uniform isotropic internal source; rows correspond to $\tau=0.1$, $1$, and $5$. Colours denote $n=1$, $3$, $6$, and $10$, while solid and dashed curves denote escape through the upper and lower boundaries, respectively. Positive polarization corresponds to an electric vector preferentially parallel to the slab surface and negative polarization to one preferentially parallel to the projected slab normal. } 
    \label{fig:finite_order_polarization}
\end{figure*}

Figure~\ref{fig:finite_order_polarization} shows the signed linear-polarization fractions $\Pi_n^\pm(\mu)$. The polarization is generally more sensitive to scattering order than the normalized angular distribution. Although all three seed-photon sources considered here are initially unpolarized, Thomson scattering generates a non-zero linear polarization. The polarization calculations have also been checked against independent polarized MCRT simulations. For every scattering order, $\Pi_n^\pm(\mu)\rightarrow0$ as $\mu\rightarrow1$, as required by axial symmetry.

The polarization also changes systematically with optical depth. In optically thin media, the high-order polarization evolves toward the negative values in our sign convention and can reach a large magnitude, whereas in optically thick media the high-order state is much more weakly polarized. At a fixed viewing direction and scattering order, the polarization can change sign as the optical depth increases. Similar to the normalized angular distribution, the differences among the three source configurations decrease systematically with scattering order. The simultaneous convergence of the angular distribution and polarization indicates that photons surviving many scatterings are governed by a common high-order Stokes eigenstate.

\section{High-order asymptotic polarized transport}
\label{sec:high_order_polarization}

Paper I has shown that the scalar scattering-order recursion approaches a well-defined asymptotic state at high scattering orders. An analogous high-order eigenmode also exists for the Stokes-vector recursion developed in this work. Although the low-order radiation retains information about the spatial, angular, and polarization properties of the seed photons, repeated application of the recursion operator progressively drives the radiation toward a universal high-order Stokes eigenstate that is independent of the detailed properties of the seed-photon source.

\subsection{Polarized eigenmodes and the high-order limit}
\label{sec:polarized_eigenmodes}

Eigenfunction expansions have long been used to describe photon diffusion and the asymptotic escape behaviour in Comptonizing media \citep{1980A&A....86..121S}. In particular, \citet{1985A&A...143..374S} applied a spectral decomposition of the polarized transfer operator to investigate the high-scattering-order angular and polarization distributions. Here we establish an explicit weighted self-adjoint structure for the finite-slab Stokes recursion operator, which provides a convenient basis for analysing its complete eigenvalue spectrum.

The polarized recursion operator $\mathbfss K_{\rm pol}$ given by (\ref{eq:polarized_depth_kernel}) can be written in the form
\begin{equation}
    \mathbfss K_{\rm pol}
    =
    \frac{3}{16}
    \left(
        \mathcal K_1\mathbfss A_1
        +
        \mathcal K_3\mathbfss A_3
        +
        \mathcal K_5\mathbfss A_5
    \right),
    \label{eq:polarized_operator_decomposition}
\end{equation}
where
\begin{equation}
    \mathbfss A_1
    =
    \begin{pmatrix}
        3 & -1\\
        -1 & 3
    \end{pmatrix},
    \quad
    \mathbfss A_3
    =
    \begin{pmatrix}
        -1 & 5\\
        3 & -7
    \end{pmatrix},
    \quad
    \mathbfss A_5
    =
    \begin{pmatrix}
        0 & -2\\
        0 & 6
    \end{pmatrix}.
    \label{eq:polarized_stokes_matrices}
\end{equation}

The scalar depth operators $\mathcal K_m$ have symmetric kernels that depend only on $|t-t'|$, and are therefore self-adjoint on the finite slab. The matrices in (\ref{eq:polarized_stokes_matrices}) all possess the common positive symmetrizer
\begin{equation}
    \mathbfss W
    =
    \begin{pmatrix}
        1 & 1/3\\
        1/3 & 1
    \end{pmatrix},
    \label{eq:W_matrix}
\end{equation}
which satisfies
\begin{equation}
    \mathbfss W\mathbfss A_m
    =
    \mathbfss A_m^{\rm T}\mathbfss W,
    \qquad
    m=1,3,5.
    \label{eq:W_symmetrization}
\end{equation}
The weighted inner product of two-component depth functions therefore is defined as
\begin{equation}
    \left\langle
        \bm{u},\bm{v}
    \right\rangle
    =
    \int_0^\tau
    \bm{u}^{\rm T}(t)
    \mathbfss W
    \bm{v}(t)
    \,{\rm d}t .
    \label{eq:W_inner_product}
\end{equation}

$\mathbfss K_{\rm pol}$ is a compact self-adjoint operator in the $\mathbfss W$-weighted function space, i.e., 
\begin{equation}
    \langle\bm{u},\mathbfss K_{\rm pol}\bm{v}\rangle = \langle\mathbfss K_{\rm pol}\bm{u},\bm{v}\rangle,
\end{equation}
so that its spectrum is real, and its eigenfunctions may be chosen to form a $\mathbfss W$-orthonormal basis. We denote these eigenfunctions by $\bm{u}_j(t;\tau)$,
\begin{equation}
    \mathbfss K_{\rm pol}\bm{u}_j
    =
    \lambda_j\bm{u}_j,
    \label{eq:polarized_eigenvalue_problem}
\end{equation}
with
\begin{equation}
    \left\langle
        \bm{u}_i,\bm{u}_j
    \right\rangle
    =
    \delta_{ij}.
    \label{eq:polarized_W_orthogonality}
\end{equation}

Since all depth kernels depend only on $|t-t'|$ and both surfaces obey the same free-escape boundary condition, the system is symmetric under reflection about the slab mid-plane and the eigenfunctions of $\mathbfss K_{\rm pol}$ may be chosen to have definite parity, i.e.,
\begin{equation}
    \bm{u}_j(\tau - t)
    =
    p_j\bm{u}_j(t),
    \qquad
    p_j=\pm1,
    \label{eq:parity_eigenvalue}
\end{equation}
where $p_j=+1$ and $-1$ denote even and odd modes, respectively.

For the finite Thomson slab, a unique eigenmode dominates the transport at high scattering orders. We denote the dominant physical eigenvalue by $\lambda_0$ and order the spectrum according to
\begin{equation}
    \lambda_0>
    |\lambda_1|
    \geq
    |\lambda_2|
    \geq
    \cdots.
    \label{eq:polarized_eigenvalue_ordering}
\end{equation}
The principal eigenvalue $\lambda_0$ is real and positive. Because photons have a non-zero probability of escaping between successive scatterings, the dominant eigenvalue is smaller than 1, i.e., $\lambda_0<1$, as in the scalar scattering-order recursion discussed by Paper I.

The first-scattering state can be expanded in the eigenfunctions as
\begin{equation}
    \bm{y}_1(t)
    =
    \sum_{j=0}^\infty
    c_j\bm{u}_j(t;\tau),
    \label{eq:first_scattering_eigenmode_expansion}
\end{equation}
where 
\begin{equation}
    c_j
    =
    \left\langle
        \bm{u}_j,\bm{y}_1
    \right\rangle
    \label{eq:eigenmode_projection_coefficient}
\end{equation}
depends on the spatial, angular, and polarization properties of the seed-photon source. After applying the recursion operator $n-1$ times, we obtain
\begin{equation}
\begin{aligned}
    \bm{y}_n(t)
    =
    \lambda_0^{n-1}
        \left[c_0\bm{u}_0(t;\tau)+
        \sum_{j\geq1}c_j
        \left(
            \frac{\lambda_j}{\lambda_0}
        \right)^{n-1}
        \bm{u}_j(t;\tau)\right].
\end{aligned}
\label{eq:dominant_mode_expansion}
\end{equation}
For any seed-photon source with a non-zero $c_0$, all subdominant contributions are progressively suppressed as the scattering order increases. Hence, at sufficiently high scattering order,
\begin{equation}
    \bm{y}_n(t)
    \simeq
    c_0\lambda_0^{n-1}
    \bm{u}_0(t;\tau),
    \qquad n\gg1.
\label{eq:high_order_dominant_mode}
\end{equation}

We write the dominant polarized eigenfunction as
\begin{equation}
    \bm{u}_0(t;\tau)
    =
    \begin{pmatrix}
        a_*(t;\tau)\\
        b_*(t;\tau)
    \end{pmatrix}.
    \label{eq:dominant_ab_eigenfunction}
\end{equation}
For the symmetric free slab considered here, the principal mode belongs to the even-parity sector and therefore satisfies
\begin{equation}
    a_*(t;\tau)
    =
    a_*(\tau-t;\tau),
    \qquad
    b_*(t;\tau)
    =
    b_*(\tau-t;\tau).
    \label{eq:dominant_mode_midplane_symmetry}
\end{equation}
Consequently, the emergent Stokes distributions associated with the principal mode are identical at the two boundaries after expressing the escaping direction by the positive cosine $0<\mu<1$.

Analogously to the escape projections in (\ref{eq:IQ_upper}) and (\ref{eq:IQ_lower}), we define the emergent Stokes functions associated with the dominant eigenmode as
\begin{equation}
    \begin{aligned}
        &I_*(\mu) = \int_0^\tau \left[a_*(t;\tau)+\mu^2 b_*(t;\tau)\right]\exp(-t/\mu) \, {\rm d} t,\\
        &Q_*(\mu) = (1-\mu^2) \int_0^\tau b_*(t;\tau)\exp(-t/\mu) \, {\rm d} t.
    \end{aligned}
    \label{eq:high-order-projection}
\end{equation}
We further define the corresponding angle-integrated intensity as
\begin{equation}
    F_* = \int_0^1 I_*(\mu) \, {\rm d}\mu = \int_0^\tau \left[a_*(t;\tau)E_2(t)+b_*(t;\tau)E_4(t)\right] \, {\rm d} t.
\end{equation}

Using (\ref{eq:high_order_dominant_mode}), we can obtain the high-order Stokes parameters of the radiation through either boundary
\begin{equation}
    \begin{aligned}
        I_n^\pm(\mu) \simeq c_0 \lambda_0^{n-1} I_\ast(\mu),\quad
        Q_n^\pm(\mu) \simeq c_0 \lambda_0^{n-1} Q_\ast(\mu)
    \end{aligned},\qquad n\gg1.
    \label{eq:high_order_IQ}
\end{equation}
The corresponding high-order escape probabilities are
\begin{equation}
    P_n^\pm \simeq c_0 \lambda_0^{n-1} F_*.
\end{equation}
The ratio between successive high-order escape probabilities therefore approaches the dominant eigenvalue \citep[e.g.,][]{1994ApJ...434..570T},
\begin{equation}
    \lim_{n\to\infty} \frac{P_n^\pm}{P_{n-1}^\pm} = \lambda_0.
    \label{eq:escape_ratio}
\end{equation}

The high-order polarization and normalized angular distribution are therefore
\begin{equation}
    \Pi_\infty^\pm(\mu)=\frac{Q_*(\mu)}{I_*(\mu)},\qquad
    \Phi_\infty^\pm(\mu)=\frac{I_*(\mu)}{F_*}.
    \label{eq:asymptotic_general}
\end{equation}
In these expressions the common factor $c_0\lambda_0^{n-1}$ cancels, so the high-order polarization and normalized angular distributions are independent of both the eigenmode normalization and the detailed seed-photon source. This source-independent high-order behaviour is consistent with the asymptotic polarized transfer solutions obtained by \citet{1985A&A...143..374S}.

\subsection{The polarized eigenmode in the optically thin limit} 
\label{sec:thin_polarized_eigenmode}

\subsubsection{Leading-order asymptotic eigenmode} 
\label{sec:thin_limit_mode} 

We first consider the limit $\tau\rightarrow0$. For two depths within the slab, $|t-t'|\lesssim\tau$, the generalized exponential integrals have the small-argument behaviour
\begin{equation}
\begin{aligned}
    &E_1(x)
    =
    -\gamma_{\rm E}
    -\ln x
    +
    \mathcal O(x),\\
    &E_3(x)
    =
    \frac{1}{2}
    +
    \mathcal O(x),
    \qquad
    E_5(x)
    =
    \frac{1}{4}
    +
    \mathcal O(x),
\end{aligned}
    \label{eq:E135_small_argument}
\end{equation}
where $\gamma_{\rm E}$ is the Euler constant. Thus, compared to $\mathcal K_3$ and $\mathcal K_5$, $\mathcal K_1$ is logarithmically enhanced. To leading logarithmic order, the polarized recursion operator reduces to
\begin{equation}
    \mathbfss K_{\rm pol}
    \simeq
    \mathcal K_1\mathbfss C,
    \label{eq:thin_polarized_operator}
\end{equation}
where
\begin{equation}
    \mathbfss C
    =
    \frac{3}{16}
    \begin{pmatrix}
        3 & -1\\
        -1 & 3
    \end{pmatrix}.
    \label{eq:thin_constant_polarization_matrix}
\end{equation}
The two eigenvalues of $\mathbfss C$ are
\begin{equation}
    \kappa_0=\frac{3}{4},
    \qquad
    \kappa_1=\frac{3}{8},
    \label{eq:thin_constant_matrix_eigenvalues}
\end{equation}
with corresponding eigenvectors proportional to
\begin{equation}
    \bm{v}_0 = \begin{pmatrix}
        1\\
        -1
    \end{pmatrix},
    \qquad
    \bm{v}_1 = \begin{pmatrix}
        1\\
        1
    \end{pmatrix},
    \label{eq:thin_constant_matrix_eigenvectors}
\end{equation}
respectively.

The dominant eigenfunction of the full polarized operator in the thin limit therefore satisfies
\begin{equation}
    b_*(t)
    =
    -a_*(t).
    \label{eq:thin_ab_relation}
\end{equation}
Substituting this relation into (\ref{eq:asymptotic_general}) gives
\begin{equation}
    \Pi_\infty^\pm(\mu)
    \longrightarrow
    -1,
    \qquad \Phi_\infty^\pm(\mu)
    \to\frac{3}{2}(1-\mu^2),
    \qquad
    \tau\rightarrow0.
\label{eq:thin_asymptotic}
\end{equation}
Thus the radiation of high scattering orders in an optically thin slab approaches complete linear polarization, consistent with the analytical thin-slab result of \citet{1985A&A...143..374S}. 

As shown below, the denominator of the approximate polarization fraction (\ref{eq:polarization_from_alpha}) can be written as $(1-\mu^2)+[1+\alpha(\tau)]\mu^2$, where $\alpha(\tau)$ is given by (\ref{eq:thin_stokes_eigendirection}). The complete-polarization limit is therefore non-uniform near the slab normal, where the two terms become comparable, corresponding to 
\begin{equation}
    1 - \mu^2 \lesssim \frac{1}{2\ln(1/\tau)}.
    \label{eq:thin_polarization_boundary_layer}
\end{equation}

\subsubsection{Logarithmic corrections to the thin-limit eigenmode}
\label{sec:log_correction_thin}

The strict $\tau\rightarrow0$ solution provides a useful physical result, but it is not quantitatively accurate over much of the finite-optical-depth range relevant to slab-corona applications. In particular, for $0.1\lesssim\tau\lesssim1$, the contributions associated with $\mathcal K_3$ and $\mathcal K_5$, as well as the finite-$\tau$ part of $\mathcal K_1$, can no longer be neglected when determining the polarization eigenvector. We therefore construct a finite-optical-depth extension that retains the contributions of all three depth operators.

For simplicity, we introduce the projection operator
\begin{equation}
(\mathcal P f)(t')
=
\frac{1}{\tau}
\int_0^\tau
f(t)\,{\rm d}t,
\label{eq:constant_depth_projector_tau}
\end{equation}
and the finite logarithmic operator
\begin{equation}
(\mathcal H f)(t')
=
\frac{1}{\tau}
\int_0^\tau
\left[
-\gamma_{\rm E}
-\ln\left(\frac{|t'-t|}{\tau}\right)
\right]
f(t)\,{\rm d}t.
\label{eq:H_tau_operator}
\end{equation}
The three depth operators then have the asymptotic forms
\begin{equation}
\begin{aligned}
    &\mathcal K_1
=
\tau
\left[
\ln\left(\frac{1}{\tau}\right)\mathcal P
+
\mathcal H
\right]
+
\mathcal O(\tau^2),\\
&\mathcal K_3
=
\frac{\tau}{2}\mathcal P
+
\mathcal O(\tau^2),
\qquad
\mathcal K_5
=
\frac{\tau}{4}\mathcal P
+
\mathcal O(\tau^2).
\end{aligned}
\label{eq:K135_thin_projected_expansion}
\end{equation}
The full polarized recursion operator in the optically thin limit can therefore be written as
\begin{equation}
\mathbfss K_{\rm pol}
=
\tau
\left[
\ln\left(\frac{1}{\tau}\right)
\mathbfss C\mathcal P
+
\mathbfss C\mathcal H
+
\mathbfss B\mathcal P
\right]
+
\mathcal O(\tau^2),
\label{eq:thin_operator_first_correction}
\end{equation}
where
\begin{equation}
\mathbfss B
=
\frac{3}{16}
\begin{pmatrix}
-1/2 & 2\\
3/2 & -2
\end{pmatrix}.
\label{eq:thin_B_matrix}
\end{equation}
In (\ref{eq:thin_operator_first_correction}), the first term is the logarithmically enhanced contribution from $\mathcal K_1$, the second term contains its finite part, while the third term collects the leading contributions from $\mathcal K_3$ and $\mathcal K_5$.

Because the logarithmically dominant depth operator is proportional to $\mathcal P$, whose eigenfunction is constant, we obtain $\bm{u}_0(t;\tau)\simeq \bm{v}_0$ at leading logarithmic order. (\ref{eq:thin_operator_first_correction}) naturally suggests expanding the principal eigenpair in inverse powers of $\ln(1/\tau)$. We therefore seek the principal eigenfunction and eigenvalue of the full polarized operator in the forms
\begin{equation}
\bm u_0(t;\tau)
=
\bm v_0
+
\frac{1}{\ln(1/\tau)}
\bm u_0^{(1)}(t;\tau)
+
\mathcal O\left[\frac{1}{\ln^2(1/\tau)}\right],
\label{eq:thin_eigenfunction_expansion}
\end{equation}
and
\begin{equation}
\lambda_0
=
\tau\ln\left(\frac{1}{\tau}\right)
\left[
\kappa_0
+
\frac{1}{\ln(1/\tau)}\lambda_0^{(1)}
+
\mathcal O\left(\frac{1}{\ln^2(1/\tau)}\right)
\right].
\label{eq:thin_eigenvalue_expansion}
\end{equation}

Substituting (\ref{eq:thin_eigenfunction_expansion}) and (\ref{eq:thin_eigenvalue_expansion}) into the eigenvalue equation for (\ref{eq:thin_operator_first_correction}) yields
\begin{equation}
\mathbfss C\mathcal P\bm u_0^{(1)}
+
\mathbfss C\mathcal H\bm v_0
+
\mathbfss B\bm v_0
=
\kappa_0\bm u_0^{(1)}
+
\lambda_0^{(1)}\bm v_0.
\label{eq:first_order_eigenfunction_equation}
\end{equation}
Since $\bm v_0$ is independent of depth, it is sufficient to evaluate $\mathcal H1$:
\begin{equation}
(\mathcal H1)(t) = 1-\gamma_{\rm E}
-\frac{t}{\tau}
\ln\left(\frac{t}{\tau}\right)
-\left(1-\frac{t}{\tau}\right)
\ln\left(1-\frac{t}{\tau}\right),
\label{eq:H_on_constant}
\end{equation}
with depth average
\begin{equation}
\mathcal P(\mathcal H1)
=
\frac{3}{2}
-\gamma_{\rm E}.
\label{eq:H_constant_average}
\end{equation}
Solving (\ref{eq:first_order_eigenfunction_equation}) and fixing the otherwise arbitrary normalization of the perturbed eigenfunction by requiring $\mathcal P\phi=0$, we obtain
\begin{equation}
\lambda_0^{(1)}
=
\frac{3}{4}
\left(
\frac{3}{4}
-\gamma_{\rm E}
\right).
\label{eq:first_order_eigenvalue_correction}
\end{equation}
and
\begin{equation}
\bm u_0^{(1)}(t;\tau)
=
\phi(t)\bm v_0
+
\frac{1}{4}\bm v_1,
\label{eq:first_order_eigenfunction_correction}
\end{equation}
in which 
\begin{equation}
\phi(t)=(\mathcal H1)(t)-\frac{3}{2}+\gamma_{\rm E}.
\label{eq:phi_tau_definition}
\end{equation}

The dominant polarized eigenfunction is therefore
\begin{equation}
\bm u_0(t;\tau)
=
\bm v_0
+
\frac{1}{\ln(1/\tau)}
\left[
\phi(t)\bm v_0
+
\frac{1}{4}\bm v_1
\right]
+
\mathcal O
\left[
\frac{1}{\ln^2(1/\tau)}
\right].
\label{eq:first_order_full_polarized_eigenfunction}
\end{equation}
The corresponding dominant eigenvalue is
\begin{equation}
\lambda_0
=
\frac{3\tau}{4}
\left[
\ln\left(\frac{1}{\tau}\right)
+
\frac{3}{4}
-\gamma_{\rm E}
+
\mathcal O
\left(
\frac{1}{\ln(1/\tau)}
\right)
\right].
\label{eq:first_order_full_polarized_eigenvalue}
\end{equation}

Using the definitions of $\bm v_0$ and $\bm v_1$, (\ref{eq:first_order_full_polarized_eigenfunction}) gives
\begin{align}
&a_*(t;\tau)
=
1
+
\frac{
\phi(t)+1/4
}{
\ln(1/\tau)
}
+
\mathcal O
\left[
\frac{1}{\ln^2(1/\tau)}
\right],
\label{eq:first_order_a_star}\\
&b_*(t;\tau)
=
-1
+
\frac{
-\phi(t)+1/4
}{
\ln(1/\tau)
}
+
\mathcal O
\left[
\frac{1}{\ln^2(1/\tau)}
\right].
\label{eq:first_order_b_star}
\end{align}
Although both coefficients acquire a depth-dependent correction through $\phi(t)$, this function cancels from their ratio at first logarithmic order. Taking the ratio of (\ref{eq:first_order_a_star}) and (\ref{eq:first_order_b_star}) and expanding in powers of $1/\ln(1/\tau)$, we obtain
\begin{equation}
\frac{b_*(t;\tau)}{a_*(t;\tau)}
=
-1
+
\frac{1}{
2\ln(1/\tau)
}
+
\mathcal O
\left[
\frac{1}{\ln^2(1/\tau)}
\right].
\label{eq:first_order_stokes_ratio}
\end{equation}

This property motivates writing the dominant eigenfunction in the factorized
form
\begin{equation}
\bm u_0(t;\tau)
\simeq
A(t;\tau)
\begin{pmatrix}
1\\
\alpha(\tau)
\end{pmatrix},
\label{eq:factorized_dominant_eigenfunction}
\end{equation}
where
\begin{equation}
A(t;\tau)
=
1
+
\frac{
\phi(t)+1/4
}{
\ln(1/\tau)
}
+
\mathcal O
\left[
\frac{1}{\ln^2(1/\tau)}
\right]
\label{eq:thin_common_amplitude}
\end{equation}
and
\begin{equation}
\alpha(\tau)
=
-1
+
\frac{1}{
2\ln(1/\tau)
}
+
\mathcal O
\left[
\frac{1}{\ln^2(1/\tau)}
\right]
\label{eq:thin_stokes_eigendirection}
\end{equation}
to first logarithmic order.

\subsubsection{Finite-optical-depth Galerkin approximation}
\label{sec:galerkin_thin}

The depth-independent Stokes eigendirection in (\ref{eq:factorized_dominant_eigenfunction}) provides a useful starting point for a finite-optical-depth approximation. The leading inverse-logarithmic expression (\ref{eq:thin_stokes_eigendirection}), however, becomes inaccurate as $\tau$ approaches unity. We therefore retain the constant-ratio structure and project the polarized operator onto the constant depth mode, following the Galerkin principle \citep[e.g.,][]{Atkinson_1997}. 

The exact slab-averaged kernel coefficients are given by
\begin{equation}
    L_m(\tau)=\mathcal P(\mathcal K_m1)
    =\frac{2}{\tau}\left[
        \frac{\tau}{m}-\frac{1}{m+1}+E_{m+2}(\tau)
    \right],
    \label{eq:finite_tau_Lm_exact}
\end{equation}
for $m=1,3,5$. The eigendirection of the projected operator is
\begin{equation}
    \alpha(\tau)
    =\frac{3(L_3-L_5)-\sqrt{\Delta_L}}{L_1-5L_3+2L_5},
    \label{eq:alpha_dominant_branch}
\end{equation}
where
\begin{equation}
    \Delta_L=L_1^2-8L_1L_3+2L_1L_5
              +24L_3^2-24L_3L_5+9L_5^2.
    \label{eq:alpha_discriminant}
\end{equation}
The branch is selected by the condition $\alpha\rightarrow-1$ as $\tau\rightarrow0$ and reproduces the first logarithmic correction in (\ref{eq:first_order_stokes_ratio}). The projected eigenproblem and the suppression of amplitude-weighting corrections are derived in Appendix~\ref{app:galerkin_thin}.

Within the factorized form (\ref{eq:factorized_dominant_eigenfunction}), the depth amplitude cancels from the polarization fraction, giving
\begin{equation}
    \Pi_\infty^\pm(\mu)
    =\frac{\alpha(\tau)(1-\mu^2)}{1+\alpha(\tau)\mu^2}.
    \label{eq:polarization_from_alpha}
\end{equation}
The angular distribution retains a dependence on the depth amplitude. Applying the constant-depth approximation also to the escape projection gives
\begin{equation}
    \Phi_{\infty,{\rm G}}^\pm(\mu)
    =\frac{[1+\alpha(\tau)\mu^2]\mu(1-e^{-\tau/\mu})}
    {[1/2-E_3(\tau)]+\alpha(\tau)[1/4-E_5(\tau)]}.
    \label{eq:thin_angular_galerkin}
\end{equation}
The amplitude-related error in this expression is estimated in Appendix~\ref{app:galerkin_escape}. This estimate applies for $\tau\rightarrow0$ with $\mu\gg\tau$.

\begin{figure*}
    \includegraphics[width=0.75\textwidth]{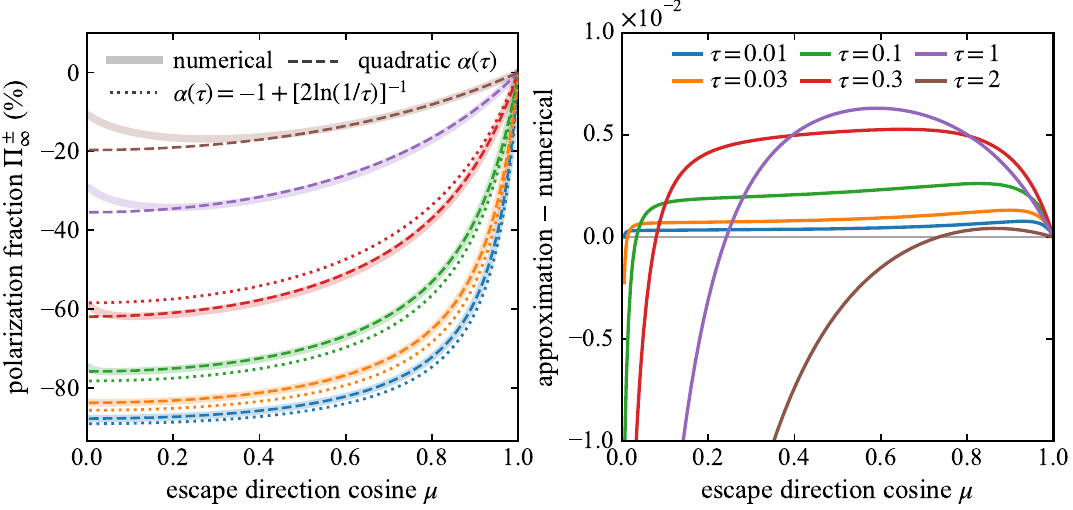}
    \caption{High-order signed linear-polarization fraction $\Pi_\infty^\pm(\mu)$ for optically thin slabs. Left: numerical     principal-eigenmode results compared with (\ref{eq:polarization_from_alpha}) evaluated using the finite-$\tau$ Galerkin eigendirection (\ref{eq:alpha_dominant_branch}) and the leading inverse-logarithmic approximation (\ref{eq:thin_stokes_eigendirection}). Right: the residual between Galerkin approximation and the numerical solution.}
    \label{fig:thin_polarization_comparison}
\end{figure*}

Figure~\ref{fig:thin_polarization_comparison} compares the numerical high-order polarization with (\ref{eq:polarization_from_alpha}) evaluated using the finite-$\tau$ Galerkin eigendirection and the leading inverse-logarithmic approximation. The Galerkin approximation is substantially more accurate and remains close to the numerical solution over much of the angular range even for $\tau\sim1$, well beyond the regime in which the leading inverse-logarithmic expression is reliable. At the larger optical depths shown, its largest deviations occur toward grazing emergence, $\mu\rightarrow0$. Here, the escape factor $e^{-t/\mu}$ selectively samples a narrow surface layer, where the depth dependence of $b_*(t;\tau)/a_*(t;\tau)$ makes a single slab-averaged eigendirection less accurate.

As $\tau\rightarrow0$, the polarization approaches the complete-polarization limit $\Pi_\infty^\pm=-1$ for every fixed $0<\mu<1$. This convergence is non-uniform near the slab normal, where axial symmetry requires $\Pi_\infty^\pm(1)=0$ at every finite optical depth. In sufficiently thin slabs, the polarization therefore rises rapidly to zero within a narrow angular boundary layer near $\mu=1$, whose width decreases as $\tau\rightarrow0$, as quantified by (\ref{eq:thin_polarization_boundary_layer}).

\begin{figure*}
    \includegraphics[width=0.75\textwidth]{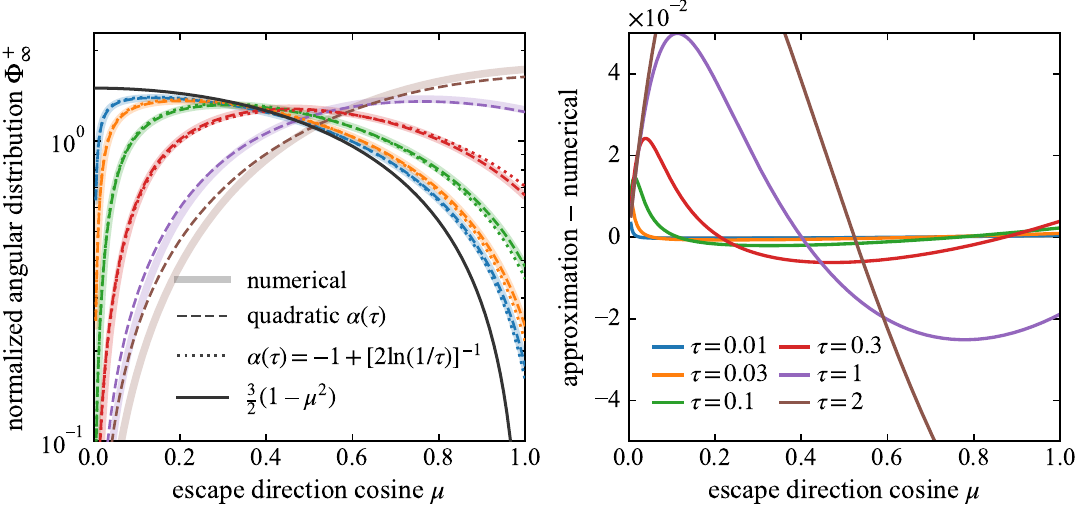}
    \caption{Normalized high-order angular distribution $\Phi_\infty^\pm(\mu)$ for optically thin slabs. Left: numerical principal-eigenmode results compared with (\ref{eq:thin_angular_galerkin}) using the Galerkin eigendirection and the leading inverse-logarithmic approximation. The strict thin-slab limit $3(1-\mu^2)/2$ is also shown. Right: the residual between Galerkin approximation and the numerical solution.}
    \label{fig:thin_angular_comparison}
\end{figure*}

Figure~\ref{fig:thin_angular_comparison} shows the corresponding
normalized high-order angular distributions. The finite-$\tau$
Galerkin expression (\ref{eq:thin_angular_galerkin}) reproduces the
numerical distributions very accurately throughout the optically thin
regime. As $\tau$ decreases, the curves approach the strict thin-slab
limit (\ref{eq:thin_asymptotic}) over an increasingly broad
range of escape directions. At finite optical depth, the angular
distribution tends to zero as $\mu\rightarrow0$, because grazing
escape predominantly samples a surface layer whose vertical optical
thickness scales as $\mu$. The Galerkin approximation remains
substantially more accurate than the leading inverse-logarithmic
approximation once $\tau$ is no longer asymptotically small.

\subsection{The polarized eigenmode in the optically thick limit}
\label{sec:thick_limit}

\subsubsection{Bulk Stokes eigenmodes in the diffusion limit}
\label{sec:thick_bulk}

To derive the asymptotic behaviour in the optically thick limit, we first consider the bulk region in the slab interior,
\begin{equation}
    t \gg 1,
    \qquad
    \tau-t \gg 1,
\end{equation}
where both free surfaces are separated from the local radiation field by many mean free paths.  In a thick slab, the longest-lived spatial mode varies on a scale comparable to the total optical depth, so that its characteristic wavenumber satisfies $k=O(\tau^{-1})\ll 1$.  This motivates a long-wavelength expansion of the bulk transport operator, closely related to the classical diffusion-eigenmode treatments of photon transport in optically thick media \citep{1985A&A...143..374S}.

To isolate the bulk behaviour, we write the depth operators as
\begin{equation}
    (\mathcal{K}_m f)(t')
    =
    \int_{-t'}^{\tau-t'}
    E_m(|s|)f(t'+s)\,{\rm d}s .
\end{equation}
Since the kernels $E_m(|s|)$ decay exponentially for $|s|\gg1$, the finite interval can be extended to an infinite medium in the bulk:
\begin{equation}
    (\mathcal{K}_{m, \infty} f)(t')
    \simeq
    \int_{-\infty}^{+\infty}
    E_m(|s|)f(t'+s)\,{\rm d}s .
\label{eq:bulk_infinite_convolution}
\end{equation}
We note that (\ref{eq:bulk_infinite_convolution}) is a local approximation and is not applicable within an optical depth of order unity from either free surface.

In the bulk, the kernels depend only on the depth separation $t'-t$. Fourier modes are therefore natural eigenfunctions of the corresponding depth operators. We consequently seek a bulk mode of the form
\begin{equation}
    \bm{u}_k(t)
    =
    \bm{v}(k)e^{ikt}.
\end{equation}
For each depth operator $\mathcal K_{\rm m}$, the action on the Fourier mode is equivalent to the Fourier transform:
\begin{equation}
    \mathcal{K}_m e^{ikt}
    =
    \widehat{K}_m(k)e^{ikt}.
\end{equation}
The corresponding Fourier symbol is
\begin{equation}
    \widehat{K}_m(k)
    =
    \int_{-\infty}^{+\infty}
    E_m(|s|)e^{iks}\,{\rm d}s =
    2\int_0^1
    \frac{u^{m-1}}{1+k^2u^2}\,{\rm d}u .
    \label{eq:Khat_exact}
\end{equation}
Thus the depth-integral eigenvalue equation separates into an algebraic problem for the two Stokes coefficients,
\begin{equation}
    \widehat{\mathbfss{K}}_{\rm pol}(k)\bm{v}(k)
    =
    \lambda(k)\bm{v}(k),
    \label{eq:bulk_matrix_eigenproblem}
\end{equation}
where
\begin{equation}
\widehat{\mathbfss{K}}_{\rm pol}(k)
=
\frac{3}{16}
\begin{pmatrix}
3\widehat{K}_1-\widehat{K}_3 &
-\widehat{K}_1+5\widehat{K}_3-2\widehat{K}_5
\\
-\widehat{K}_1+3\widehat{K}_3 &
3\widehat{K}_1-7\widehat{K}_3+6\widehat{K}_5
\end{pmatrix}.
\label{eq:Mk_exact}
\end{equation}

For the long-wavelength mode considered here, the Fourier symbols have the expansion
\begin{equation}
    \widehat{K}_{\rm m}(k)
    =
    2\left[
        \frac{1}{m}
        -\frac{k^2}{m+2}
        +O(k^4)
    \right].
    \label{eq:Khat_smallk}
\end{equation}
Substitution into (\ref{eq:Mk_exact}) gives
\begin{equation}
\widehat{\mathbfss{K}}_{\rm pol}(k)
=
\begin{pmatrix}
1-3k^2/10 &
{1}/{10}-{k^2}/{7}\\
-{k^2}/{10} &
{7}/{10}-{6k^2}/{35}
\end{pmatrix}
+O(k^4).
\label{eq:Mk_smallk}
\end{equation}

The matrix $\widehat{\mathbfss K}_{\rm pol}(k)$ has two eigenvalue
branches. Expanding the two eigenvalues about $k=0$ gives
\begin{equation}
    \lambda_{\rm D}(k)=
    1-\frac{k^2}{3}
    +O(k^4),\quad
    \lambda_{\rm P}(k)=
    \frac{7}{10}
    -\frac{29}{210}k^2
    +O(k^4).
\label{eq:lambda_bulk}
\end{equation}
The corresponding eigenvectors can also be expanded about $k=0$, giving
\begin{equation}
\bm{v}_{\rm D} \propto \begin{pmatrix}
        1\\
        -k^2/3
    \end{pmatrix}+\mathcal{O}(k^4),\qquad
    \bm{v}_{\rm P} \propto \begin{pmatrix}
        1 - 8k^2/9\\
        -3
    \end{pmatrix} + \mathcal{O}(k^4).
\label{eq:diffusion_decay_mode}
\end{equation}
In the limit $k\to0$, the first branch has
\begin{equation}
    \lambda_{\rm D}(0)=1,
    \qquad
    \bm{v}_{\rm D}(0)
    \propto
    \begin{pmatrix}
        1\\
        0
    \end{pmatrix},
\end{equation}
and therefore corresponds to an isotropic and unpolarized state in the strict long-wavelength limit.  This is the conservative diffusion branch.  The second branch has
\begin{equation}
    \lambda_{\rm P}(0)=\frac{7}{10},
    \qquad
    \bm{v}_{\rm P}(0)
    \propto
    \begin{pmatrix}
        1\\
        -3
    \end{pmatrix},
\end{equation}
and represents a rapidly decaying anisotropic-polarization mode. Since $\lambda_{\rm P}(k)$ remains separated from unity whereas $\lambda_{\rm D}(k)\to1$ as $k\rightarrow0$, the relative contribution of the former scales as $(\lambda_{\rm P}/\lambda_{\rm D})^{n-1}$ and is rapidly suppressed at high scattering orders.  The dominant high-order bulk state is therefore controlled by the diffusion branch.

Since the bulk operator depends only on $k^2$, the modes with $+k$ and $-k$ are degenerate and have the same Stokes eigenvector.  A physical finite-slab mode can therefore be constructed from a real linear combination of these two Fourier solutions.  We write the dominant diffusion mode in the form
\begin{equation}
    \bm{u}_{\rm D}(t)
    \simeq
    A_{\rm D}(t)\bm{v}_{\rm D}(k),
    \label{eq:finite_bulk_mode}
\end{equation}
where the slowly varying spatial envelope satisfies
\begin{equation}
    \frac{{\rm d}^2 A_{\rm D}}{{\rm d}t^2}
    +k^2A_{\rm D}
    =0 .
    \label{eq:bulk_diffusion_wave}
\end{equation}

Within an optical depth of order unity from either free surface, the infinite-medium approximation is no longer valid and the radiation field develops a surface layer. At leading diffusion order, the influence of the surface layer is characterized by the zero-wavenumber extrapolation length $z_0$, as in classical diffusion treatments \citep{1985A&A...143..374S,1994ApJ...434..570T}. The corresponding boundary conditions for the bulk envelope can be written as
\begin{equation}
    A_{\rm D}(0)-z_0 A_{\rm D}'(0) =0,\qquad
    A_{\rm D}(\tau)+z_0 A_{\rm D}'(\tau) =0.
\label{eq:effective_bc}
\end{equation}
The value of $z_0$ is determined by matching the interior diffusion mode to the polarized surface-transfer problem. Higher-order corrections are incorporated through a wavenumber-dependent surface phase length introduced below.

Because of the symmetry about the slab mid-plane, we can write
\begin{equation}
    A_{\rm D}(t)
    =
    A_0
    \cos\left[
        k\left(t-\frac{\tau}{2}\right)
    \right].
    \label{eq:principal_cosine_mode}
\end{equation}
Substitution into (\ref{eq:effective_bc}) gives the wavenumber condition
\begin{equation}
    \cos\left(\frac{k\tau}{2}\right)
    -
    kz_0
    \sin\left(\frac{k\tau}{2}\right)
    =0.
    \label{eq:finite_k_quantization}
\end{equation}
The smallest positive root corresponds to the longest-lived spatial mode.  In the limit $k\ll1$, this root is
\begin{equation}
    k_1\simeq
    \frac{\pi}{\tau+2z_0}
    +O(\tau^{-4}).
    \label{eq:k1_thick}
\end{equation}

The principal scattering-order eigenvalue of the finite slab is obtained by evaluating the bulk diffusion branch at $k=k_1$,
\begin{equation}
    \lambda_0(\tau)
    =
    \lambda_{\rm D}(k_1)=
    1- \frac{k_1^2}{3}
    +O(k_1^{4}).
    \label{eq:finite_lambda_from_bulk}
\end{equation}
Thus the departure of the principal eigenvalue from unity scales as $\tau^{-2}$, as expected for diffusive escape from an optically thick medium.

\subsubsection{Surface-layer asymptotics and the Chandrasekhar limit}
\label{sec:thick_surface}

The emergent radiation predominantly samples the region $t=O(1)$, where the infinite-medium bulk approximation is no longer valid. Reflection symmetry allows us to consider the lower surface at $t=0$. For an optically thick slab, the local surface structure is described by the half-space operator $\mathbfss K_{\rm H}$, obtained by extending the depth integrals from $[0,\tau]$ to $[0,\infty)$. 

The principal eigenfunction of the half-space surface problem can be expanded in powers of the principal wavenumber $k_1$:
\begin{equation}
    \bm u_0(t;\tau)
    =
    \bm U_0(t)
    +k_1^2\bm U_2(t)
    +k_1^4\bm U_4(t)
    +O(k_1^6).
    \label{eq:surface_eigenfunction_expansion_k4}
\end{equation}
The first two coefficient functions satisfy
\begin{equation}
    (\mathbfss I-\mathbfss K_{\rm H})\bm U_0=0,\qquad
    (\mathbfss I-\mathbfss K_{\rm H})\bm U_2=\frac13\bm U_0.
    \label{eq:polarized_milne}
\end{equation}
The first equation is the conservative polarized Milne problem, while the second describes the leading finite-thickness correction.

To match these surface solutions to the bulk diffusion mode in the overlap region $1\ll t\ll\tau$, we introduce a wavenumber-dependent surface phase length,
\begin{equation}
    z(k)=z_0+k^2z_2+O(k^4).
    \label{eq:surface_phase_length}
\end{equation}
The bulk envelope near the lower surface can then be represented in the form
\begin{equation}
    A_{\rm sur}(t;k)
    =
    \frac{\sin[k(t+z(k))]}{k},
    \label{eq:finite_diffusion_envelope}
\end{equation}
with the normalization that remains finite at fixed depth as $k\to0$.

Writing $X=t+z_0$, in the overlap region, we obtain
\begin{equation}
    A_{\rm sur}(t;k)
    =
    X+k^2\left(z_2-\frac{X^3}{6}\right)
    +O(k^4).
    \label{eq:surface_amplitude_expansion}
\end{equation}
Combining this expansion with the bulk Stokes eigenvector in (\ref{eq:diffusion_decay_mode}) gives the far-field matching conditions
\begin{equation}
    \bm U_0(t)\sim\bm P_0(t)
    =
    \begin{pmatrix}
        X\\
        0
    \end{pmatrix},
    \qquad
    \bm U_2(t)\sim\bm P_2(t)
    =
    \begin{pmatrix}
        z_2-X^3/6\\
        -X/3
    \end{pmatrix}.
    \label{eq:P0P2_definition}
\end{equation}
Calibration against numerical finite-slab eigenmodes gives
\begin{equation}
    z_0\simeq0.7121,\qquad
    z_2\simeq-0.2732.
    \label{eq:z0z2_numerical}
\end{equation}
The value of $z_0$ incorporates the polarized Milne surface response and improves on the Marshak approximation $z_0=2/3$ used in classical diffusion treatments \citep{1985A&A...143..374S} and Paper~I.

The half-space coefficient functions $\bm U_j$ are independent of the total slab optical depth. Once computed, they can be reused for any thick slab; within the present expansion, the dependence on $\tau$ enters through the powers of $k_1$. The matching construction, the equations for the localized boundary response, and the fourth-order extension are given in Appendix~\ref{app:surface_asymptotics}.

Projecting the surface solution onto the escaping radiation and expanding the ratio of Stokes $Q$ to $I$ gives
\begin{equation}
    \Pi_\infty^\pm(\mu;\tau)
    =\Pi_{\rm Ch}(\mu)+k_1^2\Delta\Pi_2(\mu)
    +k_1^4\Delta\Pi_4(\mu)+O(k_1^6).
    \label{eq:thick_pol_expansion}
\end{equation}
Here $\Pi_{\rm Ch}$ is the Chandrasekhar semi-infinite result, which vanishes at $\mu=1$ and approaches a positive polarization of approximately $11.7$ per cent toward the limb \citep{1960ratr.book.....C}. The normalized angular distribution has the similar expansion
\begin{equation}
    \Phi_\infty^\pm(\mu;\tau)
    =\Phi_{\rm thick}(\mu)+k_1^2\Delta\Phi_2(\mu)
    +k_1^4\Delta\Phi_4(\mu)+O(k_1^6).
    \label{eq:thick_ang_expansion}
\end{equation}
Its coefficient functions satisfy
\begin{equation}
    \int_0^1\Phi_{\rm thick}(\mu)\,{\rm d}\mu=1,\qquad \int_0^1\Delta\Phi_2(\mu)\,{\rm d}\mu
     =\int_0^1\Delta\Phi_4(\mu)\,{\rm d}\mu=0.
    \label{eq:ang_normalization_thick}
\end{equation}
The escape integrals and the expressions relating these observable coefficients to the Stokes projections are given in Appendix~\ref{app:surface_observables}.

For practical applications, we provide compact analytic representations of the universal angular functions, allowing their evaluation without reconstructing the half-space solutions. Define
\begin{equation}
    \Xi(x)=x\ln\left(1+\frac1x\right),\qquad\Xi(0)=0.
    \label{eq:Phi_surface}
\end{equation}
The polarization coefficient functions are approximated by
\begin{equation}
    \Pi_{\rm Ch}(\mu)\simeq
    0.1171(1-\mu)
    \exp\left[-\frac65\left(\Xi(\mu)+\frac\mu2\right)\right],
    \label{eq:Pi_Ch_compact_fit}
\end{equation}
\begin{equation}
    \Delta\Pi_2(\mu)\simeq
    -0.2933(1-\mu)
    \left[1+\frac9{10}\left(\mu+\frac4{15}\Xi(\mu)\right)\right],
    \label{eq:Pi2_compact_fit}
\end{equation}
and
\begin{equation}
    \Delta\Pi_4(\mu)\simeq0.2933\cdot\frac3{10}(1-\mu)\times\left[1+\frac87\left(
        \mu(1-\mu)-\frac45\mu^3+\frac5{16}\Xi(\mu)
    \right)\right].
    \label{eq:Pi4_compact_fit}
\end{equation}
The coefficients $0.1171$ and $-0.2933$ approximate the numerical grazing-direction values of $\Pi_{\rm Ch}$ and $\Delta\Pi_2$; the fourth-order fit reuses the latter amplitude through the factor $0.2933\times3/10$.

For the normalized angular distribution, we use
\begin{equation}
    \Phi_{\rm thick}(\mu)\simeq
    0.8227\mu\left(1+\frac32z_0\mu\right)
    \exp[\gamma_{\rm E}\Xi(\mu)],
    \label{eq:Phi_thick_fit}
\end{equation}
\begin{equation}
    \Delta\Phi_2(\mu)\simeq\Phi_{\rm thick}(\mu)
    \left[0.5960-\frac\mu{20}-\frac97\mu^2+\frac{\Xi(\mu)}4\right],
    \label{eq:Phi2_fit}
\end{equation}
and
\begin{equation}
    \Delta\Phi_4(\mu)\simeq\Phi_{\rm thick}(\mu)
    \left[-0.0723+\frac\mu{20}-\frac89\mu^2+\frac{13}{10}\mu^4
        +\frac{\Xi(\mu)}{21}\right].
    \label{eq:Phi4_fit}
\end{equation}
The normalization requirements in (\ref{eq:ang_normalization_thick}) determine the prefactor and the constant terms in these fits. 

\begin{figure*}
    \centering
    \includegraphics[width=0.75\textwidth]{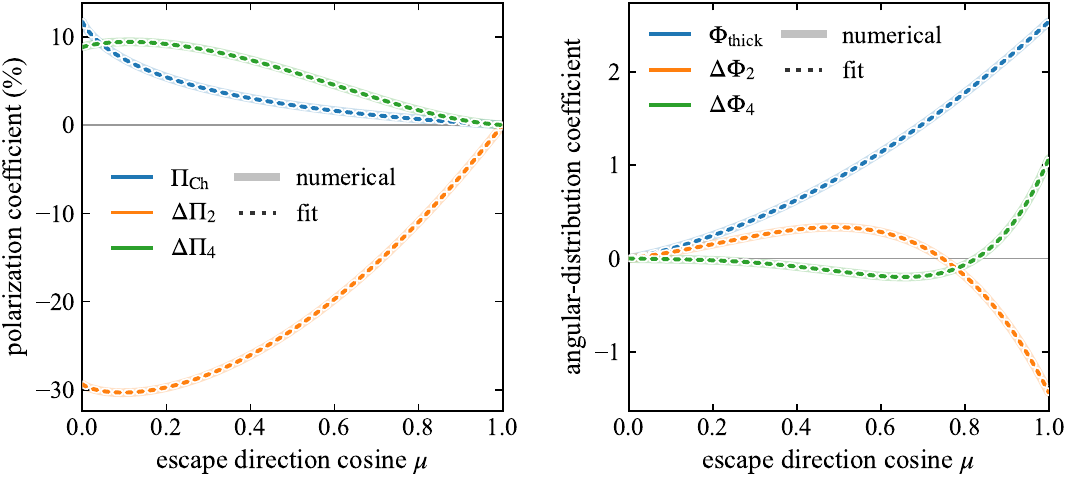}
    \caption{Universal coefficient functions entering the optically
    thick expansions of polarization (\ref{eq:thick_pol_expansion})
    and normalized angular distribution (\ref{eq:thick_ang_expansion}).
    Left: the Chandrasekhar polarization $\Pi_{\rm Ch}(\mu)$ and the
    correction coefficients $\Delta\Pi_2(\mu)$ and $\Delta\Pi_4(\mu)$,
    expressed in per cent. Right: the leading angular distribution
    $\Phi_{\rm thick}(\mu)$ and the correction coefficients
    $\Delta\Phi_2(\mu)$ and $\Delta\Phi_4(\mu)$.
    Solid curves show the numerical results, while dotted curves show
    the compact analytic representations.}
    \label{fig:thick_surface_coefficients}
\end{figure*}

The left panel of Figure~\ref{fig:thick_surface_coefficients} shows the polarization coefficient functions. The Chandrasekhar polarization decreases toward the slab normal. The negative second-order coefficient reduces this polarization as the slab becomes thinner, whereas the fourth-order coefficient is positive and has a different angular dependence. All three functions vanish at $\mu=1$, as required by axial symmetry. Although a correction coefficient can be comparable to or larger than the zeroth-order term, its actual contribution is weighted by $k_1^2$ or $k_1^4$.

The right panel shows the angular-distribution coefficient functions. The leading function $\Phi_{\rm thick}$ is positive for $0<\mu\leq1$, while the two correction functions contain both positive and negative regions. Their zero angular integrals imply a redistribution of escaping photons among directions without changing the total normalization. Their signs identify the directions enhanced or depleted at each order.

The compact analytic representations reproduce all six numerical coefficient functions with maximum absolute deviations below $3\times10^{-4}$, with the polarization coefficients expressed as dimensionless fractions. The full residual curves and numerical details are provided in the accompanying notebook.

\begin{figure*}
    \includegraphics[width=0.75\textwidth]{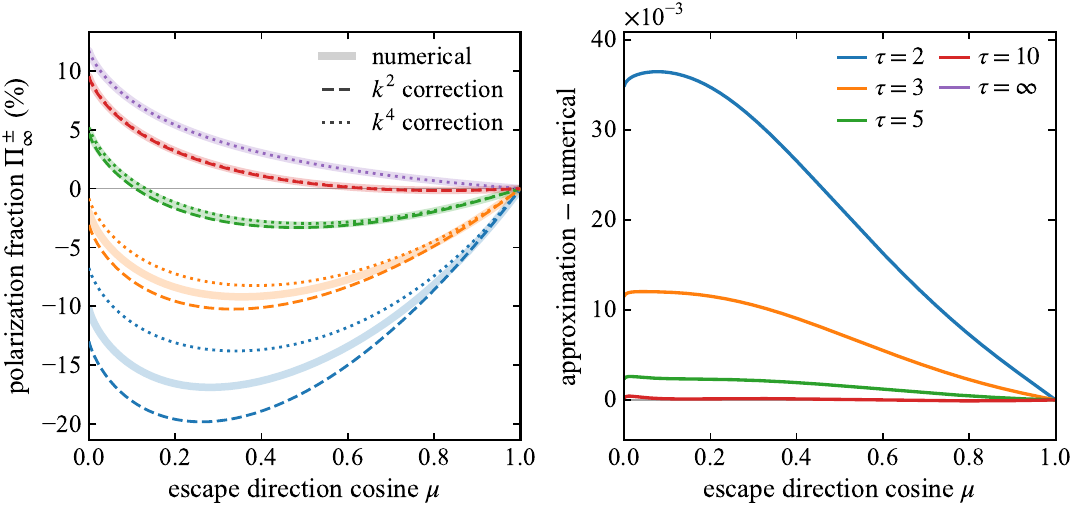}
    \caption{High-order signed linear-polarization fraction
    $\Pi_\infty^\pm(\mu)$ for optically thick slabs. Left: numerical
    principal-eigenmode results compared with
    (\ref{eq:thick_pol_expansion}) truncated after the $k_1^2$ and
    $k_1^4$ terms.
    Right: the residual between the fourth-order reconstruction and the numerical
    solution.}
    \label{fig:thick_polarization_reconstruction}
\end{figure*}

We next examine the reconstruction of a finite optically thick slab. Using the compact fits above and the leading wavenumber estimate (\ref{eq:k1_thick}), Figure~\ref{fig:thick_polarization_reconstruction} compares the numerical high-order polarization with the expansion truncated after the $k_1^2$ and $k_1^4$ terms. As $\tau$ increases, the numerical solution approaches the Chandrasekhar half-space limit. The second-order correction captures the leading finite-thickness departure, while inclusion of the fourth-order term substantially improves the agreement and provides a useful approximation for $\tau\gtrsim5$. The right panel shows the residuals of the fourth-order reconstruction.

\begin{figure*}
    \includegraphics[width=0.75\textwidth]{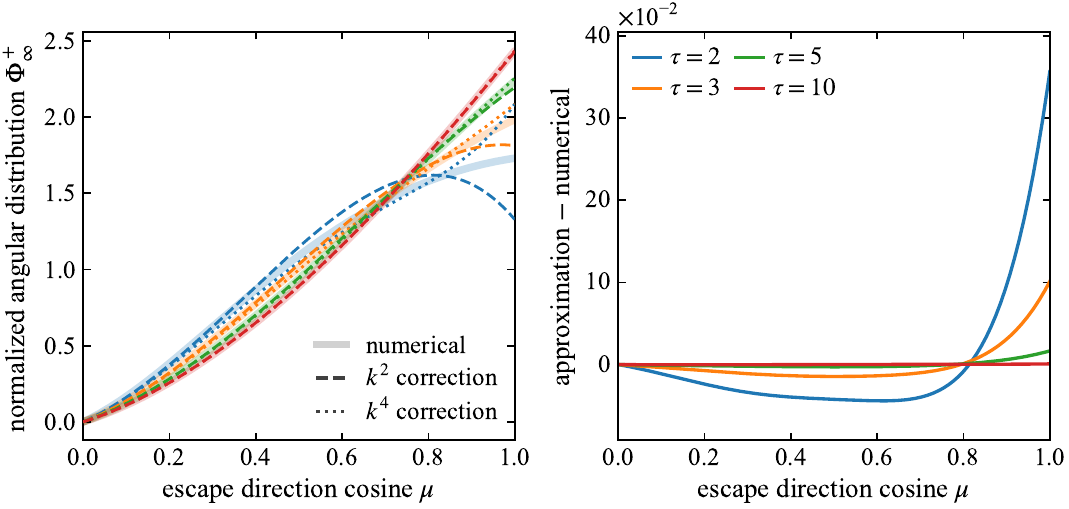}
    \caption{Normalized high-order angular distribution
    $\Phi_\infty^\pm(\mu)$ for optically thick slabs. Left: numerical
    principal-eigenmode results compared with
    (\ref{eq:thick_ang_expansion}) truncated after the $k_1^2$ and
    $k_1^4$ terms. Right: the residual between the fourth-order reconstruction and the
    numerical solution.}
    \label{fig:thick_angular_reconstruction}
\end{figure*}

Figure~\ref{fig:thick_angular_reconstruction} shows the corresponding normalized angular distributions and the residuals of the fourth-order reconstruction. The numerical distributions approach the half-space form as the slab becomes thicker. The second-order correction accounts for the leading finite-thickness modification, and the fourth-order term gives a further improvement, particularly for $\tau\sim5$.

The construction can be continued to higher orders; the accompanying notebook illustrates improvements from the $k_1^6$ correction. These algebraic corrections do not include the direct influence of the opposite surface, which is omitted by the half-space approximation and scales nearly as $O(e^{-\tau})$ up to algebraic prefactors. Once this contribution becomes comparable to the retained corrections, increasing the expansion order need not yield a systematic improvement.

\subsection{Eigenvalue channels and the approach to the high-order regime}
\label{sec:eigenvalue_channels}

\subsubsection{Global eigenvalue structure and leading spectral modes}
\label{sec:DP_channels}

The optically thin and optically thick analyses above both reveal two distinct Stokes branches of the polarized transport operator. The two branches identified in the opposite asymptotic limits can be continuously connected to one another. In this subsection, we use these asymptotic structures to classify the eigenvalue spectrum at finite optical depth.

In the optically thin limit, the leading polarized operator can be written as $\mathcal K_1\mathbfss C$. The two eigenvectors of $\mathbfss C$ are $\bm{v}_0$ and $\bm{v}_1$ given by (\ref{eq:thin_constant_matrix_eigenvectors}), with the corresponding eigenvalues $\kappa_0=3/4$ and $\kappa_1=3/8$ given by (\ref{eq:thin_constant_matrix_eigenvalues}). Let $f_j(t)$ denote the $j$-th depth eigenfunction of $\mathcal K_1$, 
\begin{equation}
    \mathcal K_1 f_j
    =
    \omega_j f_j.
\end{equation}
The corresponding thin-slab polarized modes are therefore
\begin{equation}
    \bm u_{{\rm D},j}
    \simeq
    f_j(t)\bm v_0,
    \qquad
    \bm u_{{\rm P},j}
    \simeq
    f_j(t)\bm v_1,
    \qquad
    \tau\rightarrow0,
    \label{eq:thin_DP_modes}
\end{equation}
with
\begin{equation}
    \lambda_{{\rm D},j}
    \simeq
    \frac{3}{4}\omega_j,
    \qquad
    \lambda_{{\rm P},j}
    \simeq
    \frac{3}{8}\omega_j.
    \label{eq:thin_DP_eigenvalues}
\end{equation}
The two Stokes branches identified in the optically thin limit can be continuously connected to the diffusion and  anisotropic-polarization branches of the optically thick limit in a diabatic sense. We therefore use the labels $D$ and $P$ for these two connected spectral families.

In the optically thick limit, these two connected spectral families approach the long-wavelength branches (\ref{eq:lambda_bulk}) derived in the bulk calculation, with the corresponding eigendirections given by (\ref{eq:diffusion_decay_mode}). For a fixed spatial order, the characteristic wavenumber decreases as $k=O(\tau^{-1})$ as the slab becomes optically thick. The two families therefore approach the asymptotic endpoints
\begin{equation}
    \lambda_{{\rm D},j}\to1,
    \qquad
    \lambda_{{\rm P},j}\to\frac{7}{10},
    \qquad
    \tau\rightarrow\infty.
    \label{eq:DP_thick_endpoints}
\end{equation}
The $D$ family therefore forms the sequence of increasingly long-lived diffusion modes, whereas the $P$ family remains associated with the more rapidly damped polarization-relaxation branch.

However, the $D/P$ classification is not an exact symmetry decomposition of the finite-slab operator. The matrices $\mathbfss A_1$, $\mathbfss A_3$, and $\mathbfss A_5$ in (\ref{eq:polarized_operator_decomposition}) do not share a common eigenbasis, so the full finite-slab operator cannot be decomposed into invariant $D$ and $P$ sectors.

Nevertheless, since $\mathbfss K_{\rm pol}$ depends continuously on the optical depth $\tau$, its isolated eigenvalues and associated eigenfunctions can be followed as $\tau$ varies, in accordance with the perturbation theory of self-adjoint operators \citep[e.g.,][]{Kato1995}. This allows the Stokes character identified in the optically thin limit to be traced through intermediate optical depths and connected to the corresponding $D$- and $P$-branches in the optically thick limit.

The behaviour near a degeneracy is determined by parity. Modes of opposite parity cannot couple and may therefore cross exactly, whereas modes of the same parity can mix and generally undergo avoided crossings. Near a same-parity $D_j$--$P_k$ near-degeneracy, the local spectral behaviour can be represented in the diabatic $D/P$ channel basis by
\begin{equation}
    \mathbfss H_{\rm eff}
    =
    \begin{pmatrix}
        \lambda_{{\rm D},j} & g_{jk}\\
        g_{jk} & \lambda_{{\rm P},k}
    \end{pmatrix},
    \label{eq:DP_effective_operator}
\end{equation}
where $\lambda_{{\rm D},j}$ and $\lambda_{{\rm P},k}$ are the smoothly continued $D$- and $P$-connected channel values introduced above, and $g_{jk}$ measures their local coupling. For opposite-parity modes, symmetry requires $g_{jk}=0$, allowing an exact crossing. For same-parity modes, a finite $g_{jk}$ produces an avoided crossing, across which the two exact eigenfunctions exchange their $D/P$ character.

For describing the global spectral structure, we follow the $D/P$ character diabatically through same-parity avoided crossings. This construction is particularly useful here because the numerical calculations show that the coupling between the $D$- and $P$-connected channels is always weak, so that the associated avoided-crossing gaps remain narrow. For example, the largest gap encountered along the $P_0$-connected channel is only $1.2$ per cent of the local eigenvalue.

Among the $D/P$-connected channels introduced above, the long-lived part of the scattering-order spectrum is controlled primarily by $\lambda_{{\rm D},0}$, $\lambda_{{\rm D},1}$, and $\lambda_{{\rm P},0}$. The principal eigenvalue is given by the lowest $D$-connected eigenvalue, $\lambda_0=\lambda_{{\rm D},0}$, whereas the latter two channels account for the leading subdominant behaviour. 

The $P_0$-connected mode is even, whereas the first excited $D_1$-connected mode is odd. The two modes therefore cannot couple, and their interchange in eigenvalue ordering occurs through an exact crossing at $\tau\simeq2.55$. The leading subdominant eigenvalue can consequently be represented by
\begin{equation}
    \lambda_{\rm sub}(\tau)
    =
    \max
    \left[
        \lambda_{{\rm P},0}(\tau),
        \lambda_{{\rm D},1}(\tau)
    \right].
    \label{eq:lambda_sub_DP}
\end{equation}
Below the crossing, the leading subdominant contribution is the  anisotropic-polarization mode $P_0$, whereas above it the diffusion mode $D_1$ dominates.

These three eigenvalues determine both the decay of the high-order component and the rate at which the radiation approaches the high-order regime. For practical applications, it is useful to have compact expressions for the three low-order channel eigenvalues over the full optical-depth range. Following the interpolation strategy of \citet{1994ApJ...434..570T}, we construct smooth expressions that connect the optically thin results derived in Section~\ref{sec:thin_polarized_eigenmode} to the optically thick limits obtained in Section~\ref{sec:thick_limit}.

We first introduce the following regularized thin-side expressions:
\begin{equation}
\begin{aligned}
    \lambda_{{\rm D},0}^{\rm thin}
    &=
    1-\exp\left\{
        -\frac{3\tau}{4}
        \left[
            \ln\left(1+\frac{1}{\tau}\right)
            +\frac{3}{4}
            -\gamma_{\rm E}
        \right]
    \right\},\\
    \lambda_{{\rm D},1}^{\rm thin}
    &=
    1-\exp(-s_1\tau),\\
    \lambda_{{\rm P},0}^{\rm thin}
    &=
    \frac{7}{10}
    \left[
        1-\exp\left\{
            -\frac{10}{7}\cdot\frac{3\tau}{8}
            \left[
                \ln\left(1+\frac{1}{\tau}\right)
                +\frac{7}{4}
                -\gamma_{\rm E}
            \right]
        \right\}
    \right],
\end{aligned}
\label{eq:regularized_channel_eigenvalues}
\end{equation}
where
\begin{equation}
    s_1\simeq0.587252
\end{equation}
is the leading coefficient of the first excited $D$-connected mode, $\lambda_{{\rm D},1}\sim s_1\tau$, in the optically thin limit.

On the thick side, we evaluate the eigenvalue branches $\lambda_{\rm D}(k)$ and $\lambda_{\rm P}(k)$ directly from the full Fourier-symbol matrix (\ref{eq:Mk_exact}), yielding
\begin{equation}
\begin{aligned}
    \lambda_{{\rm D},0}^{\rm thick}
    &=
    \lambda_{\rm D}(k_1),\\
    \lambda_{{\rm D},1}^{\rm thick}
    &=
    \lambda_{\rm D}(2k_1),\\
    \lambda_{{\rm P},0}^{\rm thick}
    &=
    \lambda_{\rm P}(k_1),
\end{aligned}
\label{eq:thick_channel_eigenvalues}
\end{equation}
where $k_1$ is given by (\ref{eq:k1_thick}). We then smoothly
connect the thin- and thick-side expressions in logarithmic form
\begin{equation}
\begin{aligned}
    \ln\lambda_{{\rm D},0}
    \simeq&
    \left\{
        1-
        \exp\left[
            -\left(\frac{3\tau}{8}\right)^{4/3}
        \right]
    \right\}
    \ln\lambda_{{\rm D},0}^{\rm thick}
    +
    \exp\left(-\frac{7\tau}{10}\right)
    \ln\lambda_{{\rm D},0}^{\rm thin},
    \\
    \ln\lambda_{{\rm D},1}
    \simeq&
    \left[
        1-
        \exp\left(-\frac{\tau}{3}\right)
    \right]
    \ln\lambda_{{\rm D},1}^{\rm thick}
    +
    \ln\lambda_{{\rm D},1}^{\rm thin},
    \\
    \ln\lambda_{{\rm P},0}
    \simeq&
    \left\{
        1-
        \exp\left[
            -\left(\frac{5\tau}{7}\right)^{7/6}
        \right]
    \right\}
    \ln\lambda_{{\rm P},0}^{\rm thick}+
    \exp\left[
        -\left(\frac{2\tau}{3}\right)^{9/8}
    \right]
    \ln\lambda_{{\rm P},0}^{\rm thin}.
\end{aligned}
\label{eq:fit_channel_eigenvalues}
\end{equation}
The switching factors in (\ref{eq:fit_channel_eigenvalues}) are empirical.

\begin{figure*}
	\includegraphics[width = 0.75\textwidth]{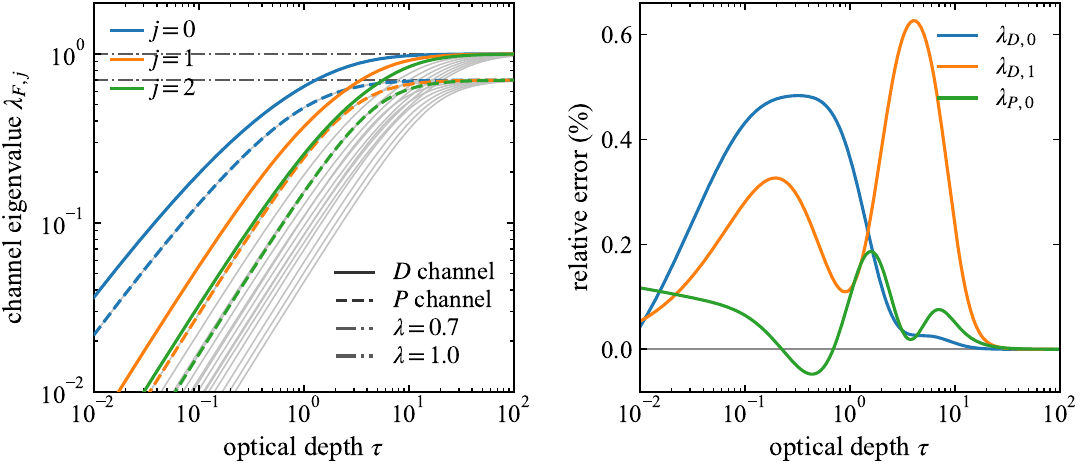}
    \caption{Eigenvalue structure of the polarized slab recursion operator. Left: the lowest-order mode-resolved $D$- and $P$-connected channels for $j=0,1,2$, shown by solid and dashed coloured curves, respectively. Thin light-grey curves show a representative subset of the eigenvalues of the full finite-slab polarized operator. The horizontal dot-dashed lines indicate the long-wavelength limits $\lambda_{\rm D}\rightarrow1$ and $\lambda_{\rm P}\rightarrow0.7$. Right: relative errors of the practical approximations to $\lambda_{{\rm D},0}$, $\lambda_{{\rm D},1}$, and $\lambda_{{\rm P},0}$ with respect to the numerical mode-resolved channel values.}
\label{fig:DP_eigenvalue_channels}
\end{figure*}

Figure~\ref{fig:DP_eigenvalue_channels} summarizes the resulting finite-slab eigenvalue structure and the accuracy of the low-order channel approximations. The left panel shows a representative subset of the exact finite-slab spectrum together with the lowest-order $D$- and $P$-connected channels. The narrow same-parity avoided crossings confirm that the smooth diabatic representation remains accurate for the low-order modes. The right panel compares the practical approximations with the numerical mode-resolved channel values and shows that the relative deviations remain below about $0.7$ per cent over a broad range of optical depths.

\begin{figure*}
	\includegraphics[width = 0.75\textwidth]{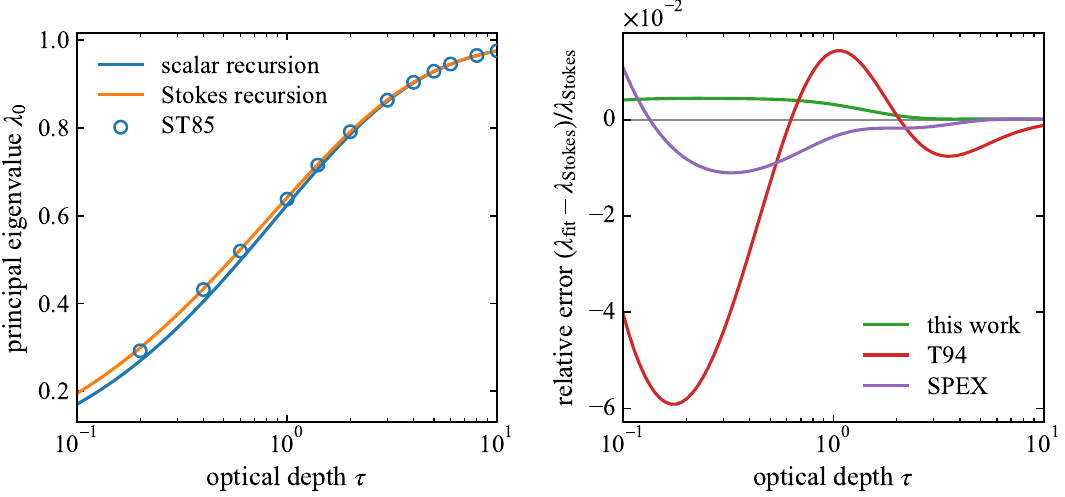}
    \caption{Principal eigenvalue of the slab scattering-order recursion as a function of the total optical depth $\tau$. Left: numerical principal eigenvalues obtained from the polarization-averaged scalar recursion and from the full Stokes recursion. Open circles show the flat-layer transfer results tabulated by \citet{1985A&A...143..374S}. Right: relative deviations from the numerical Stokes-recursion eigenvalue for the practical approximation developed in this work, the \citet{1994ApJ...434..570T} approximation, and the corresponding \texttt{SPEX} slab approximation \citep{1996uxsa.conf..411K}.}
\label{fig:principal_eigenvalue}
\end{figure*}

The principal eigenvalue $\lambda_0$ is obtained directly from the full finite-slab Stokes recursion operator and is also recovered from the high-order limit of the scattering-order escape ratio given by (\ref{eq:escape_ratio}). An independent comparison is provided by the polarized radiative transfer calculation of \citet{1985A&A...143..374S}. After expressing their results in the notation adopted here, their transfer results agree closely with the present numerical $\lambda_0$ over the available optical-depth range, as shown in Figure~\ref{fig:principal_eigenvalue}. This figure also compares the numerical result with the analytical approximation of \citet{1994ApJ...434..570T} and the modified plane-parallel form
adopted in \texttt{SPEX} \citep{1996uxsa.conf..411K}.\footnote{\url{https://spex-xray.github.io/spex-help/models/comt.html}}

In this figure, the polarization-averaged scalar and full Stokes recursions give similar but not identical principal eigenvalues, with visible difference occurring at small optical depths, $\tau\lesssim1$. This difference reflects the feedback of the polarization generated at one scattering into the angular redistribution at subsequent scatterings.

The results above therefore provide practical expressions for the principal and leading subdominant eigenvalues over the full optical-depth range. Their relative separation determines the rate at which the scattering-order distribution approaches the asymptotic high-order state, which we quantify in the following subsection.

\subsubsection{Onset of the high-order regime}
\label{sec:high_order_onset}

The eigenmode expansion derived in Section~\ref{sec:polarized_eigenmodes} provides a direct measure of how rapidly the scattering-order distribution approaches its high-order asymptotic form. Because the eigenfunctions are orthonormal under the weighted inner product introduced in Section~\ref{sec:polarized_eigenmodes}, the relative remainder with respect to the principal contribution can be
written exactly as
\begin{equation}
    \eta_n
    =
    \left[
        \sum_{j\geq1}
        \left|
            \frac{c_j}{c_0}
        \right|^2
        \left|
            \frac{\lambda_j}{\lambda_0}
        \right|^{2(n-1)}
    \right]^{1/2}.
    \label{eq:relative_remainder}
\end{equation}

For the finite-slab spectrum considered here, the slowest-decaying correction is characterized by
\begin{equation}
    \rho(\tau)
    =
    \frac{\lambda_{\rm sub}(\tau)}
         {\lambda_0(\tau)}
    =
    \frac{
        \max\left[
            \lambda_{{\rm P},0}(\tau),
            \lambda_{{\rm D},1}(\tau)
        \right]
    }{
        \lambda_{{\rm D},0}(\tau)
    }.
    \label{eq:spectral_convergence_factor}
\end{equation}
Then (\ref{eq:relative_remainder}) gives the bound
\begin{equation}
    \eta_n
    \leq
    \left[
        \sum_{j\geq1}
        \left|
            \frac{c_j}{c_0}
        \right|^2
    \right]^{1/2}\rho^{\,n-1}.
    \label{eq:remainder_bound}
\end{equation}
The coefficient prefactor contains the dependence on the seed-photon distribution, whereas $\rho$ is determined by the transport operator.

A characteristic scattering order for the onset of the high-order regime may therefore be defined by requiring the relative remainder to fall below a prescribed tolerance $\epsilon$. In many cases of interest, the source-dependent prefactor is expected to be of order unity, giving a representative estimate
\begin{equation}
    n_\epsilon
    \simeq
    1-
    \frac{
        \ln(1/\epsilon)
    }{
        \ln\rho
    }.
    \label{eq:spectral_convergence_order}
\end{equation}
This quantity provides a convenient estimate of the number of scattering orders required to suppress the leading non-principal spectral component to a fraction $\epsilon$ of the principal mode.

An important consequence of principal-mode dominance is the separation of the scattering-order dependence from the angular and Stokes structure. As discussed in Paper I and ST85, once the remainder becomes negligible, all sufficiently high scattering orders share the same spatial, angular, and polarization structure. In a Comptonization application in which photon energy is organized primarily by scattering order, this implies that the high-order spectral shape becomes asymptotically independent of viewing angle.

The limiting behaviour follows directly from the eigenvalue structure obtained above. In the optically thin limit, the leading subdominant mode is the $P_0$-connected mode. Since the $D_0$ and $P_0$ eigenvalues share the same leading depth eigenvalue, their ratio is simply $\kappa_1/\kappa_0=1/2$ and the intrinsic convergence order approaches
\begin{equation}
    n_\epsilon^{\rm thin}
    \to
    1+
    \frac{
        \ln(1/\epsilon)
    }{
        \ln2
    },
    \qquad
    \tau\to0.
    \label{eq:n_thin}
\end{equation}
Thus, the photons approach the principal high-order state after only a finite number of scattering orders in an optically thin slab.

The situation is different in the optically thick limit, where the leading subdominant mode is $D_1$, and
\begin{equation}
    \rho
    \simeq
    \frac{
        \lambda_{\rm D}(2k_1)
    }{
        \lambda_{\rm D}(k_1)
    },
    \qquad
    \tau\gg1.
    \label{eq:rho_thick}
\end{equation}
Using the long-wavelength diffusion branch (\ref{eq:lambda_bulk}), we obtain
\begin{equation}
    -\ln\rho
    \simeq
    k_1^2
    +O(k_1^4).
    \label{eq:rho_thick_expansion}
\end{equation}
The intrinsic convergence order becomes
\begin{equation}
    n_\epsilon^{\rm thick}
    \simeq
    1+
    \frac{
        (\tau+2z_0)^2
    }{
        \pi^2
    }
    \ln\left({1}/{\epsilon}\right),
    \qquad
    \tau\gg1.
    \label{eq:n_thick}
\end{equation}
The onset of the principal high-order state therefore scales as $n_\epsilon\propto\tau^2$ in an optically thick slab.

(\ref{eq:relative_remainder}) also indicates that the precise convergence for a particular seed-photon distribution depends on its modal projections. In particular, a source that is exactly symmetric about the slab mid-plane has zero projection onto the odd $D_1$ mode, as is the case for the isotropic internal source considered here. In that case the leading source-dependent correction is supplied by the next excited mode with non-zero projection, and the actual approach to the high-order state can be faster than the generic spectral estimate above.

\section{Energy-resolved polarization: validity and application to 4U 1624-49}
\label{sec:band_polarization}

\subsection{From scattering order to energy-resolved Stokes spectra}
\label{sec:band_polarization_formalism}

The scattering-order-resolved quantities derived in Section~\ref{sec:stokes_recursion} provide a direct description of how the angular distribution and polarization evolve during repeated scattering. Individual scattering orders, however, are not directly observable. An X-ray polarimeter typically measures the Stokes parameters in a finite energy interval, within which photons escaping after different numbers of scatterings contribute simultaneously. Since photon energy in thermal Comptonization is closely related to scattering history, the scattering-order representation provides a natural bridge between the transport calculation developed above and band-limited X-ray polarization.

Let $N_{I,n}^{\pm}(E,\mu)$ and $N_{Q,n}^{\pm}(E,\mu)$ denote the photon-number Stokes spectra escaping through the corresponding boundary after exactly $n$ scatterings, normalized such that
\begin{equation}
    \int_0^\infty
    N_{I,n}^{\pm}(E,\mu)\,{\rm d}E
    =
    I_n^\pm(\mu),
    \qquad
    \int_0^\infty
    N_{Q,n}^{\pm}(E,\mu)\,{\rm d}E
    =
    Q_n^\pm(\mu).
    \label{eq:order_resolved_spectral_stokes}
\end{equation}
Here $I_n^\pm(\mu)$ and $Q_n^\pm(\mu)$ are given by (\ref{eq:IQ_upper}) and (\ref{eq:IQ_lower}). Using (\ref{eq:order_resolved_escape_polarization}) and (\ref{eq:normalized_angular}), they may be written in the more useful form
\begin{equation}
    I_n^\pm(\mu)
    =
    P_n^\pm\Phi_n^\pm(\mu),
    \qquad
    Q_n^\pm(\mu)
    =
    I_n^\pm(\mu)\Pi_n^\pm(\mu).
    \label{eq:simple_IQ}
\end{equation}

For an observational energy band $B=[E_1,E_2]$, the exact band-integrated Stokes parameters are therefore
\begin{equation}
    \begin{aligned}
    &I_B^\pm(\mu)
    =
    \sum_{n=0}^{\infty}
    \int_{E_1}^{E_2}
    N_{I,n}^{\pm}(E,\mu)\,{\rm d}E ,\\
    &Q_B^\pm(\mu)
    =
    \sum_{n=0}^{\infty}
    \int_{E_1}^{E_2}
    N_{Q,n}^{\pm}(E,\mu)\,{\rm d}E .
\end{aligned}
\label{eq:band_IQ_exact}
\end{equation}
The corresponding signed polarization fraction is
\begin{equation}
    \Pi_B^\pm(\mu)
    =
    \frac{Q_B^\pm(\mu)}{I_B^\pm(\mu)} .
    \label{eq:band_polarization_exact}
\end{equation}

(\ref{eq:band_IQ_exact}) and (\ref{eq:band_polarization_exact}) are general, but their direct evaluation requires a redistribution calculation that simultaneously retains the energy, direction, polarization, and scattering-order dependences, as in full successive-scattering Comptonization calculations \citep[e.g.,][]{1996ApJ...470..249P}. The present Thomson recursion explicitly follows the angular and polarization distributions but does not itself evolve the photon energy. A fully energy-dependent extension of the Stokes recursion will be considered for future work. 

Paper I showed that the scattering-order statistics obtained in the Thomson framework can be combined with an order-dependent energy redistribution prescription to reproduce the broad properties of Comptonized spectra. We adopt the same strategy here. Let $S_0(E)$ denote the normalized seed-photon spectrum, and let $\mathcal R(E|E')$ be an angle-averaged single-scattering energy redistribution kernel satisfying
\begin{equation}
    \int_0^\infty
    \mathcal R(E|E')\,{\rm d}E=1.
\end{equation}
The normalized energy distribution after $n$ scatterings is then obtained recursively from
\begin{equation}
    S_n(E)
    =
    \int_0^\infty
    S_{n-1}(E')
    \mathcal R(E|E')\,{\rm d}E',
    \qquad n\geq1,
    \label{eq:energy_redistribution_recursion}
\end{equation}
with
\begin{equation}
    \int_0^\infty S_n(E)\,{\rm d}E=1.
    \label{eq:normalized_order_energy_distribution}
\end{equation}

To combine this energy redistribution with the present Thomson transport calculation, we adopt an energy--angle separability approximation at fixed scattering order,
\begin{equation}
    N_{I,n}^{\pm}(E,\mu)
    \simeq
    S_n(E)\,I_n^\pm(\mu),\quad N_{Q,n}^{\pm}(E,\mu)
    \simeq
    S_n(E)\,Q_n^\pm(\mu).
    \label{eq:energy_angle_factorization}
\end{equation}
Summing over all scattering orders gives the energy-resolved Stokes spectra,
\begin{equation}
    \begin{aligned}
    &I_E^\pm(E,\mu)
    =E
    \sum_{n=0}^{\infty} N_{I,n}^{\pm}(E,\mu)
    \simeq E
    \sum_{n=0}^{\infty} S_n(E)\,I_n^\pm(\mu),
    \\
    &Q_E^\pm(E,\mu)
    =E
    \sum_{n=0}^{\infty} N_{Q,n}^{\pm}(E,\mu)
    \simeq E
    \sum_{n=0}^{\infty} S_n(E)\,Q_n^\pm(\mu).
    \end{aligned}
    \label{eq:total_spectrum}
\end{equation}

For simplicity, we define the fraction of the $n$-th order spectrum contained in the band $B$ by
\begin{equation}
    w_{n,B}
    =
    \int_{E_1}^{E_2}
    S_n(E)\,{\rm d}E.
    \label{eq:band_order_weight}
\end{equation}
The band polarization can therefore be written directly in terms of the scattering-order quantities,
\begin{equation}
    \Pi_B^\pm(\mu)
    \simeq
    \frac{
        \displaystyle
        \sum_{n=0}^{\infty}
        w_{n,B}
        P_n^\pm
        \Phi_n^\pm(\mu)
        \Pi_n^\pm(\mu)
    }{
        \displaystyle
        \sum_{n=0}^{\infty}
        w_{n,B}
        P_n^\pm
        \Phi_n^\pm(\mu)
    } .
    \label{eq:band_polarization_order_sum}
\end{equation}
The polarization fractions of individual scattering orders must therefore not be averaged directly. Their Stokes intensities must first be weighted by the spectral contribution of each scattering order and summed before forming the ratio $Q_B^\pm/I_B^\pm$.

This factorization neglects correlations between energy redistribution and polarized angular transport. In addition, thermal electron motions modify the angular and polarization redistribution relative to the stationary-electron Thomson phase matrix adopted here, even when scattering remains in the Thomson regime in the electron rest frame \citep{1996ApJ...470..249P}. Nevertheless, Paper I showed that once the scattering-order and angular escape distributions are determined accurately, combining them with a comparatively simple order-dependent energy prescription already recovers much of the broad shape of the scalar Comptonized intensity spectrum.

The same level of robustness cannot be assumed for the polarized spectrum. The Stokes parameter $Q$ is a signed difference between orthogonally polarized intensity components and may also change sign with photon energy. Consequently, relatively small differential changes in the angular and energy redistribution can produce a substantial change in $Q_E(E)$ while leaving the total intensity $I_E(E)$ almost unaffected. A more accurate treatment of the energy redistribution is therefore required before the validity of (\ref{eq:total_spectrum}) can be assessed for polarized Comptonization.

In this paper, we adopt the angle-averaged Compton redistribution originally derived by \citet{1968PhRv..167.1159J}, using the corrected exact expression given by \citet[Appendix~A1]{1990MNRAS.245..453C}. We denote by $P_{\rm J}(E|E',\gamma)$ the normalized probability density for a photon of incident energy $E'$ to emerge with energy $E$ after scattering from an electron with Lorentz factor $\gamma$. For an isotropic thermal electron population, the redistribution kernel entering (\ref{eq:energy_redistribution_recursion}) is obtained by weighting the fixed-electron redistribution by the corresponding angle-averaged Compton scattering rate,
\begin{equation}
    {\cal R}_{\rm J}(E|E';\theta_{\rm e})
    =
    \frac{
    \displaystyle
    \int_1^\infty
    f_{\rm MJ}(\gamma;\theta_{\rm e})
    R_{\rm sc}(E',\gamma)
    P_{\rm J}(E|E',\gamma)\,
    {\rm d}\gamma
    }{
    \displaystyle
    \int_1^\infty
    f_{\rm MJ}(\gamma;\theta_{\rm e})
    R_{\rm sc}(E',\gamma)\,
    {\rm d}\gamma
    },
    \label{eq:thermal_jones_kernel}
\end{equation}
where $f_{\rm MJ}$ is the Maxwell--J\"uttner distribution \citep{1911AnP...339..856J}, $\theta_{\rm e}=kT_{\rm e}/m_{\rm e}c^2$ is the dimensionless electron temperature, and $R_{\rm sc}$ is the angle-averaged Compton scattering rate for an electron of Lorentz factor $\gamma$.

To assess the validity of the energy--angle separability approximation, we compare the resulting energy-resolved Stokes spectra with two complementary radiative-transfer calculations that retain the coupled energy, angular, and polarization redistribution of thermal Compton scattering: the polarized MCRT calculation described in Appendix~\ref{app:mcrt} and the polarized extension of \texttt{compPSc}.

\texttt{compPSc} is the convolution implementation of the \texttt{compPS} Comptonization code developed by \citet{2024ApJ...962..101Z} and employed as a spectral reference in Paper~I. However, the original \texttt{compPSc} implementation does not provide the energy-resolved polarized Stokes spectra. We therefore extended the source code to propagate the two-component Stokes vector $(I,Q)^{\rm T}$ in the plane-parallel slab. The extension follows the relativistic polarized Compton scattering matrix for an isotropic electron gas derived by \citet{1993A&A...275..325N} and the Stokes-vector iterative transfer formulation of \citet{1996ApJ...470..249P}. The scattering iteration in the native \texttt{compPSc} code is replaced by a two-component Stokes-vector iteration based on the thermal polarized Compton redistribution matrix, so that $I$ and $Q$ are evolved self-consistently through successive scatterings. The polarized MCRT calculation is described in Appendix~\ref{app:mcrt}. The MCRT and the extended \texttt{compPSc} calculations provide independent tests of the energy-resolved Stokes spectra reconstructed from the present recursion.

\begin{figure*}
	\includegraphics[width = 0.75\textwidth]{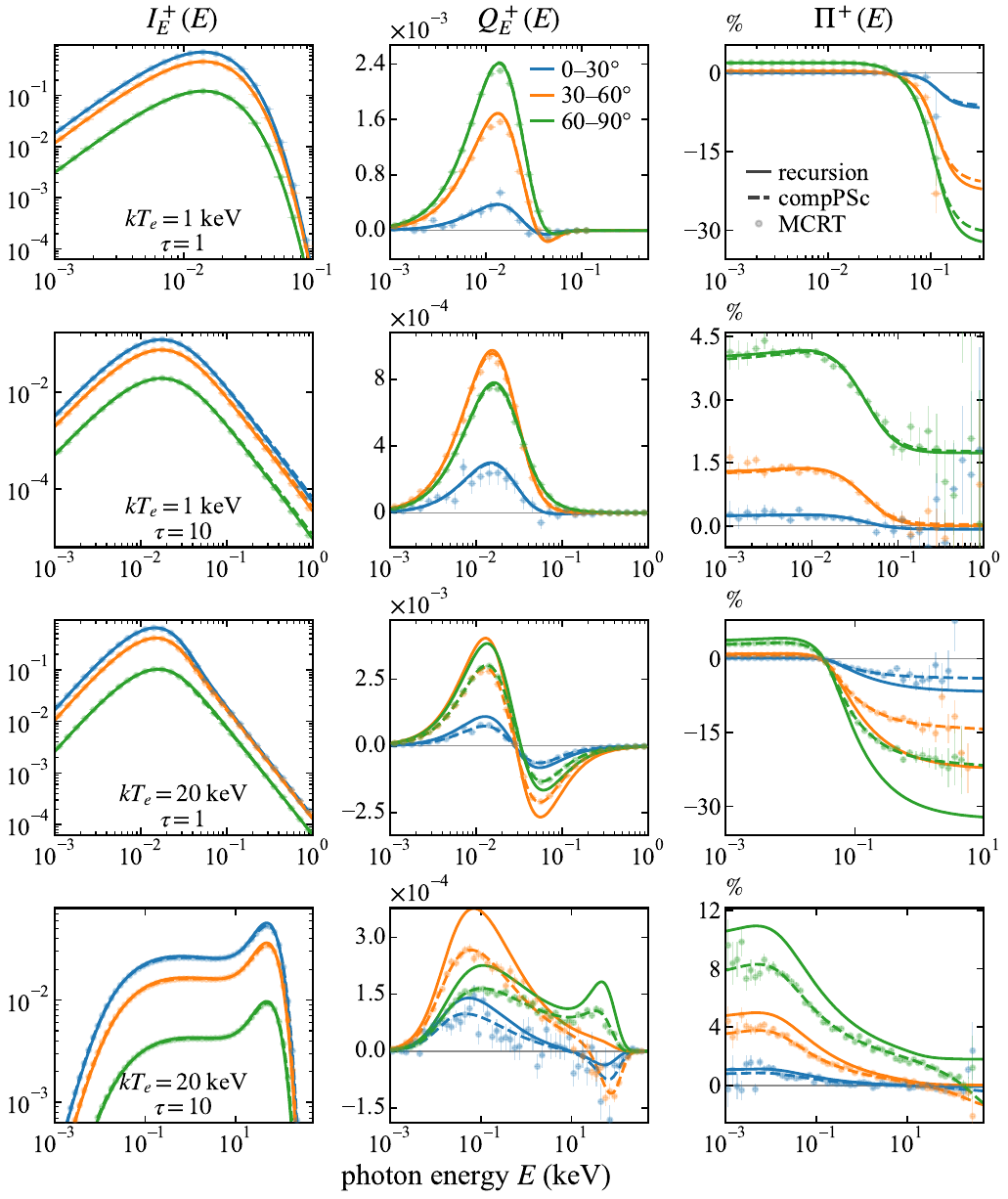}
    \caption{
    Comparison of the energy-resolved Stokes spectra emerging through the upper surface of the slab, calculated with the present scattering-order recursion combined with the Jones energy-redistribution kernel (solid curves), the polarized \texttt{compPSc} calculation (dashed curves), and MCRT simulations (points with statistical uncertainties). From left to right, the columns show the azimuth-integrated intensity spectrum $I_E^{+}(E)$, the signed linear-Stokes spectrum $Q_E^{+}(E)$, and the corresponding signed polarization fraction $\Pi^{+}(E)=Q_E^{+}(E)/I_E^{+}(E)$. The different rows correspond to different $kT_{\rm e}$ and $\tau$. Curves with different colours denote photons escaping within different inclination intervals. All calculations adopt the same initially unpolarized Lambert-law seed-photon injection through the lower boundary.
    }
    \label{fig:energy_resolved_stokes_benchmark}
\end{figure*}

Figure~\ref{fig:energy_resolved_stokes_benchmark} compares the energy-resolved Stokes spectra obtained with the present recursive approximation, the polarized \texttt{compPSc} calculation, and the polarized MCRT simulation. In all cases we adopt unpolarized blackbody seed photons with $kT_{\rm bb}=5\,{\rm eV}$, injected from the lower boundary with a Lambert-law angular distribution. The escaping radiation is divided into three inclination bins, $0^\circ$--$30^\circ$, $30^\circ$--$60^\circ$, and $60^\circ$--$90^\circ$, and the plotted spectra represent the angular averages within each bin. The MCRT spectra are accumulated from 400 independent realizations, each containing $10^6$ injected photons. The photon energy range $10^{-4}$--$10^3\,{\rm keV}$ is divided into 280 bins, and bins containing fewer than 50 photons are excluded from the displayed Monte Carlo spectra.

After replacing the peak-shift treatment of Paper~I by the Jones energy redistribution kernel, the intensity spectra $I_E^+(E)$ are reproduced remarkably well over the full range of viewing angles and energies shown in the figure. At low coronal temperature, $kT_{\rm e}=1\,{\rm keV}$, the same approximation also reproduces the linear-Stokes spectra $Q_E^+(E)$ well, in good agreement with both the polarized \texttt{compPSc} and MCRT calculations. By contrast, when the coronal temperature is increased to $kT_{\rm e}=20\,{\rm keV}$, substantial discrepancies appear in $Q_E^+(E)$ even though the intensity spectra remain in close agreement. These discrepancies highlight the limitations of combining stationary-electron Thomson angular transport with a separate, angle-averaged thermal energy redistribution.

The behaviour of the polarization spectrum $\Pi^+=Q_E^+/I_E^+$ further clarifies the regime of applicability of the approximation. In the cool and optically thick case ($kT_{\rm e}=1\,{\rm keV}$, $\tau=10$), all three calculations agree closely, and the polarization fraction approaches a nearly energy-independent asymptotic level at high energies. In the cool case with $\tau=1$, the three calculations remain consistent at low energies, while at higher energies the MCRT statistics become too sparse for a stringent comparison. Both the recursive approximation and the polarized \texttt{compPSc} calculation approach stable high-energy polarization levels, but these levels do not coincide exactly, indicating that the separable reconstruction already shows small departures from the full Compton calculation.

In the hot cases ($kT_{\rm e}=20\,{\rm keV}$), these departures become substantial: the recursive approximation still reproduces the broad qualitative trend, but fails to recover the detailed $\Pi^+(E)$ distributions given by the two full Compton reference calculations. 

To examine the validity of the separable treatment, we repeated the comparison between the recursive calculation and the polarized MCRT simulations over a grid in electron temperature and optical depth. For each model, we compare the escaping Stokes spectra in two inclination intervals, $0^\circ$--$45^\circ$ and $45^\circ$--$90^\circ$, and define
\begin{equation}
    \frac{\chi_X^2}{N_X}
    =
    \frac{1}{N_X}
    \sum_{j=1}^{N_X}
    \frac{
    \left[
    X_{\rm rec}(E_j)-X_{\rm MC}(E_j)
    \right]^2
    }{
    \sigma_{X,{\rm MC}}^2(E_j)
    },
    \qquad X=I,Q,
    \label{eq:mcrt_reduced_chi2}
\end{equation}
where $N_X$ is the number of spectral bins included in the comparison and $\sigma_{X,{\rm MC}}$ is the statistical uncertainty
of the MCRT spectrum.

\begin{figure*}
	\includegraphics[width = 0.75\textwidth]{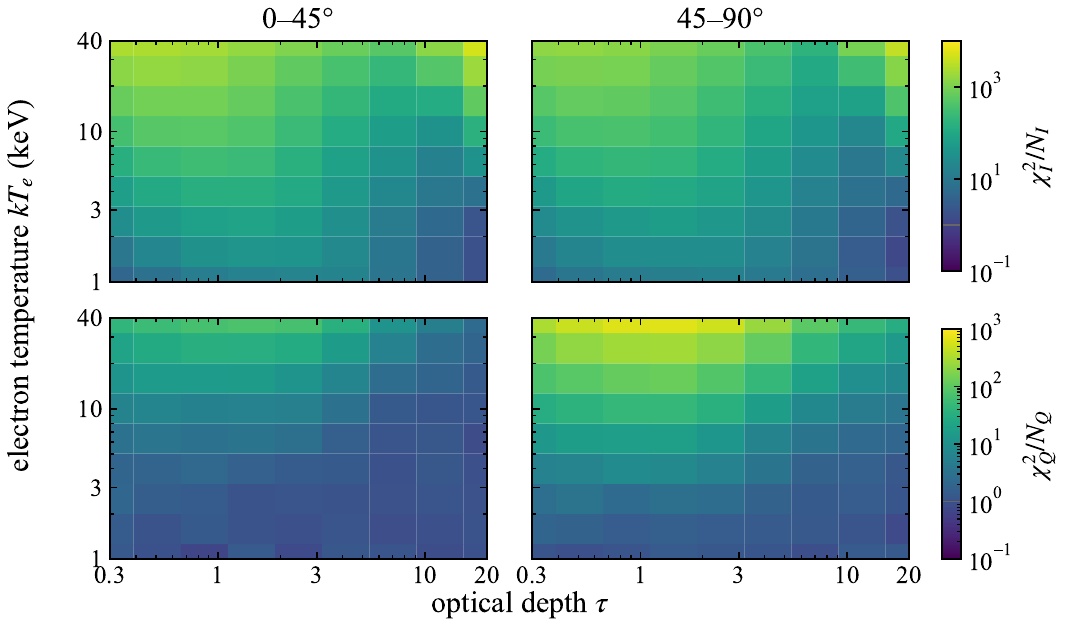}
\caption{
Maps of the fixed-model discrepancy statistic $\chi_X^2/N_X$ between the energy-resolved spectra from the recursive approximation and polarized MCRT calculations, as a function of electron temperature $kT_{\rm e}$ and total optical depth $\tau$. The upper and lower rows show the comparisons for Stokes $I$ and $Q$, respectively, while the left and right columns correspond to photons escaping through the upper surface within the inclination ranges $0^\circ$--$45^\circ$ and $45^\circ$--$90^\circ$. The colour scale shows the statistic defined in (\ref{eq:mcrt_reduced_chi2}).}
\label{fig:jones_mcrt_chi2_map} 
\end{figure*}

Figure~\ref{fig:jones_mcrt_chi2_map} shows the fixed-model discrepancy statistics $\chi_X^2/N_X$. Their magnitudes depend on both the model differences and the MCRT uncertainties. As shown in Figure~\ref{fig:energy_resolved_stokes_benchmark}, the reconstructed intensity spectra remain in close agreement with the MCRT results over a broad range of the representative parameters, so small intensity discrepancies can produce large values. We therefore use these maps to assess variations across parameter space within each Stokes component. They are not direct measures of fractional error, and the values for $I$ and $Q$ should not be compared on a common scale.

Figure~\ref{fig:jones_mcrt_chi2_map} reveals a clear dependence of the agreement on both electron temperature and optical depth. For both Stokes components, the discrepancy generally increases towards higher electron temperatures, consistent with the growing importance of thermal corrections to polarized angular transport and its coupling to energy redistribution. At a given electron temperature, the agreement generally improves towards larger $\tau$, and this contrast between optically thin and optically thick slabs becomes particularly evident at high $kT_{\rm e}$. Thus, the approximation remains substantially more robust in the multiple-scattering regime than in hot, optically thin coronae. The polarized spectrum additionally shows a strong viewing-angle dependence, with the discrepancies becoming especially prominent for photons escaping at large inclinations in the hot, optically thin regime.

\subsection{Joint spectro-polarimetric modelling of 4U 1624--49}
\label{sec:4u1624_application}

The preceding benchmarks indicate that the present separable treatment is best suited to polarized spectra from relatively cool, optically thick Comptonizing media. Such conditions are commonly encountered in neutron-star low-mass XRBs (NS-LMXBs), particularly in their soft states \citep[e.g.,][]{2001AdSpR..28..307B,2007A&ARv..15....1D}. By contrast, many black-hole XRBs and AGNs have hotter and generally more optically thin coronae \citep[e.g.,][]{2007A&ARv..15....1D,2015MNRAS.451.4375F}, for which quantitative polarization predictions require a fully coupled energy-, angle-, and polarization-dependent Compton treatment.

The dipping NS-LMXB 4U~1624--49 provides a particularly suitable astrophysical application of the present model. \citet{2024ApJ...963..133S} reported a highly significant polarization detection during the non-dipping intervals, with an average polarization degree of $3.1\pm0.7$ per cent and a polarization angle of $81^\circ\pm6^\circ$ in the 2--8~keV band. The polarization degree was also found to increase with energy, reaching approximately $6$ per cent in the 6--8~keV band, suggesting that the polarized signal is closely associated with the Comptonized component. A subsequent joint analysis of simultaneous \textit{IXPE}, \textit{NuSTAR}, and \textit{NICER} observations by \citet{2024A&A...690A.230G} reached a similar conclusion, finding that the Comptonized emission provides the dominant contribution to the observed polarization, although a non-negligible contribution from disc reflection may also be present.

In NS-LMXB systems, the interpretation of their continua is commonly discussed in terms of the Eastern and Western models. In an Eastern-like decomposition, the multitemperature disc emission is directly observed, while photons from the neutron-star surface are Comptonized in a boundary or spreading layer \citep[e.g.,][]{1984PASJ...36..741M, 1989PASJ...41...97M}. In a Western-like decomposition, the neutron-star blackbody is directly visible and the disc photons provide the seeds for Comptonization \citep[e.g.,][]{1988ApJ...324..363W}. Here we adopt a Western-like decomposition for the non-dipping emission of 4U~1624--49 and further approximate the disc-seeded Comptonizing region as a plane-parallel slab, allowing its polarized transport to be calculated using the framework developed above. In the broadband Western-like fit of \citet{2024A&A...690A.230G}, the non-dipping spectrum requires a cool, optically thick Comptonizing medium with $kT_{\rm e}\simeq2.6$~keV and a large optical depth in the \texttt{thcomp} parametrization. The disc seed temperature was fixed at $kT_{\rm in}=0.6$~keV. 

We perform an independent reanalysis of the non-dipping observation, using the \textit{IXPE} Stokes $I$, $Q$, and $U$ spectra together with the simultaneous \textit{NuSTAR} and \textit{NICER} intensity spectra. The fitting ranges are 2--8~keV for \textit{IXPE}, 4.5--20~keV for \textit{NuSTAR}, and 1.5--10~keV for \textit{NICER}. The three \textit{IXPE} detector units (DUs) are retained separately, with their corresponding $I$, $Q$, and $U$ spectra and instrumental responses entering the fit simultaneously. Our analysis uses the same observational data set and follows the non-dipping selection adopted in \citet{2024A&A...690A.230G}.

Table~\ref{tab:4u1624_polarization_comparison} compares the band-integrated polarization degrees obtained in our non-dipping reanalysis with those reported by \citet{2024ApJ...963..133S}. Their agreement provides a consistency check on the data selection and polarization extraction.

\begin{table}
    \centering
    \caption{Band-integrated polarization degree of 4U~1624--49 during the non-dipping intervals, comparing the reanalysis performed in this work with the measurements of \citet{2024ApJ...963..133S}. The values reported for this work are obtained from the standard PCUBE \citep{2022SoftX..1901194B} analysis of the source region using the same energy intervals as \citet{2024ApJ...963..133S}.}
    \label{tab:4u1624_polarization_comparison}
    \begin{tabular}{ccc}
        \hline
        Energy band
        & This work
        & \citet{2024ApJ...963..133S}
        \\
        (keV)
        & $\Pi$ (\%)
        & $\Pi$ (\%)
        \\
        \hline
        2--4
        & $1.59 \pm 0.77$
        & $1.6 \pm 0.8$
        \\
        4--6
        & $2.90 \pm 0.89$
        & $2.9 \pm 0.9$
        \\
        6--8
        & $6.58 \pm 2.17$
        & $6.4 \pm 2.0$
        \\
        4--8
        & $4.21 \pm 1.01$
        & $4.1 \pm 0.9$
        \\
        2--8
        & $3.20 \pm 0.75$
        & $3.1 \pm 0.7$
        \\
        \hline
    \end{tabular}
\end{table}

For the present analysis, the Comptonized disc component of the Western-like model is replaced by the energy-resolved slab calculation constructed from the Stokes recursion and the thermal Jones redistribution kernel described above. The remaining absorption and reflection components are retained in the broadband spectral model. In the polarization treatment adopted here, only the Comptonizing slab contributes to Stokes $Q$ and $U$; the directly visible blackbody component and the reflection component are taken to be unpolarized. 

Because the slab is axisymmetric, the intrinsic Comptonized radiation has $U_{\rm C}=0$ in the Stokes basis defined with respect to the projected slab normal. To compare the model with the \textit{IXPE} Stokes spectra, we rotate this intrinsic basis into the sky reference frame by a position angle $\psi$. With the convention adopted here, the predicted Stokes spectra are
\begin{equation}
\begin{aligned}
    &I_{\rm mod}(E)
    =
    {\cal A}(E)
    \left[
        I_{\rm C}(E)
        + I_{\rm BB}(E)
        + I_{\rm refl}(E)
    \right],
    \\
    &Q_{\rm mod}(E)
    =
    {\cal A}(E)\,
    Q_{\rm C}(E)\cos 2\psi,
    \\
    &U_{\rm mod}(E)
    =
    {\cal A}(E)\,
    Q_{\rm C}(E)\sin 2\psi ,
\end{aligned}
    \label{eq:4u1624_IQU_model}
\end{equation}
where ${\cal A}(E)$ denotes the multiplicative absorption terms, while $I_{\rm C}$ and $Q_{\rm C}$ are the intensity and signed linear-Stokes spectra predicted self-consistently by the slab model. The \textit{IXPE} $I$, $Q$, and $U$ spectra are then fitted simultaneously together with the broadband \textit{NuSTAR} and \textit{NICER} intensity spectra, so that the coronal parameters are constrained by the total spectral--polarimetric statistic 
\begin{equation}
    \chi^2_{\rm tot}
    =
    \chi^2_I+\chi^2_Q+\chi^2_U .
    \label{eq:joint_stokes_chi2}
\end{equation}

For the broadband intensity modelling, we adopt the same
Western-like spectral setup and fixed-parameter choices as \citet{2024A&A...690A.230G}. In \texttt{xspec}, the baseline intensity model is written schematically as
\begin{equation}
    \begin{aligned}
        &{\tt constant*TBabs*edge*CLOUDY}\\
    &{\tt *(slabseed+bbodyrad+relxillNS)} ,
    \end{aligned}
    \label{eq:4u1624_xspec_model}
\end{equation}
where \texttt{slabseed} denotes the energy-resolved Comptonized spectrum generated by the Stokes recursion combined with the Jones redistribution kernel. The seed spectrum of \texttt{slabseed} is a multitemperature disc blackbody with inner temperature fixed at $kT_{\rm in}=0.6$~keV, following the Western fit of \citet{2024A&A...690A.230G}. The electron temperature $kT_{\rm e}$, vertical Thomson depth $\tau$, viewing inclination, normalization, and sky position angle of the coronal component are allowed to vary. The coronal inclination is tied to the inclination of \texttt{relxillNS}, while the illuminating blackbody temperature of the reflection component is tied to the directly visible \texttt{bbodyrad} temperature. Parameters that were fixed in the reference Western model, including the emissivity index, neutron-star spin, inner reflection radius, iron abundance, and disc density, are kept fixed at the same values, whereas the absorption parameters, component normalizations, and instrumental cross-calibration parameters are optimized together with the coronal parameters. For each \textit{IXPE} detector unit, the gain correction is shared by its $I$, $Q$, and $U$ spectra.

Starting from the Western-like solution as a physically motivated reference point, we explored the joint $I/Q/U$ likelihood from multiple initial conditions. The optimization revealed two statistically acceptable solution branches. The low-reflection branch retains the basic Western-like decomposition, with a directly visible thermal component and a weak reflection contribution. By contrast, high-reflection branch favours stronger reflection and a higher inclination, with the direct-blackbody normalization approaching zero; this component is therefore omitted from its final fit. We retain both branches in the following analysis.

The best-fitting parameters of the two solution branches are listed in Table~\ref{tab:4u1624_fit_parameters}. The two branches yield statistically consistent electron temperatures but favour different optical depths and viewing inclinations, reflecting a degeneracy between the spectral decomposition and the geometry of the polarized Comptonizing medium. In the high-reflection branch, the best-fitting inclination reaches the upper model boundary of $87^\circ$, and the upper confidence limit on $\tau$ remains unconstrained within the model range $\tau\leq20$. These constraints are therefore limited by the adopted model boundaries.

\begin{table}
    \centering
    \caption{
    Best-fitting parameters of the low- and high-reflection solutions obtained from the joint $I/Q/U$ fit of the     non-dipping observations of 4U~1624--49. The quoted uncertainties are the $1\sigma$ profile-confidence intervals.
    }
    \label{tab:4u1624_fit_parameters}
    \renewcommand{\arraystretch}{1.18}
    \begin{tabular}{lcc}
        \hline
        Parameter
        & Low reflection
        & High reflection
        \\
        \hline

        $kT_{\rm e}$ (keV)
        & $2.645^{+0.058}_{-0.050}$
        & $2.654^{+0.059}_{-0.053}$
        \\

        $\tau$
        & $13.10^{+1.14}_{-0.98}$
        & $17.39^{+\mathrm{B}}_{-2.46}$
        \\

        $i$ (deg)
        & $81.6^{+2.8}_{-5.6}$
        & $87.0^{+\mathrm{B}}_{-1.1}$
        \\

        $\psi$ (deg)
        & $78.9^{+5.9}_{-5.9}$
        & $80.1^{+6.2}_{-6.2}$
        \\

        $N_{\rm C}$
        & $220^{+120}_{-100}$
        & $360^{+100}_{-110}$
        \\

        $kT_{\rm BB}$ (keV)
        & $1.214^{+0.015}_{-0.013}$
        & --
        \\

        $N_{\rm BB}$
        & $32.0^{+5.3}_{-4.5}$
        & --
        \\

        $kT_{\rm refl}$ (keV)
        & $=kT_{\rm BB}$
        & $1.326^{+0.011}_{-0.012}$
        \\

        $\log\xi_{\rm refl}$
        & $2.64^{+0.21}_{-0.26}$
        & $3.897^{+0.052}_{-0.058}$
        \\

        $N_{\rm refl}~(10^{-3})$
        & $2.8^{+1.6}_{-1.2}$
        & $21.4^{+0.9}_{-6.0}$
        \\
        \hline

        $\chi_I^2$
        & 699.14
        & 699.09
        \\

        $\chi_Q^2$
        & 419.58
        & 416.96
        \\

        $\chi_U^2$
        & 451.16
        & 451.54
        \\

        $\chi^2/{\rm dof}$
        & $1569.88/1572$
        & $1567.58/1573$
        \\

        $P_{\rm null}$
        & 0.51
        & 0.53
        \\

        \hline
    \end{tabular}

    \begin{minipage}{\columnwidth}
        \footnotesize
        \textit{Notes.}
        The symbol $\mathrm{B}$ indicates that the corresponding confidence boundary is not measured because the profile is truncated by the adopted model boundary. $N_{\rm C}$ denotes the normalization of the polarized slab Comptonization component, $N_{\rm BB}$ that of the directly visible blackbody, and $N_{\rm refl}$ the \texttt{relxillNS} normalization. The high-reflection solution does not include a directly visible blackbody component. The disc seed temperature of the Comptonizing slab is fixed at $kT_{\rm in}=0.6$~keV and the covering fraction at 0.99 in both branches. The \texttt{relxillNS} emissivity index, dimensionless spin, inner radius, iron abundance, and density are fixed at $q_{\rm em}=1.8$, $a=0.1$, $R_{\rm in}=1.5R_{\rm ISCO}$, $A_{\rm Fe}=1$, and $\log(n_{\rm e}/{\rm cm^{-3}})=16$, respectively.
    \end{minipage}
\end{table}

We additionally tested the coronal transfer calculation at the best-fitting parameters of both reflection branches using independent polarized MCRT simulations. These calculations adopted the same multitemperature disc seed spectrum. The reconstructed $I(E)$, $Q(E)$, and $\Pi(E)$ spectra agree well with the MCRT results, supporting the validity of the separable approximation for the present spectro-polarimetric analysis.

\begin{figure*}
	\includegraphics[width = 0.75\textwidth]{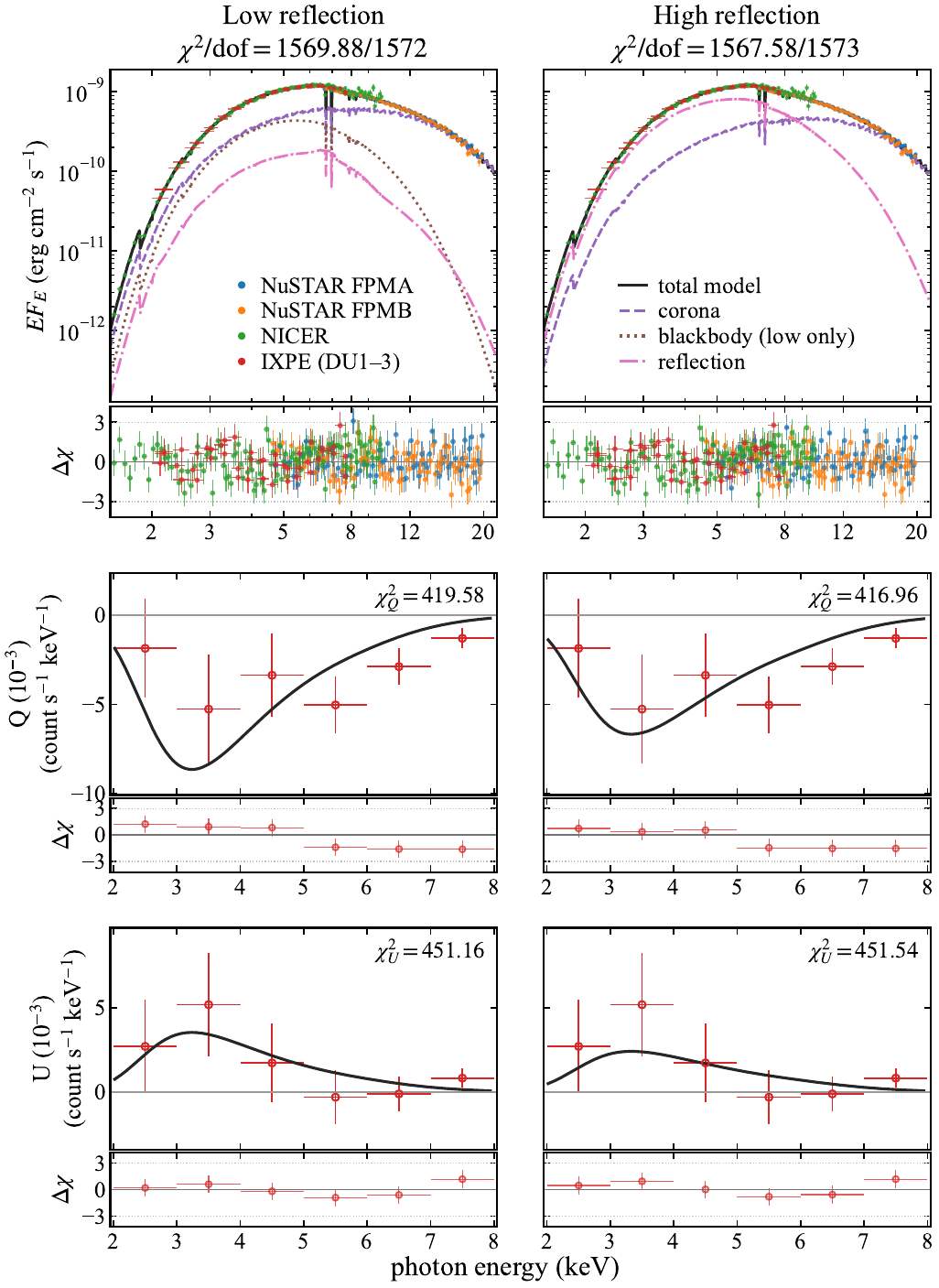}
    \caption{Joint spectro-polarimetric fits of the Western-like model to the non-dipping observations of 4U~1624--49. The left and right columns show the low- and high-reflection solutions, respectively. The upper panels display the unfolded intensity spectra from \textit{NuSTAR} FPMA/FPMB, \textit{NICER}, and the three \textit{IXPE} detector units (DUs), together with the total model and its Comptonized-corona, direct-blackbody, and reflection components; the blackbody component is negligible in the high-reflection branch. The middle and lower panels show the corresponding \textit{IXPE} Stokes $Q$ and $U$ spectra. For clarity, the three DUs are combined for display after folding the same source model through their individual responses and applying their respective calibration corrections.}
\label{fig:4u1624_joint_iqu}
\end{figure*}

Figure~\ref{fig:4u1624_joint_iqu} shows the joint $I/Q/U$ fits for the two solution branches. Despite the absence of polarized reflection or blackbody emission in the current model, the overall energy dependence of both Stokes $Q$ and $U$ is well captured by the Comptonization-only polarization model. However, the energy-dependent $Q/U$ spectrum is sensitive to the slab inclination, optical depth, and dilution by unpolarized spectral components. The persistence of both minima nevertheless shows that the current data do not yet uniquely determine the decomposition of the continuum or the corresponding coronal geometry.

\section{Discussion}
\label{sec:discussion}

In this work, we developed the Stokes recursion for scattering-order-resolved photon transport in a finite Thomson slab. The present results clarify how repeated scattering drives the radiation from a source-dependent low-order state towards a universal high-order polarized state, and under which conditions this transport information can be combined with an energy-redistribution kernel to predict observable Comptonized polarization. 

\subsection{Polarization feedback and asymptotic transport}
\label{sec:discussion_eigenmodes}

For an unpolarized radiation field, the Thomson scattering generates Stokes $Q$, which subsequently enters the redistribution of Stokes $I$ through the polarized phase matrix. As a result, the scattering-order escape probabilities, angular distributions, eigenvalues, and mean scattering numbers are close to, but not identical to, those obtained from the polarization-averaged scalar recursion.

The differences remain modest because the total Thomson cross section and photon-number conservation are unchanged. Polarization modifies the differential angular redistribution rather than introducing an additional photon source. This explains why the scalar formalism of Paper~I remains a useful approximation for integrated transport quantities, even though it cannot reproduce the polarization properties. The mean scattering number provides a particularly clear example. The Stokes calculation gives small but systematic corrections for normally incident and internally generated photons, whereas the exact Lambert-law result $\langle N\rangle=2\tau$ remains unchanged.

Paper~I showed that repeated scattering progressively erases the detailed memory of the source. At low scattering orders, the radiation retains information about the seed-photon angular distribution, injection boundary, and initial polarization. At sufficiently high scattering orders, however, the principal eigenmode dominates as contributions from all subdominant modes are progressively suppressed and the high-order radiation approaches a universal polarized transport state determined primarily by the slab optical depth and geometry. The same mode then determines the asymptotic survival probability, normalized angular distribution, and linear polarization distribution.  

The subdominant spectrum further specifies how rapidly this universal state is approached. The ratio $\rho$ defined in (\ref{eq:spectral_convergence_factor}) measures the rate at which the non-principal components decay, and therefore provides a natural estimate of the scattering order required to reach the asymptotic regime. In optically thin slabs this transition occurs after only several scatterings, whereas in optically thick media the required order increases approximately as $\tau^2$. Therefore, the required scattering order for entering the high-order asymptotic region may be much greater than the mean scattering number in thick slabs.

\subsection{Finite-optical-depth extensions of classical asymptotic limits}
\label{sec:discussion_asymptotics}

The optically thin and optically thick limits derived above also improve two classical results of polarized radiative transfer. In the optically thin limit, we recover the complete-polarization result of \citet{1985A&A...143..374S}, but the finite-slab analysis shows that the convergence is logarithmically slow and non-uniform in viewing angle. In particular, (\ref{eq:thin_polarization_boundary_layer}) identifies a narrow angular region around the slab normal in which the complete-polarization limit fails. More importantly for practical applications, (\ref{eq:polarization_from_alpha}) and (\ref{eq:alpha_dominant_branch}) provide a finite-$\tau$ approximation that remains useful for $\tau\lesssim1$.

The angular distribution contains a physically distinct non-uniform limit.  For fixed $\mu>0$, we obtain $\Phi_\infty\rightarrow3(1-\mu^2)/2$ as $\tau\rightarrow0$. However, at any finite optical depth the escape probability must turn over towards grazing emergence. The polarization and intensity therefore possess different angular boundary layers: the former occurs close to the slab normal as a consequence of axial symmetry, whereas the latter occurs close to the surface as a consequence of finite escape geometry. The finite-$\tau$ construction captures these departures while extending the usefulness of the strict thin-limit solution towards optical depths of order unity.

In the optically thick regime, the finite-thickness corrections are obtained by matching a long-wavelength bulk expansion to a polarized Milne surface layer, where the direct influence of the opposite boundary is neglected. This construction extends the Chandrasekhar semi-infinite result to finite slab thickness while retaining a controlled asymptotic interpretation. A useful feature of this construction is that the angular correction functions are universal, while the dependence on the total optical depth enters only through the finite-slab eigenmode scale. Numerically, the finite-thickness correction captures the departure from the half-space solution well for $\tau\gtrsim5$.

The sign of the leading correction is also physically informative.  The classical half-space solution produces a positive polarization that reaches approximately $11.7$ per cent towards the limb, whereas the leading finite-thickness correction is negative over the full angular range. A finite slab therefore reduces the Chandrasekhar polarization before the system leaves the thick-medium regime.  For the normalized intensity, the corresponding correction integrates to zero and therefore redistributes the radiation among escape directions without changing the total escaping probability. The leading finite-thickness correction is positive over most of the angular range and becomes negative only near the slab normal, implying a redistribution of escaping photons from near-normal directions toward more inclined directions.

\subsection{Energy-resolved polarized spectra}
\label{sec:discussion_energy}

The scattering-order decomposition provides a natural bridge from radiative transport to the emergent spectrum. Once the escape probability, angular distribution, and polarization are known at each scattering order, approximate energy- and angle-resolved spectra can be reconstructed by coupling the transport calculation to an energy-redistribution prescription, provided that energy redistribution and angular transport can be treated as separable. 

(\ref{eq:band_polarization_order_sum}) allows us to examine how the polarization fraction in a given energy band approaches its high-order asymptotic value. The contribution from each scattering order is weighted by $w_{n,B}P_n^\pm\Phi_n^\pm(\mu)$. In the high-order regime, the escape probability decreases as $P_n^\pm\propto\lambda_0^{n-1}$, whereas the normalized angular distribution approaches a limiting form. The relative importance of different scattering orders is therefore controlled primarily by the competition between this decreasing escape probability and the energy weight $w_{n,B}$. Within the present stationary-electron, separable description, the polarization approaches $\Pi_\infty^\pm(\mu)$ if the energy band is dominated by scattering orders that have already converged to the principal eigenmode.

Paper~I used a characteristic energy shift for each scattering order to construct the approximate Comptonized intensity spectrum. In the present work, we replace the peak-shift prescription by the angle-averaged thermal Compton redistribution kernel of \citet{1968PhRv..167.1159J}. This substantially improves the reconstruction of the intensity spectrum and, in cool and optically thick media, also provides a useful approximation to the Stokes-$Q$ spectrum. The latter is more demanding because $Q$ is a signed quantity and is sensitive to partial cancellation between photons with different scattering histories.

For the hotter cases tested here, the substantial discrepancies in the Stokes-$Q$ spectra highlight the limitations of stationary-electron angular transport combined with separable energy redistribution. The present separable treatment is therefore particularly well suited to cool, optically thick coronae, as commonly found in soft-state NS-LMXBs, where electron temperatures are typically a few keV and optical depths are relatively large. Hotter and more optically thin coronae in black-hole XRBs, AGNs and hard-state NS-LMXBs will require a fully coupled energy-, angle-, and polarization-dependent Compton redistribution operator for quantitative polarimetric modelling.

These limitations also have implications for inferring coronal optical depths from X-ray polarization. \citet{2025A&A...704A..87T} applied high-order Thomson polarization curves to \textit{IXPE} measurements of black-hole XRBs and AGNs, estimating optical depths from the measured polarization and source inclination. Our comparisons, presented in Section~\ref{sec:band_polarization_formalism} and the accompanying notebook, show that a weak energy dependence of polarization alone is insufficient to establish the applicability of these curves to a hot plasma. Such parameter estimates therefore require an assessment of the approximation over the relevant temperature, optical-depth, and energy ranges.

A possible extension is to replace the stationary-electron Thomson phase matrix with a thermally averaged angular redistribution matrix that retains the Lorentz transformations of photon directions and polarization bases while treating each scattering in the electron-rest-frame Thomson limit. This would incorporate the omitted thermal angular and polarization corrections and could improve the separable reconstruction where these corrections dominate. It would not, however, provide a full Klein--Nishina transport treatment.

Another natural application of this order-resolved description is radiative transfer through accretion-disc atmospheres. The thermal emission of an optically thick disc is not, in general, a perfect blackbody at the local effective temperature \citep[e.g.,][]{1973A&A....24..337S}. In a scattering-dominated atmosphere, photons are thermalized at depths where the effective optical depth is of order unity and subsequently propagate through an overlying electron-scattering layer before escape. The resulting continuum is therefore spectrally hardened and anisotropic \citep[e.g.,][]{1995ApJ...445..780S,2005ApJ...621..372D}.

A self-consistent disc-atmosphere calculation would additionally require absorption and emission together with the vertical structure of the atmosphere. The present Stokes recursion instead provides the scattering-transport component of such a calculation. In an X-ray band sampling the high-energy tail well above the characteristic disc temperature, the observed disc high-energy photons may be dominated by the multiply scattered photons and the high-energy tail can become dominated by scattering orders that have already approached the principal eigenmode. In this case, the asymptotic angular distribution and polarization provide a natural approximation to the atmosphere-modified disc X-ray emission.

\subsection{Application to 4U~1624--49 and astrophysical implications}
\label{sec:discussion_4u1624}

The non-dipping emission of 4U~1624--49 provides a useful observational application of the present framework. Our joint spectro-polarimetric analysis of 4U~1624--49 shows that a plane-parallel Comptonizing region can reproduce the overall energy dependence of the observed \textit{IXPE} Stokes $Q$ and $U$ spectra while  simultaneously fitting the broadband intensity spectrum. Both fitting branches favour electron temperatures near $kT_{\rm e}\simeq2.65$~keV and relatively large optical depths, placing the Comptonizing medium in the regime where our separable reconstruction is most reliable. These results show that polarization generated within a flattened Comptonizing region can account for the observed polarization of 4U 1624--49.

This interpretation is consistent with previous polarized radiative-transfer calculations. Using relativistic Monte Carlo simulations, \citet{2022MNRAS.514.2561G} compared an idealized spherical-shell corona surrounding the neutron star with a slab corona above the accretion disc, which may be regarded as idealized Comptonizing geometries associated with the Eastern-like and Western-like spectral scenarios, respectively. They found that the slab geometry can produce several-per-cent polarization at higher X-ray energies, whereas the polarization generated in the spherical-shell configuration is largely suppressed. In addition, \citet{2024A&A...684A..62F} found that an optically thick equatorial boundary layer produces a net polarization below approximately $0.5$ per cent under representative soft-state conditions. Taken together, these studies support a flattened Comptonizing region as a physically plausible source of the observed polarization. 

The low- and high-reflection solutions both provide statistically acceptable joint fits, but imply different continuum decompositions, optical depths, and viewing inclinations. An important limitation of the present analysis is that only the Comptonizing slab is allowed to contribute intrinsic polarization, while the directly visible blackbody and reflection components are treated as unpolarized. Including the intrinsic polarization of the reflected radiation may help to break this degeneracy.

In particular, our results do not rule out an Eastern-like interpretation. As discussed in Section~\ref{sec:discussion_energy}, the directly visible disc may carry intrinsic polarization generated in its atmosphere \citep[e.g.,][]{2009ApJ...691..847L}, while reflection of radiation from the neutron-star surface or boundary layer by the accretion disc may provide an additional polarized component \citep[e.g.,][]{1985MNRAS.217..291L}. A discriminating comparison between alternative spectral and geometrical interpretations will therefore require the Stokes spectra of the Comptonized, thermal-disc, and reflected components to be treated together.

The present framework provides the energy-, incidence-angle-, and polarization-resolved radiation field illuminating the disc in a sandwich disc--corona configuration, offering a more complete boundary condition for reflection calculations \citep[e.g.,][]{1991MNRAS.249..352G,1995MNRAS.273..837M,1996MNRAS.283..892P}, since the reflected Stokes spectrum depends not only on the incident flux but also on the angular distribution and polarization state of the illuminating radiation. Once a reflection calculation provides the Stokes spectrum at the lower boundary of the Comptonizing layer, its subsequent propagation and re-Comptonization through a sufficiently cool corona can again be treated with the same slab transport framework \citep[e.g.,][]{1991ApJ...380L..51H,1993ApJ...413..507H, 2024A&A...684A..62F}, provided that the photon energies remain in the regime in which the Thomson approximation is adequate. The current transport calculation can thus serve as the coronal transport component of a future self-consistent polarized reflection model for NS-LMXBs.

Although the present analysis focuses on plane-parallel slabs, the scattering-order recursion can in principle be extended to other geometries, including spherical coronae, with appropriate modifications to the propagation kernels and angular representations. Furthermore, the eigenmode theory developed here provides a foundation for efficient hybrid Comptonization models that treat low scattering orders explicitly and employ diffusion approximation in the asymptotic high-order regime. Future work will extend the present framework to different geometries and incorporate fully coupled energy-, angle-, and polarization-dependent Compton redistribution, enabling more general spectro-polarimetric modelling of astrophysical systems.

\section{Conclusions}
\label{sec:conclusions}

In this paper, we have developed a scattering-order-resolved Stokes-vector framework for polarized transport in a finite plane-parallel Thomson slab. This work extends the scalar recursion of Paper~I to a self-consistent polarized treatment and connects the resulting transport theory to energy-resolved Comptonization and X-ray spectro-polarimetry. Our main results are summarized as follows.

\begin{enumerate}

\item
For axisymmetric Thomson scattering, the reduced Stokes state $(I,Q)^{\rm T}$ closes exactly within two depth-dependent coefficients. The resulting recursion efficiently yields the escape probability, normalized angular distribution, and signed linear polarization at each scattering order. Although photon number remains exactly conserved, polarization generated at one scattering feeds back into the subsequent redistribution of the intensity. The scalar and Stokes calculations therefore give similar, but not identical, scattering-order probabilities, mean scattering numbers and other escape statistics.

\item
We established the eigenmode structure of the finite-slab Stokes operator. At high scattering order, the transport is controlled by a principal eigenmode, while the corresponding eigenfunction determines the limiting angular and polarization distributions. We further identified diffusion-like and anisotropic-polarization spectral families and used the leading subdominant eigenvalue to estimate the scattering order at which the source-dependent radiation approaches the universal high-order state.

\item
In the optically thin limit, the dominant Stokes eigendirection approaches $(1,-1)^{\rm T}$, recovering the classical result $\Pi_\infty\rightarrow-1$ \citep[e.g.,][]{1985A&A...143..374S}. We derived the finite-$\tau$ corrections, identified the non-uniform angular layer near the slab normal, and obtained a finite-$\tau$ approximation for the high-order normalized angular distribution, whose strict limit is $\Phi_\infty(\mu)=3(1-\mu^2)/2$. 

\item
In the optically thick regime, matching the bulk diffusion solution to the polarized Milne surface layer yields the extrapolation length $z_0\simeq0.7121$ and finite-optical-depth corrections to the high-order polarization and normalized angular distribution, with the classical Chandrasekhar semi-infinite atmosphere recovered as $\tau\rightarrow\infty$.

\item
Combining the order-resolved Stokes transport with an angle-averaged thermal-Compton redistribution kernel gives accurate intensity spectra over a broad range of $kT_{\rm e}$ and $\tau$. Polarization spectra are more restricted to cool, optically thick media, because energy, angle, and polarization become strongly coupled in hot, optically thin coronae. Applied to the non-dipping observations of 4U~1624--49, the model provides acceptable joint fits to the broadband intensity and \textit{IXPE} Stokes spectra. Both allowed branches favour $kT_{\rm e}\simeq2.65$~keV, a large $\tau$, and a highly inclined slab, although the continuum decomposition and coronal geometry remain degenerate. 
\end{enumerate}

\section*{Acknowledgements}

I am grateful to Prof. Xinwu Cao for his continuous support and helpful discussions throughout this work. Part of the simulations, calculations, and analysis presented in this article were carried out on the SilkRiver Supercomputer of Zhejiang University, located at the Zhejiang University Information Center.  

I used ChatGPT and Codex (OpenAI) to assist with code development, data analysis, and selected mathematical derivations, as well as to translate the original Chinese manuscript into English and improve its language and presentation. All AI-assisted analyses and derivations were reviewed and validated by me. I take full responsibility for the conclusions and final content of this work.

%%%%%%%%%%%%%%%%%%%%%%%%%%%%%%%%%%%%%%%%%%%%%%%%%%
\section*{Data Availability}

The numerical data and notebooks supporting this work are available at \url{https://github.com/XU-Haichao/slab-polarization}. The repository includes Monte Carlo validation results and numerical outputs for this work, and derived products from our analysis of 4U 1624-49. The Monte Carlo code and the recursion-based Comptonization code are available at \url{https://github.com/XU-Haichao/MCRT} and \url{https://github.com/XU-Haichao/slab-comptonization}, respectively. Requests for access to the modified \texttt{compPSc} source may be directed to the corresponding author.



\bibliographystyle{mnras}
\bibliography{example} % if your bibtex file is called example.bib

%\begin{thebibliography}{99}
%\end{thebibliography}

%%%%%%%%%%%%%%%%%%%%%%%%%%%%%%%%%%%%%%%%%%%%%%%%%%

%%%%%%%%%%%%%%%%% APPENDICES %%%%%%%%%%%%%%%%%%%%%

\appendix

\section{Polarized Monte Carlo implementation}
\label{app:mcrt}

\subsection{Monte Carlo photon transport and polarized scattering}
\label{app:mcrt_method}

We developed an independent polarized Monte Carlo radiative-transfer (MCRT) code to validate the Stokes-vector recursion derived in this paper. The photon propagation and thermal Compton scattering are treated event by event following the standard Monte Carlo constructions of \citet{1983ASPRv...2..189P} and \citet{1984AcA....34..141G}, with the polarization state explicitly tracked through the Stokes vector. 

Each Monte Carlo photon packet is characterized by its position, propagation direction $\hat{\bm k}$, dimensionless photon energy $x={E}/{m_{\rm e}c^2}$ and Stokes vector $\bm s=(i,q,u,v)^{\rm T}$ defined with respect to an orthonormal polarization basis transverse to $\hat{\bm k}$. The seed-photon position, direction, and energy are sampled from the same source configurations used in the recursive calculations. Unless stated otherwise, the injected photons are initially unpolarized, so that the initial Stokes vector is taken to be $\bm s_0=(1,0,0,0)^{\rm T}$.

For a photon interacting with an electron of Lorentz factor $\gamma$ and dimensionless speed $\beta$, the corresponding photon energy in the electron rest frame (ERF) is
\begin{equation}
    x_*=\gamma x(1-\beta\mu_{\rm e}),
    \label{eq:mcrt_erf_energy}
\end{equation}
where $\mu_{\rm e}$ is the cosine of the angle between the incident photon direction and the electron velocity. Throughout this appendix, the $*$ subscript denotes a quantity evaluated in the ERF.

For cold electrons in the Thomson limit, the interaction cross section is
simply $\sigma_{\rm T}$. For a thermal electron population, the interaction
rate must instead be averaged over both the electron velocity distribution
and the relative photon--electron flux
\citep[e.g.][]{1990MNRAS.245..453C}, i.e.,
\begin{equation}
    \overline{\sigma}(x,\theta_{\rm e})
    =
    \int_1^\infty
    f_{\rm MJ}(\gamma;\theta_{\rm e})\,{\rm d}\gamma
    \int_{-1}^{1}
    \frac{{\rm d}\mu_{\rm e}}{2}
    (1-\beta\mu_{\rm e})
    \sigma_{\rm KN}(x_*).
\label{eq:mcrt_thermal_opacity}
\end{equation}
Here the factor $(1-\beta\mu_{\rm e})$ accounts for the relative photon--electron flux, and $\sigma_{\rm KN}(x_*)$ is the total Klein--Nishina cross section evaluated at the photon energy in the ERF \citep[e.g.,][]{1929ZPhy...52..853K,1970RvMP...42..237B}. The integral in (\ref{eq:mcrt_thermal_opacity}) is precomputed on a grid of $x$ and $\theta_{\rm e}$ and interpolated during photon transport.

For each photon, the distance to the next interaction is sampled as
\begin{equation}
    \ell
    =
    -\frac{\ln\xi}
    {n_{\rm e}\,\overline{\sigma}(x,\theta_{\rm e})},
    \qquad
    0<\xi<1.
    \label{eq:mcrt_free_path}
\end{equation}
The sampled free path is compared with the distance to the slab boundary along the current propagation direction. If the boundary is reached first, the photon escapes without another interaction; otherwise, it propagates by the distance $\ell$ and undergoes a scattering event. The exact scattering order is recorded for every photon history.

Once a scattering event occurs, the electron participating in the interaction cannot be sampled directly from the isotropic Maxwell--J\"uttner distribution, because electrons with different velocities and directions contribute different collision rates. Instead, the electron Lorentz factor and incident-angle cosine are drawn from the collision-rate-weighted distribution
\begin{equation}
p_{\rm coll}(\gamma,\mu_{\rm e}\mid x)
    =\frac{
        f_{\rm MJ}(\gamma;\theta_{\rm e})
        (1-\beta\mu_{\rm e})
        \sigma_{\rm KN}(x_*)
    }{
        2\,\overline{\sigma}(x,\theta_{\rm e})
    },
    \label{eq:mcrt_collision_electron}
\end{equation}
where the electron azimuth about the incident photon direction is uniformly distributed. The three factors in the numerator account for the thermal electron distribution, the relative photon--electron flux, and the energy dependence of the Klein--Nishina scattering cross section, respectively \citep{1983ASPRv...2..189P,1984AcA....34..141G}. 

With the scattering electron specified, the photon four-momentum and polarization basis are Lorentz-transformed to the ERF. In this frame, the outgoing photon energy is given by
\begin{equation}
    x_*'
    =
    \frac{x_*}
    {1+x_*(1-\cos\Theta_*)},
    \label{eq:mcrt_compton_recoil}
\end{equation}
where $\Theta_*$ is the scattering angle in the ERF, and a prime denotes a post-scattering quantity. Defining the recoil factor
\begin{equation}
    r=\frac{x_*'}{x_*}
    =
    \frac{1}{1+x_*(1-\cos\Theta_*)},
\end{equation}
the polarization dependence of the scattering process is described by the Fano--Mueller matrix \citep{1957RvMP...29...74F,1961RvMP...33....8M},
\begin{equation}
    \mathbfss M_{\rm F}=\frac{3}{16\pi} r^2
    \begin{pmatrix}
      F_0 & F_1 & 0 & 0\\
      F_1 & F_{11} & 0 & 0\\
      0 & 0 & F_{22} & 0\\
      0 & 0 & 0 & F_{33}
    \end{pmatrix},
    \label{eq:mcrt_fano_matrix}
\end{equation}
with
\begin{equation}
\begin{aligned}
    &F_0 = r+r^{-1}-\sin^2\Theta_*,
    &F_1 = \sin^2\Theta_*,\\
    &F_{11} =1+\cos^2\Theta_*,
    &F_{22}=2\cos\Theta_*,\\
    &F_{33} =(r+r^{-1})\cos\Theta_*.
\end{aligned}
\label{eq:mcrt_fano_coefficients}
\end{equation}
In the Thomson limit $x_*\ll1$, the Fano matrix reduces to the Thomson Mueller matrix in Section~\ref{sec:stokes_recursion}.

For each candidate outgoing direction, the corresponding scattering plane is constructed and the incident Stokes vector is rotated into the associated scattering-plane basis. In this basis, the polarized differential cross section is
\begin{equation}
    \frac{{\rm d}\sigma}{{\rm d}\Omega_*}
    =
    \frac{3\sigma_{\rm T}}{16\pi}
    r^2
    \left(
        F_0+F_1\frac{q_*}{i_*}
    \right).
    \label{eq:mcrt_polarized_dsigma}
\end{equation} 
The outgoing direction is sampled according to this angular distribution. For the selected direction, the Fano--Mueller matrix determines the polarization state of the scattered photon. Since the photon packets are propagated with equal weight, the outgoing Stokes vector is normalized to preserve its intensity,
\begin{equation}
    \bm s_*'
    =
    \frac{\sigma_{\rm T}}
    {{\rm d}\sigma/{\rm d}\Omega_*}
    \mathbfss M_{\rm F}\bm s_*,
    \label{eq:mcrt_stokes_update}
\end{equation}
which ensures $i_*'=i_*$.

After the scattering event, the outgoing photon momentum and polarization basis are transformed back to the laboratory frame. The Stokes parameters are finally rotated to the meridian basis associated with the slab normal and the new propagation direction. The propagation procedure is repeated until the photon escapes through one of the two slab boundaries.

\subsection{Monte Carlo estimators and statistical uncertainties}
\label{app:mcrt_observables}

For each injected photon, the Monte Carlo calculation records the escape boundary, the exact number of scatterings, the escaping photon energy and direction, and the final Stokes vector. When a photon escapes through either slab boundary, its Stokes vector is rotated into the reference basis adopted throughout this work before being accumulated into the corresponding observables.

We denote the total number of injected photon histories in a simulation by $N_{\rm tot}$. Let $N_n^\pm$ be the number of photons escaping through the upper or lower boundary after exactly $n$ scatterings. The corresponding Monte Carlo estimators of the scattering-order-resolved escape probabilities are
\begin{equation}
    \widehat P_n^\pm
    =
    \frac{N_n^\pm}{N_{\rm tot}},
    \qquad
    \widehat P_n
    =
    \widehat P_n^+ + \widehat P_n^- .
    \label{eq:mcrt_escape_probability}
\end{equation}
The standard error of $\widehat P_n^\pm$ is estimated from the binomial variance,
\begin{equation}
    \sigma(\widehat P_n^\pm)
    =
    \left[
        \frac{
        \widehat P_n^\pm
        \left(1-\widehat P_n^\pm\right)
        }{N_{\rm tot}}
    \right]^{1/2}.
    \label{eq:mcrt_escape_probability_error}
\end{equation}
The same expression applies to $\widehat P_n$.

To construct the angular distributions, the escaping photons are binned in the direction cosine $\mu$. Let $N_{n,j}^\pm$ denote the number of photons in the $j$-th angular bin with width $\Delta\mu_j=\mu_{j,+}-\mu_{j,-}$. The differential escape probability is estimated as
\begin{equation}
    \left(
    \frac{{\rm d}P_n^\pm}{{\rm d}\mu}
    \right)_j
    =
    \frac{
        N_{n,j}^\pm
    }{
        N_{\rm tot}\Delta\mu_j
    },
    \label{eq:mcrt_angular_escape}
\end{equation}
and the corresponding normalized angular distribution is
\begin{equation}
    \widehat\Phi_{n,j}^\pm
    =
    \frac{
        N_{n,j}^\pm
    }{
        N_n^\pm\Delta\mu_j
    }.
    \label{eq:mcrt_phi_estimator}
\end{equation}

For the same $n,j$ channel, the Monte Carlo estimator of the scattering-order-resolved Stokes distribution is
\begin{equation}
    \widehat{\bm S}_{n,j}^{\pm}
    =
    \frac{\sum
    \bm s_k^\pm}{N_{\rm tot}\Delta\mu_j}
    =
    \left(
        \widehat I_{n,j}^{\pm},
        \widehat Q_{n,j}^{\pm},
        \widehat U_{n,j}^{\pm},
        \widehat V_{n,j}^{\pm}
    \right)^{\rm T},
    \label{eq:mcrt_stokes_estimator}
\end{equation}
where the sum runs over all photon packets escaping in the specified $n, j$ channel, and $\bm s_k^\pm$ is the Stokes vector carried by the $k$-th escaping photon packet. For the equal-weight scheme adopted here, $i_k=1$, and hence
\begin{equation}
    \widehat I_{n,j}^{\pm}
    =
    \frac{N_{n,j}^{\pm}}
         {N_{\rm tot}\Delta\mu_j},
    \label{eq:mcrt_intensity_estimator}
\end{equation}
which is identical to the binned differential escape-probability estimator in (\ref{eq:mcrt_angular_escape}). 

For the axisymmetric configurations considered here, the MCRT results satisfy the expected symmetry: $\widehat U$ is consistent with zero within the Monte Carlo statistical uncertainties, while $\widehat V$ remains zero all the time. The signed polarization fraction is therefore
\begin{equation}
    \widehat\Pi_{n,j}^{\pm}
    =
    \frac{\widehat Q_{n,j}^{\pm}}
         {\widehat I_{n,j}^{\pm}}
    =
    \frac{\sum q_k^\pm}{N_{n,j}^{\pm}}.
    \label{eq:mcrt_signed_polarization}
\end{equation}
Since $\widehat\Pi_{n,j}^{\pm}$ is the mean packet polarization in a given channel, its standard error is estimated directly from the packet-to-packet dispersion,
\begin{equation}
    \sigma_{\widehat\Pi_{n,j}^{\pm}}
    =
    \left[
        \frac{
            \sum
            \left(
                q_k^\pm-\widehat\Pi_{n,j}^{\pm}
            \right)^2
        }{
            N_{n,j}^{\pm}
            \left(N_{n,j}^{\pm}-1\right)
        }
    \right]^{1/2}.
    \label{eq:mcrt_pi_error}
\end{equation}

The energy-resolved estimators are constructed analogously. For an energy bin $[E_{m,-},E_{m,+}]$ of width $\Delta E_m$ and a viewing-angle bin $\Delta \mu_j$, the photon-number Stokes spectrum is
\begin{equation}
    \widehat{\bm N}_{m,j}^{\pm}
    =
    \frac{\sum \bm s_k^\pm}{
        N_{\rm tot}\Delta E_m\Delta\mu_j
    }
    =
    \left(
        \widehat N_{I,mj}^{\pm},
        \widehat N_{Q,mj}^{\pm},
        \widehat N_{U,mj}^{\pm},
        \widehat N_{V,mj}^{\pm}
    \right)^{\rm T},
    \label{eq:mcrt_energy_stokes_estimator}
\end{equation}
where the sum now runs over all photons escaping in the specified energy--angle bin, irrespective of scattering order.

Consistent with the notation of Section~\ref{sec:band_polarization}, the corresponding energy-weighted Stokes spectra are
\begin{equation}
    \widehat I_E^\pm(E_m,\Omega_j)
    =
    E_m\,\widehat N_{I,mj}^{\pm},
    \qquad
    \widehat Q_E^\pm(E_m,\Omega_j)
    =
    E_m\,\widehat N_{Q,mj}^{\pm},
    \label{eq:mcrt_energy_spectra}
\end{equation}
and the signed polarization fraction is
\begin{equation}
    \widehat\Pi_E^\pm(E_m,\Omega_j)
    =
    \frac{\widehat Q_E^\pm}
         {\widehat I_E^\pm}
        =
    \frac{
        \widehat N_{Q,mj}^\pm
    }{
        \widehat N_{I,mj}^\pm
    }.
    \label{eq:mcrt_energy_polarization}
\end{equation}

For the high-statistics thermal-Compton calculations, each production simulation is divided into $B$ statistically independent batches of equal size. The reported energy-resolved Stokes spectra are obtained by averaging the corresponding batch spectra,
\begin{equation}
    \overline I_E
    =
    \frac{1}{B}\sum_{b=1}^{B} I_{E,b},
    \qquad
    \overline Q_E
    =
    \frac{1}{B}\sum_{b=1}^{B} Q_{E,b}.
    \label{eq:mcrt_batch_mean}
\end{equation}
Their standard errors are estimated from the batch-to-batch dispersion,
\begin{equation}
\begin{aligned}
        &\sigma_{\overline I_E}
    =
    \left[
        \frac{1}{B(B-1)}
        \sum_{b=1}^{B}
        \left(I_{E,b}-\overline I_E\right)^2
    \right]^{1/2},\\
    &\sigma_{\overline Q_E}
    =
    \left[
        \frac{1}{B(B-1)}
        \sum_{b=1}^{B}
        \left(Q_{E,b}-\overline Q_E\right)^2
    \right]^{1/2}.
\end{aligned}
\end{equation}
The covariance between $I_E$ and $Q_E$ is retained in the polarization uncertainty. With $\widehat\Pi_E=\overline Q_E/\overline I_E$, we use
\begin{equation}
    \sigma_{\widehat\Pi_E}
    =
    \left[
        \frac{1}{B(B-1)}
        \sum_{b=1}^{B}
        \left(
            \frac{
                Q_{E,b}-\widehat\Pi_E I_{E,b}
            }{
                \overline I_E
            }
        \right)^2
    \right]^{1/2}.
    \label{eq:mcrt_energy_pi_error}
\end{equation}
All energy-resolved quantities represent averages over finite energy and viewing-angle bins and are therefore compared with theoretical spectra integrated over the same intervals.

\section{Scattering-number statistics in the Stokes recursion}
\label{app:mean_number}

Paper I developed the scattering-number statistics for the scalar Thomson recursion, including the scattering-order survival probabilities and the mean number of scatterings before escape. In this appendix, we extend that framework to the present Stokes-vector recursion and examine how polarization feedback modifies these statistics while preserving photon-number conservation.

\subsection{Survival probabilities and the mean scattering number}
\label{app:survival_probability}

Following Paper I, we define the scattering-order survival probability as
\begin{equation}
    M_n
    =
    \int_0^\tau {\rm d}t
    \int_{-1}^{1}{\rm d}\mu\,
    \Psi_{I,n}(t,\mu),\qquad n\geq 1,
    \label{eq:Mn_stokes}
\end{equation}
which is the probability that a photon undergoes at least $n$ scatterings before escaping the slab. Only the intensity component enters $M_n$, since $\Psi_{I,n}$ represents the scattering-order-resolved photon-number density and is the natural Stokes-vector generalization of the scalar distribution $\Psi_n$ introduced in Paper I.

Integrating the Stokes recursion (\ref{eq:polarized_recursive_relation}) over the next scattering depth and outgoing direction, and using the photon-number normalization of the intensity component, the polarization-dependent terms cancel from the angular integral. The remaining propagation integral gives the probability that the photon undergoes another scattering before reaching a free boundary. Hence,
\begin{equation}
    M_n=P_n+M_{n+1},
    \qquad n\geq1.
    \label{eq:Mn_probability_balance}
\end{equation}
Thus, after the $n$-th scattering, a photon either escapes before the next interaction, contributing to $P_n$, or survives to undergo the $(n+1)$-th scattering, contributing to $M_{n+1}$.

For a unit-normalized seed-photon population, the same argument applied before the first scattering gives
\begin{equation}
    M_1=1-P_0.
    \label{eq:M1_probability_balance}
\end{equation}
Combining (\ref{eq:Mn_probability_balance}) and (\ref{eq:M1_probability_balance}) gives
\begin{equation}
    \sum_{n=0}^{N}P_n
    =
    1-M_{N+1}.
    \label{eq:finite_probability_sum}
\end{equation}
Section~\ref{sec:polarized_eigenmodes} shows that repeated application of the finite-slab transport operator drives the post-scattering state toward the dominant eigenmode, as given by (\ref{eq:high_order_dominant_mode}). Since the principal eigenvalue satisfies $\lambda_0<1$ for a finite slab, the amplitude of the surviving post-scattering population decreases geometrically with scattering order, so that $M_n\rightarrow0$ as $n\to\infty$. Therefore (\ref{eq:Mn_probability_balance}) gives
\begin{equation}
    \sum_{n=0}^{\infty}P_n=1,
    \label{eq:total_probability_normalization}
\end{equation}
showing explicitly that the Stokes-vector recursion remains photon-number conserving.

The mean number of scatterings before escape is 
\begin{equation}
    \langle N\rangle
    =
    \sum_{n=0}^{\infty} nP_n.
    \label{eq:mean_scattering_definition}
\end{equation}
Using $P_n=M_n-M_{n+1}$ and rearranging the convergent series gives
\begin{equation}
    \langle N\rangle
    =
    \sum_{n=1}^{\infty}M_n.
    \label{eq:mean_scattering_survival_sum}
\end{equation}
The mean scattering number is therefore the accumulated survival probability over scattering order.

This relation also permits the infinite scattering-order sum to be evaluated directly in operator form. For a general coefficient vector $\bm y(t)=[a(t),b(t)]^{\rm T}$, we define the survival projection operator
\begin{equation}
    (\mathcal{M}\bm y)(t')
    =
    2\int_0^\tau
    \left[
        a(t)+\frac{1}{3}b(t)
    \right]{\rm d}t.
    \label{eq:survival_functional}
\end{equation}
The scattering-order survival probability can then be written as
\begin{equation}
    M_n=\mathcal{M}\bm y_n.
    \label{eq:Mn_ab_stokes}
\end{equation}
Therefore (\ref{eq:mean_scattering_survival_sum}) can be expressed as 
\begin{equation}
    \langle N\rangle
    =
    \mathcal{M}
    \left[
        \sum_{n=1}^{\infty}
        \mathbfss K_{\rm pol}^{\,n-1}\bm y_1
    \right]=
    \mathcal{M}
    \left[
        (\mathbfss I-\mathbfss K_{\rm pol})^{-1}
        \bm y_1
    \right].
    \label{eq:mean_scattering_resolvent_survival}
\end{equation}

An equivalent representation follows directly from the scattering-order escape probabilities. We define the escape projection operator
\begin{equation}
    \begin{aligned}
    \mathcal{E}[\bm y]
    =
    \int_0^\tau {\rm d}t\,
    \Big\{
        &a(t)\left[
            E_2(t)+E_2(\tau-t)
        \right]
        \\
        &+
        b(t)\left[
            E_4(t)+E_4(\tau-t)
        \right]
    \Big\},
\end{aligned}
\label{eq:escape_functional}
\end{equation}
so that
\begin{equation}
    P_n
    =
    \mathcal{E}[\bm y_n].
\end{equation}
The mean scattering number in (\ref{eq:mean_scattering_definition}) can alternatively be written as
\begin{equation}
    \langle N\rangle
    =
    \mathcal{E}
    \left[
        (\mathbfss I-\mathbfss K_{\rm pol})^{-2}
        \bm y_1
    \right].
    \label{eq:mean_scattering_resolvent_escape}
\end{equation}

(\ref{eq:mean_scattering_resolvent_survival}) and (\ref{eq:mean_scattering_resolvent_escape}) are identical. The first form follows naturally from the survival-probability representation, whereas the second is the continuous-operator counterpart of the infinite-order escape-probability summation used in Paper I.

Upon discretization in depth, the same relation becomes
\begin{equation}
    \langle N\rangle
    =
    \bm m^{\rm T}
    (\mathbfss I-\mathbfss K_{\rm pol})^{-1}\bm U_1
    =
    \bm e_{\rm pol}^{\rm T}
    (\mathbfss I-\mathbfss K_{\rm pol})^{-2}\bm U_1,
    \label{eq:mean_scattering_discrete_resolvent}
\end{equation}
where $\bm m$ and $\bm e_{\rm pol}$ are the discrete survival and
escape projections, respectively. This provides a direct numerical
evaluation of $\langle N\rangle$ without explicitly iterating the
recursion to arbitrarily high scattering order.

\begin{figure*}
	\includegraphics[width = 0.75\textwidth]{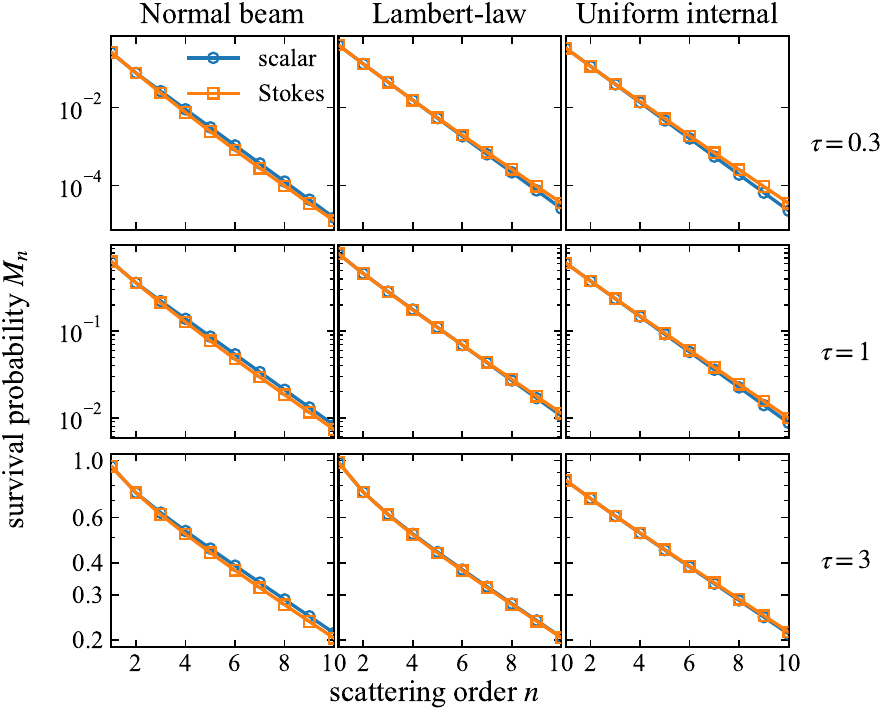}
\caption{Scattering-order survival probabilities $M_n$ obtained as functions of scattering order for the normally incident beam, Lambert-law boundary source, and vertically uniform isotropic internal source. Rows correspond to slab optical depths $\tau=0.3$, $1$, and $3$. Curves with different colours correspond to different recursive schemes.}
\label{fig:Mn_scalar_stokes}
\end{figure*}

Figure~\ref{fig:Mn_scalar_stokes} compares $M_n$ obtained from the present Stokes recursion with the corresponding scalar results of Paper I for the three source configurations considered in this work. The two calculations remain close, but are not generally identical, with the differences becoming most apparent in the high-order tail. For initially unpolarized seed photons, polarization generated during scattering subsequently feeds back into the angular redistribution of the intensity and thereby modifies the probability of surviving to later scattering orders. The increasing relative separation in the high-order tail is also consistent with the slightly different principal eigenvalues of the scalar and Stokes recursion operators discussed in Section~\ref{sec:high_order_polarization}.

\subsection{Adjoint Stokes response and the exact Lambert-law identity}
\label{app:adjoint_stokes}

The resolvent representation derived above provides a direct evaluation of the mean scattering number for any prescribed first-scattering state. A complementary and source-independent interpretation was introduced in Paper I through an adjoint transport formulation. Here we generalize this construction to the Stokes-vector recursion by introducing the source-independent adjoint mean-scattering response
\begin{equation}
    \bm{\mathcal X}(t,\mu)
    =
    \begin{pmatrix}
        \mathcal X_I(t,\mu)\\
        \mathcal X_Q(t,\mu)
    \end{pmatrix},
\end{equation}
which is the Stokes-vector generalization of the scalar adjoint response $Q$ introduced in Paper I, with the same interpretation as the source-independent response governing the expected number of future scatterings. We also introduce the intensity basis vector
\begin{equation}
    \bm e_I
    =
    \begin{pmatrix}
        1\\
        0
    \end{pmatrix}.
\end{equation}

A one-step backward decomposition gives the integral equation
\begin{align}
    \bm{\mathcal X}(t,\mu)
    =
    \int_0^\tau {\rm d}t'\,
    G(t',t;\mu)
    \Bigg[
        \bm e_I
        +
        \int_{-1}^{1}{\rm d}\mu'\,
        \mathbfss P^{\rm T}(\mu',\mu)
        \bm{\mathcal X}(t',\mu')
    \Bigg].
    \label{eq:adjoint_stokes_integral}
\end{align}
The first term inside the brackets counts the next scattering event, whereas the second term gives the expected number of subsequent scatterings after the Stokes state has been redistributed into the outgoing direction $\mu'$.

Differentiating (\ref{eq:adjoint_stokes_integral}) along the propagation characteristic gives the adjoint Stokes transport equation
\begin{equation}
    \mu\frac{\partial\bm{\mathcal X}}{\partial t}
    =
    \bm{\mathcal X}
    -
    \left[\bm e_I
    +
    \int_{-1}^{1}{\rm d}\mu'\,
    \mathbfss P^{\rm T}(\mu',\mu)
    \bm{\mathcal X}(t,\mu')\right].
    \label{eq:adjoint_stokes_transport}
\end{equation}
The corresponding free-escape boundary conditions are
\begin{equation}
    \begin{aligned}
        &\bm{\mathcal X}(0,\mu)=0,
    \qquad \mu<0,\\
    &\bm{\mathcal X}(\tau,\mu)=0,
    \qquad \mu>0.
    \end{aligned}
    \label{eq:adjoint_stokes_boundary}
\end{equation}
A photon located at either boundary and propagating outward escapes immediately and therefore undergoes no further scatterings.

For a general normalized seed-photon source $\bm\Psi_0(t,\mu)$, the mean scattering number can be expressed as the adjoint source projection
\begin{equation}
    \langle N\rangle
    =
    \int_0^\tau {\rm d}t
    \int_{-1}^{1}{\rm d}\mu\,
    \bm{\mathcal X}^{\rm T}(t,\mu)
    \bm\Psi_0(t,\mu).
    \label{eq:mean_scattering_adjoint_source}
\end{equation}
For the initially unpolarized sources considered in this work, only $\mathcal X_I$ contributes directly to the source projection. 

Taking the intensity component of (\ref{eq:adjoint_stokes_transport}) gives
\begin{equation}
\begin{aligned}
\mu\frac{\partial\mathcal X_I}{\partial t}
    =&
    \mathcal X_I-1
    -\int_{-1}^{1}
    P_{II}(\mu',\mu)\mathcal X_I(t,\mu')\,{\rm d}\mu'\\
    &-
    \int_{-1}^{1}
    P_{QI}(\mu',\mu)\mathcal X_Q(t,\mu')\,{\rm d}\mu'.
\end{aligned}
\label{eq:adjoint_intensity_component}
\end{equation}
Integrating this equation over the full slab and over $-1<\mu<1$, and applying the free-escape boundary
conditions yield the source-independent boundary identity
\begin{equation}
    \int_0^1
    \mu\,\mathcal X_I(0,\mu)\,{\rm d}\mu
    -
    \int_{-1}^{0}
    \mu\,\mathcal X_I(\tau,\mu)\,{\rm d}\mu
    =
    2\tau.
    \label{eq:adjoint_boundary_identity}
\end{equation}

The homogeneous slab with identical free boundaries is invariant under the reflection about the slab mid-plane:
\begin{equation}
    t\rightarrow\tau-t,
    \qquad
    \mu\rightarrow-\mu.
\end{equation}
The reduced phase matrix is unchanged by this transformation, and the two free-escape boundary conditions are interchanged. The unique adjoint solution therefore satisfies
\begin{equation}
    \bm{\mathcal X}(t,\mu)
    =
    \bm{\mathcal X}(\tau-t,-\mu).
    \label{eq:adjoint_reflection_symmetry}
\end{equation}
It follows that
\begin{equation}
    \int_{-1}^{0}
    \mu\,\mathcal X_I(\tau,\mu)\,{\rm d}\mu
    =
    -
    \int_0^1
    \mu\,\mathcal X_I(0,\mu)\,{\rm d}\mu.
\end{equation}
(\ref{eq:adjoint_boundary_identity}) therefore reduces to
\begin{equation}
    \int_0^1
    \mu\,\mathcal X_I(0,\mu)\,{\rm d}\mu
    =
    \tau, \qquad \int_{-1}^0
    \mu\,\mathcal X_I(\tau,\mu)\,{\rm d}\mu
    =
    -\tau.
    \label{eq:adjoint_boundary_moment}
\end{equation}

As in the scalar recursion of Paper I, the mean scattering number for lower-boundary Lambert injection retains a simple analytic expression that is valid at arbitrary optical depth. For an initially unpolarized Lambert-law source injected through the lower boundary, the angular distribution is $p_I(\mu)=2\mu$, using (\ref{eq:mean_scattering_adjoint_source}) and (\ref{eq:adjoint_boundary_moment}), we immediately obtain
\begin{equation}
    \langle N\rangle_{\rm L}=2\tau.
    \label{eq:lambert_exact_mean_scattering}
\end{equation}
Thus the exact Lambert-law relation derived in Paper I remains unchanged in the full Stokes recursion.

The other two source geometries considered in this work project onto different parts of the adjoint response and therefore do not follow directly from the global boundary moment. For a normally incident, initially unpolarized beam entering through the lower boundary,
\begin{equation}
    \langle N\rangle_{\rm B}
    =
    \mathcal X_I(0,1).
    \label{eq:beam_mean_adjoint}
\end{equation}
For the vertically uniform and isotropic internal source, the mean scattering number is instead the depth- and angle-averaged adjoint response,
\begin{equation}
    \langle N\rangle_{\rm int}
    =
    \frac{1}{2\tau}
    \int_0^\tau {\rm d}t
    \int_{-1}^{1}{\rm d}\mu\,
    \mathcal X_I(t,\mu).
    \label{eq:internal_mean_adjoint}
\end{equation}
Unlike the Lambert-law case, neither $\mathcal X_I(0,1)$ nor the volume-averaged response is fixed by the global identity alone. Their behaviour depends on the detailed structure of the adjoint solution. However, simple expressions for the two cases also exist in the optically thin and optically thick limits, which are analysed below.

\subsubsection{Optically thin limit}
\label{app:thin_scattering_statistics}

In the optically thin limit, the mean scattering number given by (\ref{eq:mean_scattering_survival_sum}) is dominated by the first few terms of $M_n$. For the initially unpolarized seed photons considered here, the seed Stokes state has no $Q$ component. The intensity distribution produced by the first scattering is therefore identical to that obtained with the polarization-averaged scalar Thomson kernel, $\Psi_{I,1}^{\rm pol}(t,\mu)=\Psi_{1}^{\rm sc}(t,\mu)$. It follows immediately that $M_1^{\rm pol}=M_1^{\rm sc}$.

The equality also extends to the second survival probability. Using the photon-number normalization condition (\ref{eq:phase_matrix_normalization}), the probability that the first-scattered radiation undergoes one further scattering is
\begin{equation}
    M_2
    =
    \int_0^\tau {\rm d}t
    \int_{-1}^{1}{\rm d}\mu\,
    \Psi_{I,1}(t,\mu)
    \int_0^\tau {\rm d}t'\,
    G(t',t;\mu).
    \label{eq:M2_first_intensity}
\end{equation}
The polarized component generated at the first scattering therefore does not contribute directly to $M_2$. Therefore, $M_2^{\rm pol}=M_2^{\rm sc}$ also holds for arbitrary slab optical depth and for any initially unpolarized seed-photon source geometry.

This equality does not extend to higher scattering orders. The linear polarization generated at the first scattering contributes to the intensity redistribution at the second scattering. Consequently, $\Psi_{I,2}$ is generally different from its scalar counterpart, and the first difference in the survival probabilities appears at $M_3$.

For a normally incident beam, the first-scattering state is given by (\ref{eq:vertical_beam_ab}). Using (\ref{eq:Mn_ab_stokes}), we can obtain 
\begin{equation}
    M_1^{\rm B}
    =
    1-e^{-\tau}
    =
    \tau-\frac{\tau^2}{2}+O(\tau^3)
    \label{eq:beam_M1_thin}
\end{equation}
in the optically thin limit. The second survival probability for $\tau\ll1$ can also be written as
\begin{equation}
    M_2^{\rm B}
    =
    \frac{3}{8}\tau^2\ln\frac{1}{\tau}
    +
    \left(
        \frac{3}{4}
        -\frac{3\gamma_{\rm E}}{8}
    \right)\tau^2
    +O\left(\tau^3\ln\frac{1}{\tau}\right).
    \label{eq:beam_M2_thin}
\end{equation}
Considering the first two terms of $M_n$, we obtain
\begin{equation}
    \langle N\rangle_{\rm B}^{\rm pol}
    \simeq
    \tau
    +\frac{3}{8}\tau^2\ln\frac{1}{\tau}
    +
    \left(
        \frac{1}{4}
        -\frac{3\gamma_{\rm E}}{8}
    \right)\tau^2.
    \label{eq:beam_mean_thin}
\end{equation}

For the vertically uniform isotropic internal source, the first-scattering state is given by (\ref{eq:ab1_uniform_isotropic_source}). We can expand the survival probabilities in the optically thin limit and obtain
\begin{equation}
    M_1^{\rm int}
    =
    \frac{\tau}{2}
    \left(
        \ln\frac{1}{\tau}+\frac{3}{2}-\gamma_{\rm E}
    \right)
    +O(\tau^2).
    \label{eq:internal_M1_thin}
\end{equation}
and
\begin{equation}
    M_2^{\rm int}
    =
    \frac{9}{32}\tau^2\ln^2\frac{1}{\tau}
    +O(\tau^2\ln\frac{1}{\tau}).
    \label{eq:internal_M2_thin}
\end{equation}
The mean scattering number therefore gives
\begin{equation}
    \langle N\rangle_{\rm int}^{\rm pol}
    \simeq
    \frac{\tau}{2}
    \left(
        \ln\frac{1}{\tau}+\frac{3}{2}-\gamma_{\rm E}
    \right)
    +
    \frac{9}{32}\tau^2\ln^2\frac{1}{\tau}.
    \label{eq:internal_mean_thin}
\end{equation}

\subsubsection{Optically thick limit}
\label{app:thick_scattering_statistics}

In the optically thick limit, as shown in Section~\ref{sec:thick_limit}, the long-wavelength conservative branch of the Stokes transport operator approaches an isotropic and unpolarized state, with diffusion coefficient $D=1/3$. The corresponding adjoint response in the slab interior may therefore be expanded, to leading diffusion order, as
\begin{equation}
    \mathcal X_{I}(t,\mu)
    \simeq       X_0(t)+\mu X_1(t), \qquad \mathcal X_{Q}(t,\mu) \simeq 0.
    \label{eq:adjoint_bulk_diffusion_expansion}
\end{equation}

Substituting (\ref{eq:adjoint_bulk_diffusion_expansion}) into the adjoint transport equation (\ref{eq:adjoint_stokes_transport}), and using the photon-number normalization and parity of the Thomson kernel, we obtain
\begin{equation}
    \mu\frac{{\rm d}X_0}{{\rm d}t}
    +
    \mu^2\frac{{\rm d}X_1}{{\rm d}t}
    =
    \mu X_1-1.
    \label{eq:adjoint_diffusion_angular}
\end{equation}
Taking the zeroth and first angular moments yields
\begin{equation}
    \frac{1}{3}\frac{{\rm d}X_1}{{\rm d}t}
    =
    -1,
    \qquad
    \frac{{\rm d}X_0}{{\rm d}t}
    =
    X_1,
    \label{eq:adjoint_diffusion_moments}
\end{equation}
and hence
\begin{equation}
    \frac{1}{3}\frac{{\rm d}^2X_0}{{\rm d}t^2}
    =
    -1.
    \label{eq:adjoint_diffusion_equation}
\end{equation}

Reflection symmetry about the slab mid-plane requires $X_1(\tau/2)=0$. (\ref{eq:adjoint_diffusion_moments}) therefore gives
\begin{equation}
    X_1(t)
    =
    \frac{3}{2}\tau-3t, \qquad
    X_0(t)
    =
    \frac{3}{2}t(\tau-t)+C_\tau.
    \label{eq:X_thick}
\end{equation}
The integration constant $C_\tau$ is determined by matching the interior diffusion solution to the polarized surface layer. The adjoint diffusion mode has the same leading extrapolation length as the forward polarized Milne problem analysed in Section~\ref{sec:thick_limit}, i.e., $z_0\simeq0.7121$. At leading order in $\tau$, the corresponding boundary conditions may be written as
\begin{equation}
    X_0(0)-z_0X_0'(0)=O(1),
    \qquad
    X_0(\tau)+z_0X_0'(\tau)=O(1),
    \label{eq:adjoint_diffusion_boundary}
\end{equation}
which give
\begin{equation}
    C_\tau
    =
    \frac{3}{2}z_0\tau+O(1).
\end{equation}
The bulk adjoint response is therefore
\begin{equation}
    X_0(t)
    =
    \frac{3}{2}
    \left[
        t(\tau-t)+z_0\tau
    \right]
    +O(1).
    \label{eq:X0_thick_solution}
\end{equation}

For the vertically uniform isotropic internal source, the mean scattering number is the depth- and angle-averaged adjoint response given by (\ref{eq:internal_mean_adjoint}). The odd angular term $\mu X_1$ vanishes upon angular integration, and the surface-localized corrections contribute only at $O(1)$ after volume averaging. We therefore obtain 
\begin{equation}
    \langle N\rangle_{\rm int}^{\rm pol}
    =
    \frac{1}{\tau}
    \int_0^\tau X_0(t)\,{\rm d}t
    +O(1)\simeq
    \frac{\tau^2}{4}
    +
    \frac{3}{2}z_0\tau.
    \label{eq:internal_mean_thick}
\end{equation}
The leading $\tau^2/4$ term is determined entirely by bulk diffusion, whereas the $O(\tau)$ correction arises from the displacement of the effective diffusion boundary by the polarized Milne surface layer.

The normally incident beam differs qualitatively because it samples the adjoint response at a single boundary state, $\langle N\rangle_{\rm B}=\mathcal X_I(0,1)$, rather than averaging over the slab interior. The bulk diffusion expansion cannot therefore be extrapolated directly to the surface layer. In the surface matching region $1\ll t\ll\tau$, the bulk diffusion
solution gives
\begin{equation}
    \bm{\mathcal X}_{\rm bulk}(t,\mu)
    =
    \frac{3}{2}\tau
    (t+z_0+\mu)\bm e_I
    +O(1).
\end{equation}
Thus, the amplitude of the adjoint response within an $O(1)$ optical depth of a free surface should be of order $\tau$. We therefore introduce the lower-surface half-space response by 
\begin{equation}
    \bm{\mathcal X}(t,\mu;\tau)
    =
    \tau\,\bm{\mathcal X}_{\rm H}(t,\mu)
    +O(1),
    \qquad
    t=O(1),
    \label{eq:adjoint_halfspace_scaling}
\end{equation}
where $\bm{\mathcal X}_{\rm H}$ satisfies the matching condition
\begin{equation}
    \bm{\mathcal X}_{\rm H}(t,\mu)
    \sim
    \frac{3}{2}
    \left(
        t+z_0+\mu
    \right)\bm e_I,
    \qquad
    t\rightarrow\infty.
    \label{eq:adjoint_halfspace_matching}
\end{equation}

Substitution of (\ref{eq:adjoint_halfspace_scaling}) into (\ref{eq:adjoint_stokes_transport}) shows that the leading half-space response satisfies
\begin{equation}
    \mu\frac{\partial\bm{\mathcal X}_{\rm H}}{\partial t}
    =
    \bm{\mathcal X}_{\rm H}
    -
    \int_{-1}^{1}{\rm d}\mu'\,
    \mathbfss P^{\rm T}(\mu',\mu)
    \bm{\mathcal X}_{\rm H}(t,\mu'),
    \label{eq:adjoint_halfspace_equation}
\end{equation}
with the free-escape condition
\begin{equation}
    \bm{\mathcal X}_{\rm H}(0,\mu)=0,
    \qquad
    \mu<0.
    \label{eq:adjoint_halfspace_boundary}
\end{equation}

(\ref{eq:adjoint_halfspace_matching})--(\ref{eq:adjoint_halfspace_boundary}) define a source-independent half-space polarized adjoint Milne problem. The asymptotic beam mean scattering number is consequently
\begin{equation}
    \langle N\rangle_{\rm B}^{\rm pol}
    =
    C_{\rm B}^{\rm pol}\tau+O(1),
    \qquad
    C_{\rm B}^{\rm pol}
    =
    \mathcal X_{{\rm H},I}(0,1).
    \label{eq:beam_mean_thick}
\end{equation}
A numerical solution of the half-space problem gives
\begin{equation}
    C_{\rm B}^{\rm pol}\simeq 2.5388.
\end{equation}

The two source geometries therefore probe the thick-slab transport in different ways. For a uniform internal source, most photons are produced in the bulk and the leading $\tau^2/4$ behaviour is fixed by the conservative diffusion branch. By contrast, a boundary beam directly samples the polarized Milne layer, so that the surface response enters in the leading $O(\tau)$ coefficient. 

\begin{figure*}
	\includegraphics[width = 0.75\textwidth]{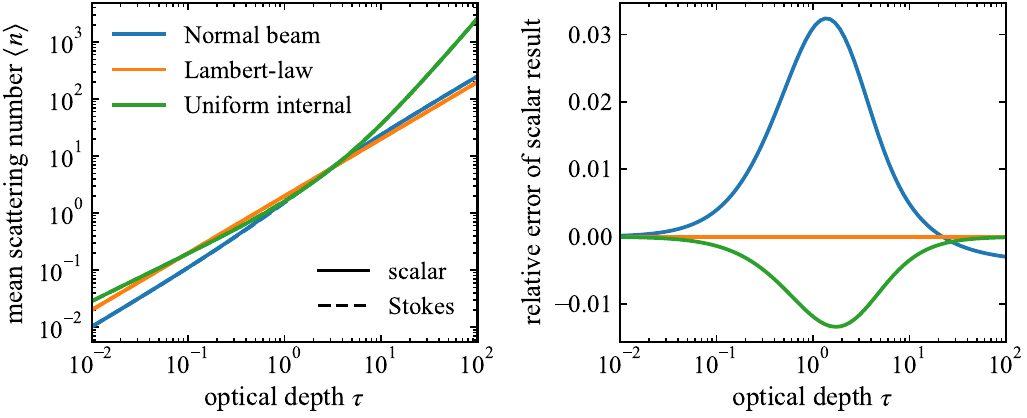}
    \caption{
Left: mean scattering number as a function of slab optical depth for normally incident beam injection, Lambert-law injection, and a vertically uniform isotropic internal source. Solid and dashed curves denote the scalar recursion of Paper I and the present Stokes-vector recursion, respectively. Right: relative error of the scalar result with respect to the Stokes result, $(\langle N\rangle_{\rm sc}-\langle N\rangle_{\rm pol})/ \langle N\rangle_{\rm pol}$.}
\label{fig:mean_scattering_scalar_stokes}
\end{figure*}

The full numerical results are summarized in Figure~\ref{fig:mean_scattering_scalar_stokes}. The left panel shows the mean scattering number over a broad range of slab optical depths, while the right panel isolates the relative difference between the scalar and Stokes treatments. The numerical solutions recover the optically thin and thick asymptotic behaviours derived above. Although the inclusion of polarization introduces small but systematic differences in the mean scattering number, the scalar recursion of Paper I still reproduces the overall dependence of $\langle N\rangle$ on optical depth remarkably well for all three source geometries.

\section{Derivation and accuracy of the finite-optical-depth Galerkin approximation}
\label{app:galerkin_thin}

This appendix derives the projected eigenproblem used in
Section~\ref{sec:galerkin_thin} and estimates the corrections associated
with the non-uniform depth amplitude. We first obtain the constant-depth
Galerkin operator, then examine amplitude weighting and the escape
projection for the normalized angular distribution.

\subsection{Constant-depth projection}
\label{app:galerkin_projection}

Using the depth-average operator defined in
(\ref{eq:constant_depth_projector_tau}), the constant-depth matrix
element of each kernel is
\begin{equation}
    L_m(\tau)=\mathcal P(\mathcal K_m1)
    =\frac{1}{\tau}\int_0^\tau{\rm d}t'
    \int_0^\tau{\rm d}t\,E_m(|t'-t|).
    \label{eq:Lm_definition}
\end{equation}
Evaluating these integrals gives (\ref{eq:finite_tau_Lm_exact}).
Replacing the depth operators by their constant-mode matrix elements
yields
\begin{equation}
    \overline{\mathbfss K}_{\rm pol}
    =\frac{3}{16}
    \begin{pmatrix}
        3L_1-L_3 & -L_1+5L_3-2L_5\\
        -L_1+3L_3 & 3L_1-7L_3+6L_5
    \end{pmatrix}.
    \label{eq:projected_Kpol}
\end{equation}
The projected dominant eigenmode satisfies
\begin{equation}
    \overline{\mathbfss K}_{\rm pol}
    \begin{pmatrix}1\\\alpha(\tau)\end{pmatrix}
    =\lambda_{\rm G}
    \begin{pmatrix}1\\\alpha(\tau)\end{pmatrix},
    \label{eq:projected_eigenproblem}
\end{equation}
where $\lambda_{\rm G}$ is the Galerkin eigenvalue. Eliminating it gives
\begin{equation}
(-L_1+5L_3-2L_5)\alpha^2(\tau)+6(L_3-L_5)\alpha(\tau)+L_1-3L_3=0.
    \label{eq:alpha_quadratic}
\end{equation}
The root that approaches $-1$ as $\tau\rightarrow0$ is the branch
given in (\ref{eq:alpha_dominant_branch}). Expanding this Galerkin
branch for $\tau\to0$ gives
\begin{equation}
    \alpha(\tau)=-1+\frac{1}{2\ln(1/\tau)}
    +\frac{\gamma_{\rm E}}{2\ln^2(1/\tau)}+\mathcal O\!\left[\frac{1}{\ln^3(1/\tau)}\right].
    \label{eq:qU_thin_expansion}
\end{equation}
Its first logarithmic correction agrees with
(\ref{eq:first_order_stokes_ratio}), establishing consistency with the
thin-slab eigenmode to that order.

\subsection{Amplitude-weighting corrections}
\label{app:galerkin_amplitude}

For the factorized eigenfunction
(\ref{eq:factorized_dominant_eigenfunction}), define the local
amplitude-weighted response
\begin{equation}
    \ell_m(t;\tau)=\frac{(\mathcal K_m A)(t;\tau)}{\mathcal P A},
    \label{eq:local_weighted_kernel}
\end{equation}
and its depth average
\begin{equation}
    \widehat L_m(\tau)=\mathcal P\ell_m
    =\frac{\mathcal P(\mathcal K_m A)}{\mathcal P A}.
    \label{eq:Lm_weighted_definition}
\end{equation}
The lowest-order Galerkin closure replaces $\widehat L_m$ by $L_m$.
To estimate the resulting correction, write
(\ref{eq:thin_common_amplitude}) as
\begin{equation}
    A(t;\tau)=A_0(\tau)+\frac{\phi(t)}{\ln(1/\tau)}
    +\mathcal O\!\left[\frac{1}{\ln^2(1/\tau)}\right],
    \label{eq:A_projected_expansion}
\end{equation}
where $A_0$ contains the depth-independent part.
Substitution into (\ref{eq:Lm_weighted_definition}) gives
\begin{equation}
    \widehat L_m=L_m+
    \frac{\mathcal P(\mathcal K_m\phi)}{A_0\ln(1/\tau)}+
    \left\{\begin{aligned}
        &\mathcal O\left[\frac{\tau}{\ln^2(1/\tau)}\right],m=1,\\
        &\mathcal O\left[\frac{\tau^2}{\ln^2(1/\tau)}\right],m=3,5.
    \end{aligned}\right.
    \label{eq:Lm_weighted_correction}
\end{equation}
Together with $\mathcal P(\mathcal H\phi)=7/12-\pi^2/18\simeq0.0350$, the operator expansions (\ref{eq:K135_thin_projected_expansion}) yield
\begin{equation}
    \begin{aligned}
    \mathcal P(\mathcal K_1\phi)
    &=\tau\mathcal P(\mathcal H\phi)+\mathcal O(\tau^2)
    =\mathcal O(\tau),\\
    \mathcal P(\mathcal K_3\phi)&=\mathcal O(\tau^2),\qquad
    \mathcal P(\mathcal K_5\phi)=\mathcal O(\tau^2).
    \end{aligned}
\end{equation}
Consequently,
\begin{equation}
    \frac{\widehat L_1-L_1}{L_1}
    =\mathcal O\!\left[\frac{1}{\ln^2(1/\tau)}\right],\qquad
    \frac{\widehat L_{3,5}-L_{3,5}}{L_{3,5}}
    =\mathcal O\!\left[\frac{\tau}{\ln(1/\tau)}\right].
\end{equation}
Thus the projected kernel coefficients can be more accurate than an
estimate based on the spatial variation of $A$ alone would suggest.

\begin{figure*}
    \includegraphics[width=0.75\textwidth]{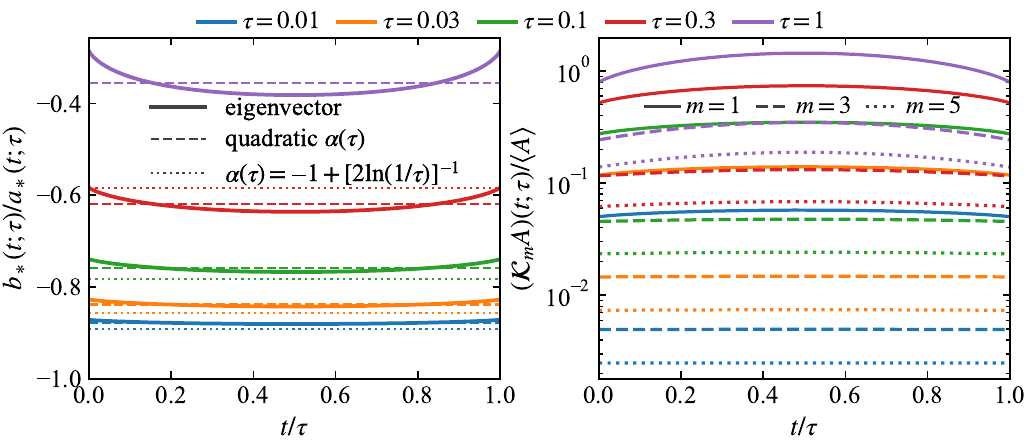}
    \caption{Structure of the principal polarized eigenmode and the
    amplitude-weighted depth-operator responses in the optically thin
    regime. Left: numerical ratio $b_*(t;\tau)/a_*(t;\tau)$ as a function
    of normalized depth $t/\tau$, compared with the Galerkin
    eigendirection (\ref{eq:alpha_dominant_branch}) and the leading
    inverse-logarithmic approximation
    (\ref{eq:thin_stokes_eigendirection}). Right: local responses
    $\ell_m(t;\tau)=(\mathcal K_m A)(t;\tau)/(\mathcal P A)$ for
    $m=1,3,5$. These profiles test the depth dependence underlying the
    constant-mode Galerkin approximation.}
    \label{fig:thin_eigenmode_comparison}
\end{figure*}

The left panel of Figure~\ref{fig:thin_eigenmode_comparison} compares
the numerical ratio $b_*(t;\tau)/a_*(t;\tau)$ with the Galerkin
eigendirection (\ref{eq:alpha_dominant_branch}) and the leading
inverse-logarithmic approximation
(\ref{eq:thin_stokes_eigendirection}). For sufficiently small optical
depth, the numerical ratio is nearly independent of depth and
approaches $-1$. The Galerkin result remains close to the numerical
eigendirection over a wider range of optical depth than the leading
inverse-logarithmic approximation.

The right panel shows $\ell_m(t;\tau)$ for $m=1,3,5$. These responses
are nearly flat in thin slabs even when $A(t;\tau)$ retains a visible
depth dependence. This behaviour provides numerical support for the
constant-depth closure and the suppression of amplitude-weighting
corrections derived above.

\subsection{Escape projection and angular-distribution error}
\label{app:galerkin_escape}

Within the factorized form
(\ref{eq:factorized_dominant_eigenfunction}), the normalized angular
distribution is
\begin{equation}
    \Phi_\infty^\pm(\mu)=
    \frac{[1+\alpha(\tau)\mu^2]
        \displaystyle\int_0^\tau A(t;\tau)e^{-t/\mu}\,{\rm d}t}
    {\displaystyle\int_0^\tau A(t;\tau)
        [E_2(t)+\alpha(\tau)E_4(t)]\,{\rm d}t}.
    \label{eq:thin_angular_general_A}
\end{equation}
For fixed $\mu>0$, $e^{-t/\mu}\rightarrow1$ uniformly across the
slab as $\tau\rightarrow0$, while $E_2(t)\rightarrow1$ and
$E_4(t)\rightarrow1/3$.

To estimate the amplitude correction, direct integration of
(\ref{eq:phi_tau_definition}) gives
\begin{equation}
    \int_0^\tau\phi(t)\,{\rm d}t=0,\qquad
    \int_0^\tau t\phi(t)\,{\rm d}t=0.
    \label{eq:phi_moment_conditions}
\end{equation}
With (\ref{eq:A_projected_expansion}), for $\mu\gg\tau$, expansion of the escape exponential yields
\begin{equation}
    \int_0^\tau A(t;\tau)e^{-t/\mu}\,{\rm d}t
    =A_0(\tau)\mu(1-e^{-\tau/\mu})+\mathcal O\left[
           \frac{\tau^3}{\mu^2\ln(1/\tau)}\right].
\label{eq:A_escape_projection_error}
\end{equation}
Similarly, the small-$t$ expansions of $E_2$ and $E_4$ give
\begin{equation}
    \begin{aligned}
    &\int_0^\tau A(t;\tau)[E_2(t)+\alpha(\tau)E_4(t)]\,{\rm d}t\\
    &=A_0(\tau)\left[
       \left(\frac12-E_3(\tau)\right)
       +\alpha(\tau)\left(\frac14-E_5(\tau)\right)\right]+\mathcal O\!\left[\frac{\tau^2}{\ln(1/\tau)}\right].
    \end{aligned}
    \label{eq:A_normalization_error}
\end{equation}
Keeping the leading terms gives the Galerkin escape distribution
(\ref{eq:thin_angular_galerkin}). For $\tau\rightarrow0$ with
$\mu\gg\tau$, the relative correction due to the non-uniform
amplitude scales as
\begin{equation}
    \frac{\Phi_\infty^\pm(\mu)-\Phi_{\infty,{\rm G}}^\pm(\mu)}
         {\Phi_{\infty,{\rm G}}^\pm(\mu)}
    =\mathcal O\!\left[
        \frac{\tau}{\ln(1/\tau)}
        +\frac{\tau^2}{\mu^2\ln(1/\tau)}\right].
    \label{eq:thin_angular_error_scaling}
\end{equation}
This estimate concerns the common depth amplitude within the
constant-eigendirection approximation. It does not include errors due
to a depth-dependent Stokes ratio, and the expansion of the escape
exponential used here is not uniform in the grazing region
$\mu\lesssim\tau$.

\section{Half-space construction of the finite-thickness surface corrections}
\label{app:surface_asymptotics}

This appendix gives the half-space construction underlying
Section~\ref{sec:thick_surface}. We derive the bulk-matching conditions,
separate the localized boundary response from the growing far-field
terms, and express the observable correction functions.

\subsection{Half-space equations and matching to the diffusion mode}
\label{app:surface_matching}

For fixed depths $t,t'=O(1)$ near the lower surface, the upper boundary
recedes to infinity as $\tau\rightarrow\infty$. The depth operators
then become
\begin{equation}
    (\mathcal K_{m,{\rm H}}f)(t')
    =\int_0^\infty E_m(|t'-t|)f(t)\,{\rm d}t.
    \label{eq:polarized_milne_source}
\end{equation}
The polarized half-space operator $\mathbfss K_{\rm H}$ is obtained by
replacing each $\mathcal K_m$ in $\mathbfss K_{\rm pol}$ with
$\mathcal K_{m,{\rm H}}$. Neglecting direct propagation from the
opposite surface, the local eigenproblem becomes
\begin{equation}
    \lambda_0(\tau)\bm u_0=\mathbfss K_{\rm H}\bm u_0,
    \label{eq:finite_surface_eigenproblem}
\end{equation}
where the bulk region and the surface layers share the same principal eigenvalue.
The local correction from the remote surface scales schematically
as $O(e^{-\tau})$, up to algebraic prefactors.

The bulk diffusion eigenvalue through fourth order is
\begin{equation}
    \lambda_{\rm D}(k)=1-\frac{k^2}{3}
    +\frac{31}{135}k^4+O(k^6).
    \label{eq:lambda_D_k4}
\end{equation}
Substituting this expression and
(\ref{eq:surface_eigenfunction_expansion_k4}) into
(\ref{eq:finite_surface_eigenproblem}) gives
(\ref{eq:polarized_milne}) at zeroth and second order.
At fourth order, we obtain
\begin{equation}
    (\mathbfss I-\mathbfss K_{\rm H})\bm U_4
    =\frac13\bm U_2-\frac{31}{135}\bm U_0.
    \label{eq:forced_polarized_milne_k4}
\end{equation}
These equations must be supplemented by matching to the bulk
diffusion solution.

The far-field matching conditions follow from the surface phase
length $z(k)$ and the normalized bulk envelope $A_{\rm sur}(t;k)$
introduced in Section~\ref{sec:thick_surface}. Combining the
expansion of (\ref{eq:surface_amplitude_expansion}) with
(\ref{eq:diffusion_decay_mode}) recovers the zeroth- and
second-order matching polynomials in
(\ref{eq:P0P2_definition}). Further retaining the fourth-order terms gives 
\begin{equation}
    \bm P_4(t)
    =
    \begin{pmatrix}
        z_4-{z_2X^2}/{2}+{X^5}/{120}\\
        -{z_2}/{3}+{X^3}/{18}
        +{20X}/{189}
    \end{pmatrix},
    \label{eq:P4_definition}
\end{equation}
where $z_4$ denotes the fourth-order coefficient
in the expansion of $z(k)$. The matching holds in the overlap
region $1\ll t\ll\tau$, with $kt\ll1$.

The surface phase coefficients can be calibrated against numerical
finite-slab eigenmodes. For each numerical principal eigenvalue,
we infer $k$ from the bulk dispersion relation and define
\begin{equation}
    z_{\rm eff}(k)
    =
    \frac{\pi}{2k}-\frac{\tau}{2}.
    \label{eq:surface_phase_calibration}
\end{equation}
Extrapolating the depth discretization and fitting
$z_{\rm eff}(k)$ in powers of $k^2$ gives the coefficients
$z_0$ and $z_2$ quoted in (\ref{eq:z0z2_numerical}).
The numerical procedure and convergence checks are described
in the accompanying notebook.

\subsection{Localized boundary response}
\label{app:surface_response}

The matching polynomials collect the non-decaying far-field terms.
Separate them from the surface-localized corrections by writing
\begin{equation}
    \bm U_j=\bm P_j+\bm R_j,
    \qquad j=0,2,4,\cdots
    \label{eq:R0R2_decomposition}
\end{equation}
where
\begin{equation}
    \bm R_j(t)\to0
    \quad\text{as }t\to\infty.
    \label{eq:R_decay}
\end{equation}
The functions $\bm R_j$ describe the localized boundary response
absent from the translationally invariant bulk solution.

To express their forcing, define the part of the infinite-medium
operator excluded from the half-space domain:
\begin{equation}
    (\mathcal K_{m,-}f)(t')
    =\int_{-\infty}^{0}E_m(|t'-t|)f(t)\,{\rm d}t.
    \label{eq:Kminus_scalar}
\end{equation}
The corresponding polarized operator $\mathbfss K_-$ is constructed
by the same replacements in $\mathbfss K_{\rm pol}$. We define the polarized infinite-space operator
$\mathbfss K_\infty=\mathbfss K_{\rm H}+\mathbfss K_-$.
The matching polynomials are generated by the infinite-medium
diffusion mode and therefore satisfy
\begin{equation}
    \begin{aligned}
    (\mathbfss I-\mathbfss K_\infty)\bm P_0
        &=0,\\
    (\mathbfss I-\mathbfss K_\infty)\bm P_2
        &=\frac13\bm P_0,\\
    (\mathbfss I-\mathbfss K_\infty)\bm P_4
        &=\frac13\bm P_2-\frac{31}{135}\bm P_0.
    \end{aligned}
    \label{eq:bulk_polynomial_identities}
\end{equation}
Substituting (\ref{eq:R0R2_decomposition}) into the half-space
equations and using these identities gives
\begin{equation}
    \begin{aligned}
    (\mathbfss I-\mathbfss K_{\rm H})\bm R_0
        &=-\mathbfss K_-\bm P_0,\\
    (\mathbfss I-\mathbfss K_{\rm H})\bm R_2
        &=\frac13\bm R_0-\mathbfss K_-\bm P_2,\\
    (\mathbfss I-\mathbfss K_{\rm H})\bm R_4
        &=\frac13\bm R_2
        -\frac{31}{135}\bm R_0
        -\mathbfss K_-\bm P_4.
    \end{aligned}
    \label{eq:localized_response_equations}
\end{equation}
Unlike the growing matching polynomials, the boundary-response
functions decay away from the surface. This formulation is
therefore well suited to numerical half-space calculations.

\subsection{Escape projections and observable coefficients}
\label{app:surface_observables}

Let $a_j^*$ and $b_j^*$ be the first and second components of
$\bm U_j(t)$, respectively. Their escape projections are
\begin{equation}
    \begin{aligned}
    &I_{j,{\rm H}}(\mu)
    =\int_0^\infty
        [a_j^*(t)+\mu^2b_j^*(t)]e^{-t/\mu}\,{\rm d}t,\\
    &Q_{j,{\rm H}}(\mu)
    =(1-\mu^2)\int_0^\infty b_j^*(t)e^{-t/\mu}\,{\rm d}t,
    \end{aligned}\qquad j=0,2,4.
    \label{eq:IQ_escape}
\end{equation}
The escaping Stokes components therefore expand as
\begin{equation}
    \begin{aligned}
    &I_*(\mu;\tau)=I_{0,{\rm H}}(\mu)
    +k_1^2I_{2,{\rm H}}(\mu)+k_1^4I_{4,{\rm H}}(\mu)+O(k_1^6),\\
    &Q_*(\mu;\tau)=Q_{0,{\rm H}}(\mu)
    +k_1^2Q_{2,{\rm H}}(\mu)+k_1^4Q_{4,{\rm H}}(\mu)+O(k_1^6).
\end{aligned}
\end{equation}
For the normalized angular distribution, define
\begin{equation}
    F_{j,{\rm H}}=\int_0^1 I_{j,{\rm H}}(\mu)\,{\rm d}\mu,
    \qquad j=0,2,4.
    \label{eq:surface_F_definition}
\end{equation}

Expanding $Q_*/I_*$ gives
(\ref{eq:thick_pol_expansion}), with
\begin{equation}
\begin{aligned}
        &\Pi_{\rm Ch}(\mu)=\Pi_0(\mu)
    =\frac{Q_{0,{\rm H}}(\mu)}{I_{0,{\rm H}}(\mu)},\\
    &\Delta\Pi_2(\mu)
    =\frac{Q_{2,{\rm H}}(\mu)}{I_{0,{\rm H}}(\mu)}
    -\Pi_{\rm Ch}(\mu)\frac{I_{2,{\rm H}}(\mu)}{I_{0,{\rm H}}(\mu)},\\
    &\Delta\Pi_4(\mu)=
    \frac{Q_{4,{\rm H}}(\mu)}{I_{0,{\rm H}}(\mu)}
    -\Pi_{\rm Ch}(\mu)\frac{I_{4,{\rm H}}(\mu)}{I_{0,{\rm H}}(\mu)}-\Delta\Pi_2(\mu)
       \frac{I_{2,{\rm H}}(\mu)}{I_{0,{\rm H}}(\mu)}.
\end{aligned}
\end{equation}

Similarly, the normalized angular distribution has the similar expansion given by
(\ref{eq:thick_ang_expansion}), with
\begin{equation}
    \begin{aligned}
    &\Phi_{\rm thick}(\mu)=\frac{I_{0,{\rm H}}(\mu)}{F_{0,{\rm H}}},\\
    &\Delta\Phi_2(\mu)=\frac{I_{2,{\rm H}}(\mu)}{F_{0,{\rm H}}}
       -\Phi_{\rm thick}(\mu)\frac{F_{2,{\rm H}}}{F_{0,{\rm H}}},\\
    &\Delta\Phi_4(\mu)=\frac{I_{4,{\rm H}}(\mu)}{F_{0,{\rm H}}}
       -\Phi_{\rm thick}(\mu)\frac{F_{4,{\rm H}}}{F_{0,{\rm H}}}-\Delta\Phi_2(\mu)\frac{F_{2,{\rm H}}}{F_{0,{\rm H}}}.
    \end{aligned}
    \label{eq:surface_angular_coefficients}
\end{equation}
Integration over $\mu$ gives the normalization identities
(\ref{eq:ang_normalization_thick}).

%%%%%%%%%%%%%%%%%%%%%%%%%%%%%%%%%%%%%%%%%%%%%%%%%%

% Don't change these lines
\bsp	% typesetting comment
\label{lastpage}
\end{document}